\documentclass[useAMS,usenatbib]{mnras}

\usepackage{color}

\usepackage{amssymb}
\usepackage{amsmath}
\usepackage{bm}
\usepackage{color}
\usepackage{mathrsfs}
\usepackage{graphicx}

\usepackage{epstopdf}

\usepackage[figuresright]{rotating}

\def\vector#1{\mbox{\boldmath $#1$}}

\renewcommand{\vec}[1]{\ensuremath{\mathbf{#1}}}

\newcommand{\Myr}{\ensuremath{\,\mathrm{Myr}}}
\newcommand{\Gyr}{\ensuremath{\,\mathrm{Gyr}}}
\newcommand{\kpc}{\ensuremath{\,\mathrm{kpc}}}
\newcommand{\pc}{\ensuremath{\,\mathrm{pc}}}
\newcommand{\kms}{\ensuremath{\,\mathrm{km\ s}^{-1}}}

\newcommand{\kmskpc}{\ensuremath{\,\mathrm{{km\ s}^{-1}\ {kpc}^{-1}}}}
\newcommand{\Omegabkmskpc}{\ensuremath{\Omega_\mathrm{bar} / (\mathrm{{km\ s}^{-1}\ {kpc}^{-1}}) }}
\newcommand{\Omegaskmskpc}{\ensuremath{\Omega_\mathrm{s} / (\mathrm{{km\ s}^{-1}\ {kpc}^{-1}}) }}
\newcommand{\Omegab}{\ensuremath{\Omega_\mathrm{bar} }}
\newcommand{\Omegas}{\ensuremath{\Omega_\mathrm{s}  }}

\newcommand{\eq}[1]{\begin{align}#1\end{align}}

\newcommand{\tform}{\ensuremath{T_{\mathrm{form}}}}

\title[${\rm [Fe/H]}$ dependence of the Hercules stream] 
{Metallicity dependence of the Hercules stream in Gaia/RAVE data -- explanation by non-closed orbits}
\author[K. Hattori et al.]  
{Kohei Hattori$^1$\thanks{khattori@umich.edu},
Naoteru Gouda$^{2,3}$, 
Hiromichi Tagawa$^{2,4}$,
Nobuyuki Sakai$^2$, 
\and Taihei Yano$^{2,3}$, 
Junichi Baba$^2$, 
Jun Kumamoto$^{2,5,6}$\\
$^{1}$University of Michigan, 1085 S. University Ave, Ann Arbor, MI 48109, USA\\
$^{2}$National Astronomical Observatory of Japan, 2-21-1, Osawa, Mitaka, Tokyo 181-8588, Japan\\
$^{3}$SOKENDAI (The Graduate University for Advanced Students), Shonan Village, Hayama, Kanagawa 240-0193, Japan\\
$^{4}$Institute of Physics, E{\"o}tv{\"o}s University, P{\'a}zm{\'a}ny P.s., Budapest, 1117, Hungary\\
$^{5}$Astronomical Institute, Tohoku University, Sendai, Miyagi 980-8578, Japan\\
$^{6}$Department of Astronomy, Graduate School of Science, The University of Tokyo, 7-3-1 Hongo, Bunkyo-ku, Tokyo, 113-0033, Japan
}

\begin{document}


\pagerange{\pageref{firstpage}--\pageref{lastpage}} \pubyear{20xx}

\maketitle
\label{firstpage}

\begin{abstract}

The origin of the Hercules stream, the most prominent velocity substructure in the Solar neighbour disc stars, is still under debate. 
Recent accurate measurements of position, velocity, and metallicity provided by Tycho Gaia Astrometric Solution (TGAS) and RAdial Velocity Experiments (RAVE) 
have revealed that the Hercules stream is most clearly seen in the metal-rich region ([Fe/H] $\gtrsim 0$), 
while it is not clearly seen in lower metallicity region ([Fe/H] $\lesssim -0.25$). 
By using a large number of chemo-dynamical 2D test-particle simulations with a rotating bar and/or spiral arms, 
we find that the observed [Fe/H] dependence of the Hercules stream 
is a natural consequence of the inside-out formation of the stellar disc 
and the existence of highly non-closed orbits in the rotating frame of the bar or spiral arms. 
Our successful models that reproduce the observed properties of the Hercules stream 
include not only fast-bar-only and fast-bar$+$spiral models, but also slow-bar$+$spiral models. 
This indicates that it is very difficult to estimate the pattern speed of the bar or spiral arms 
based only on the observations of the Hercules stream in the Solar neighbourhood. 
As a by-product of our simulations, 
we make some predictions about the locations across the Galactic plane 
where we can observe velocity bimodality that is not associated with the Hercules stream. 
These predictions can be tested by the 
Gaia Data Release 2, 
and such a test will improve our understanding of the evolution of the Milky Way stellar disc.  

\end{abstract}



\section{Introduction}

Since the discovery of the bimodal velocity distribution of the Solar neighbour disc stars \citep{Raboud1998, Dehnen1998}, 
many authors have tried to explain the origin of the secondary peak, or the Hercules stream. 
The stars in the Hercules stream are characterised by small angular momentum and radially outward mean velocity, 
and their orbits are distinct from the nearly circular orbits of stars in the primary peak (main mode). 
Since the stars in the Hercules stream show a wide range of stellar ages, metallicities, and element abundances \citep{Bensby2007}, 
this stream is believed to be formed through some dynamical process.

The pioneering work of \cite{Dehnen2000} demonstrated that 
the under-dense region between the main mode and the Hercules stream 
can arise from the outer Lindblad resonance (OLR) of a fast rotating bar. 
Based on this idea, \cite{Dehnen1999b} estimated the bar's pattern speed to be $\Omegab = (53 \pm 3) \kmskpc$. 
Dehnen's explanation of the Hercules stream is so simple, sticky, and easy to be reproduced, 
that many authors have refined the measurement of $\Omegab$ by using larger samples of nearby stars; 
and they obtained similar values of $\Omegab$ \citep{Antoja2014,Monari2017b}. 
Although the detailed explanation may differ from authors to authors (e.g., \citealt{Fux2001}), 
the fast-bar models provide the most successful explanation so far.

With the success of bar-only models, it may be natural to try spiral-only models, but none of them are as successful as bar-only models. 
For example, 
\cite{Antoja2009} investigated the effect from steady spiral arms on the local velocity distribution. 
They found that spiral-only models can sometimes produce bimodal velocity distribution, 
but the stars associated with the (Hercules-like) secondary peak have zero radial velocity on average. 
This property of the radial velocity is inconsistent with the observed Hercules stream.\footnote{ 
It may be worthwhile pointing out that some theoretical studies (e.g., \citealt{DeSimone2004, Sellwood2010})
have shown that spirals' perturbations may have created small scale velocity substructure identified in the Hipparcos data 
such as Hyades or Sirius moving groups (e.g., \citealt{Dehnen1998, Skuljan1999, Bovy2009, BovyHogg2010}).
}

Also, there have been some attempts to consider the perturbation from bar$+$spiral models to explain the Hercules stream 
\citep{Quillen2003, Chakrabarty2007, Antoja2009, Quillen2011, Monari2017a} 
or to understand the perturbed velocity distribution of disc stars in general \citep{Monari2016}. 
These models only sparsely sample the parameters such as the pattern speeds of the bar and spirals, $(\Omegab, \Omegas)$, 
yet their results roughly agree with each other that fast-rotating bar can (sometimes) reproduce a Hercules-like stream.

Although the fast-bar models are the simplest and yet most successful models to explain the Hercules stream, 
fast-bar models are inconsistent with a recent claim of the long Galactic bar (half-length of $5 \kpc$; \citealt{Wegg2015}). 
For example, based on a numerical model which is designed to reproduce the observed spatial and velocity distribution of bulge/bar stars, 
\cite{Portail2017} argued that a slow patten speed of $\Omegab = (39 \pm 3.5) \kmskpc$ is required to sustain the long bar.

This $\sim 30\%$ difference in $\Omegab$ between the fast-bar and slow-bar models is more serious than it sounds. 
For example, 
if $\Omegab \simeq 39 \kmskpc$ (slow bar), the OLR is unimportant in the Solar neighbourhood. 
This means that some other effect induces the bimodality if the slowly rotating bar is responsible for the Hercules stream. 
For example, \cite{PerezVillegas2017} analysed a realistic Milky Way model built from a $N$-body simulation 
and claimed that disc stars trapped by the co-rotation of the bar comprise the Hercules stream.
Also, \cite{HuntBovy2018} claimed that the velocity bimodality 
can be caused by a slowly rotating bar if the $m=4$ Fourier component of the bar potential is properly taken into account.  
Although the recent success of slow-bar-only models is promising, it is important 
to understand to which extent the slow-bar models are successful.\footnote{
Interestingly, none of our slow-bar-only models in this paper can reproduces a bimodal velocity distribution 
(as shown in Section \ref{section:barOnly} and interpreted in Section \ref{section:slowbar_models} of this paper). 
} 
It is worthwhile keeping in mind that the stellar disc is perturbed by not only the bar but also spiral arms. 
For example, if a slow-bar model relies on a specific resonance to reproduce the Hercules stream, 
there is a possibility that the additional perturbation from the spirals  might destroy the bar's resonance and the 
velocity
bimodality.

In this paper, we perform test-particle simulations that densely sample the parameter space of the bar and spiral arms. 
Our models include slow-bar-only, fast-bar-only, spiral-only, and bar$+$spiral models. 
One of the key features in our simulations is that 
we use the semi-analytic model proposed by \cite{SandersBinney2015} 
to take into account the chemo-dynamical evolution of the stellar disc 
such as the [Fe/H] evolution as a function of time and the Galactocentric radius, or the age-velocity dispersion relationship.

Our models are useful in understanding the rich chemo-dynamical information 
contained in the combined data of Gaia \citep{Lindegren2016} and other spectroscopic surveys such as RAdial Velocity Experiments (RAVE) \citep{Kunder2017}. 
For example, soon after the first data release of Gaia, 
it has been realised that the observed properties of the Hercules stream depends on [Fe/H],
such that the Hercules stream is more prominent in more metal-rich region \citep{PerezVillegas2017, Antoja2017, Quillen2018}. 
The interpretation of this 7D data (6D 
position-velocity
data plus 1D metallicity data) 
in terms of perturbation from bar/spirals has not been possible without sophisticated and homogeneous set of simulations. 
In this paper, 
we run a large number of ($\sim 200$) chemo-dynamical 2D test-particle models that are perturbed by the bar and/or spiral arms 
in order to investigate the origin of the [Fe/H] dependence of the Hercules stream. 

The outline of this paper is as follows.
In Section \ref{section:data}, we briefly summarise the observed velocity distribution of the Solar-neighbour disc stars revealed by Gaia and RAVE data. 
In Section \ref{section:potential}, we describe the potential models in our simulations. 
In Section \ref{section:simulation}, our prescription for the test-particle simulations is shown. 
In Section \ref{section:result}, we show the results of simulations for bar-only, spiral-only, and bar$+$spiral models. 
In Section \ref{section:closedOrbits}, we perform orbital analyses to interpret the result of the simulations. 
In Section \ref{section:discussion}, we discuss the implication from our calculations, 
and Section \ref{section:conclusions} sums up.

\section{Data} \label{section:data}

Here we explain the observed data with which our models are compared. 

First, we cross-match the sample stars in 
Tycho Gaia Astrometric Solutions (TGAS) from the Gaia Data Release 1 (DR1) \citep{Lindegren2016}
and RAdial velocity Experiments (RAVE) Data Release 5 \citep{Kunder2017}. 
Then we use the 5D astrometric data from TGAS and 
the line-of-sight velocity and [Fe/H] from RAVE 
to derive the velocity distribution of stars within 200 pc from the Sun. 
Our sample is defined by the following criteria:
(1) positive parallax ($\varpi>0$);
(2) distance cut ($1/\varpi < 200 \pc$); 
(3) small fractional error in parallax ($\delta \varpi / \varpi < 0.2$);
(4) small line-of-sight velocity error ($0 < \delta v_{\rm los}/(\kms) < 5$); 
(5) metallicity cut ($-1 \leq$ [Fe/H] $\leq 0.5$); and 
(6) large S/N ratio ($>40$) in the RAVE spectra. 
We note that the Solar velocity in the Galactic rest frame is assumed to be $(U_\odot,V_\odot)=(11.1, 230.24) \kms$ \citep{Schonrich2010, BovyRix2013}. 
The definition for $(U,V)$ is given in Section \ref{section:coordinate_system}.

\begin{figure*}
\begin{center}
 \includegraphics[width=0.65\columnwidth]{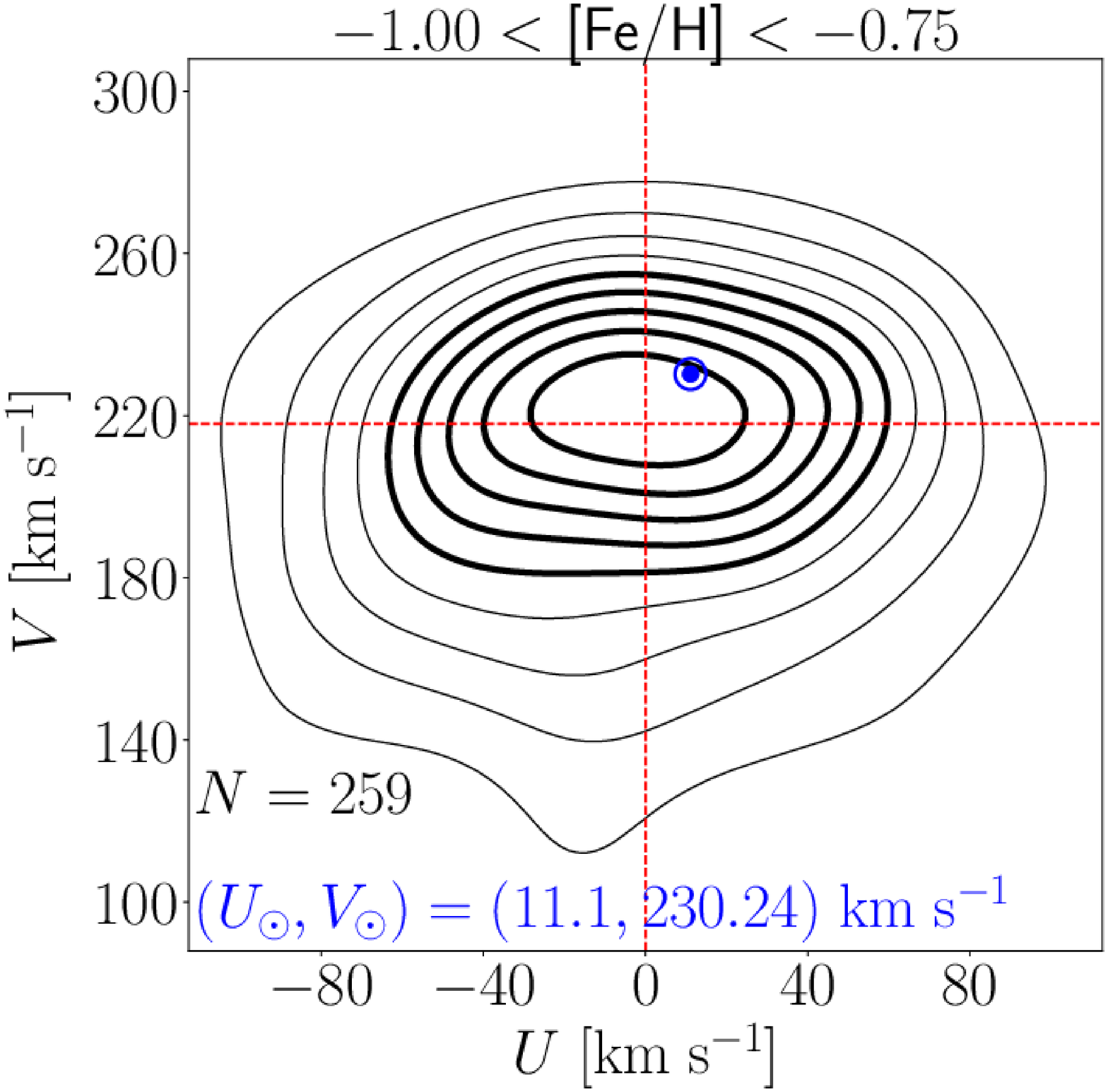} 
 \includegraphics[width=0.65\columnwidth]{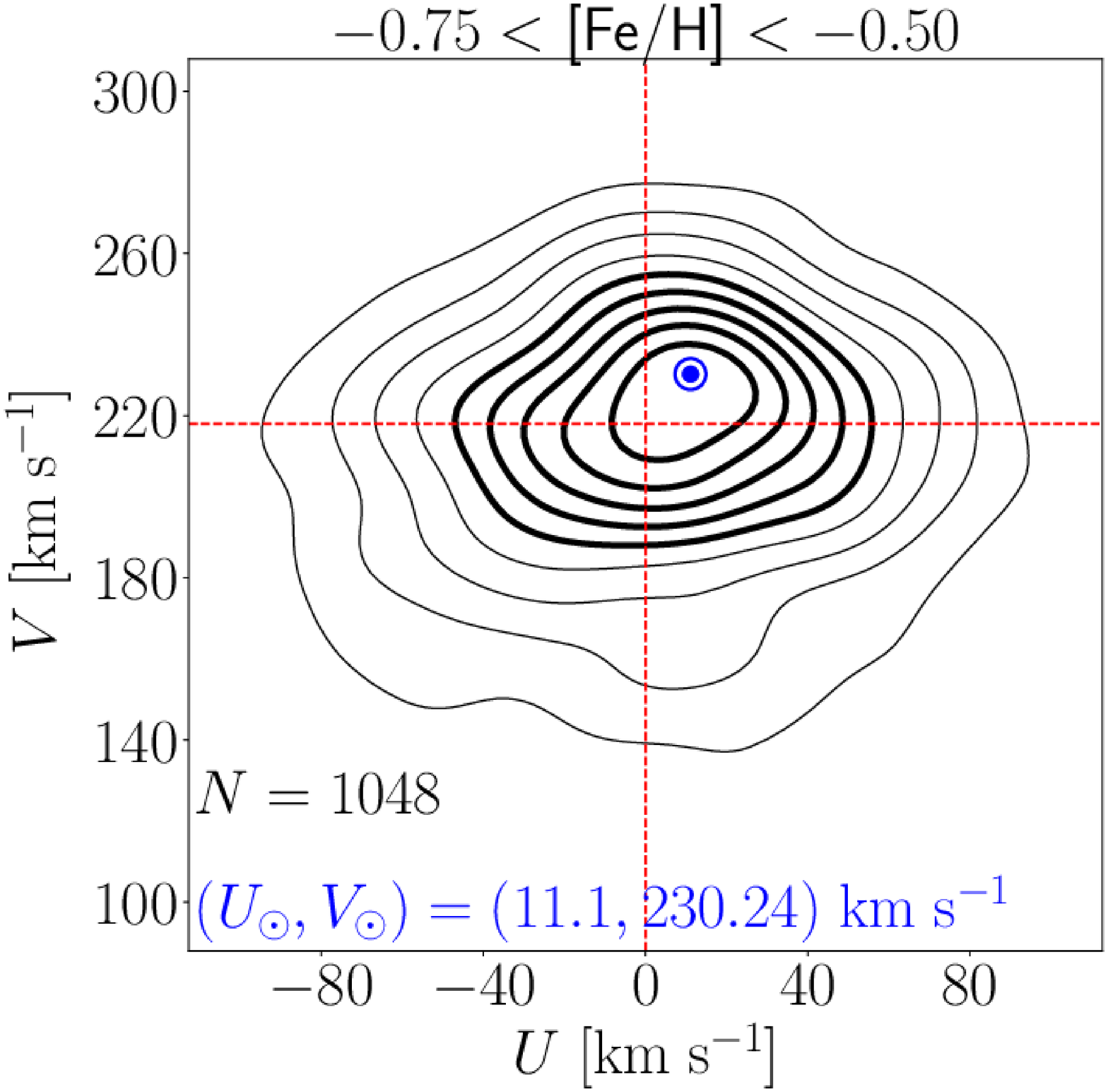}
 \includegraphics[width=0.65\columnwidth]{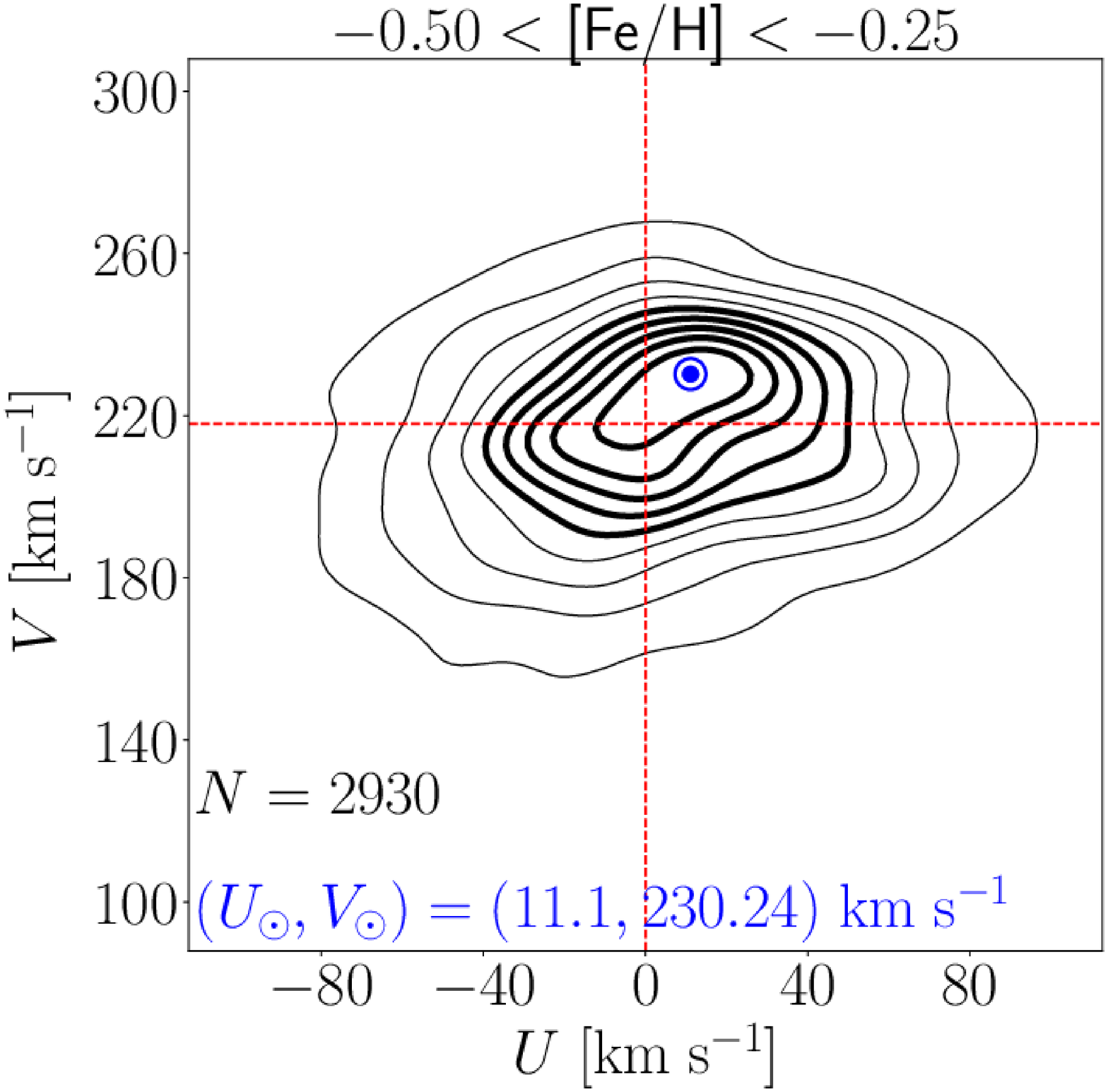} \\
 \includegraphics[width=0.65\columnwidth]{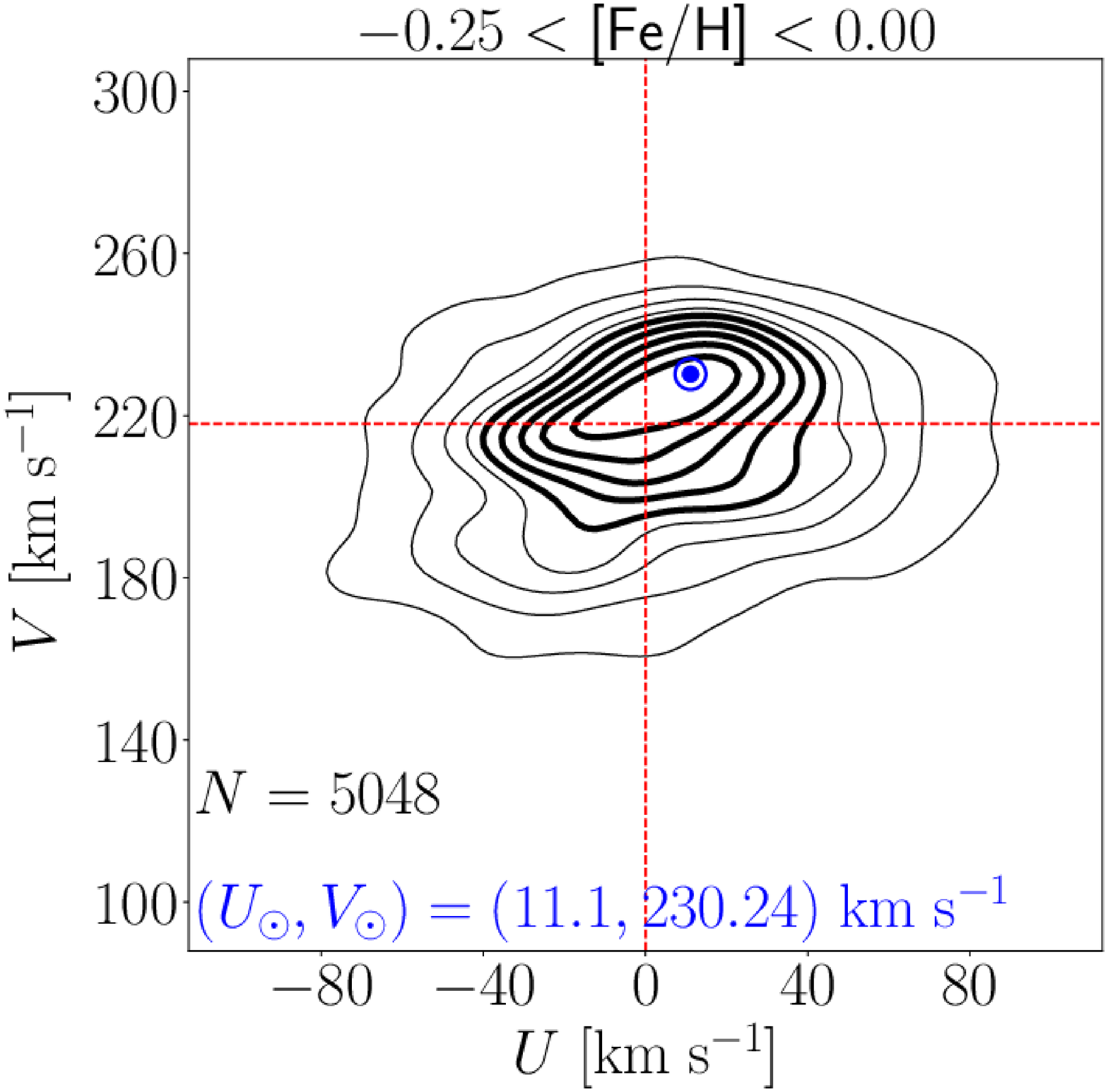} 
 \includegraphics[width=0.65\columnwidth]{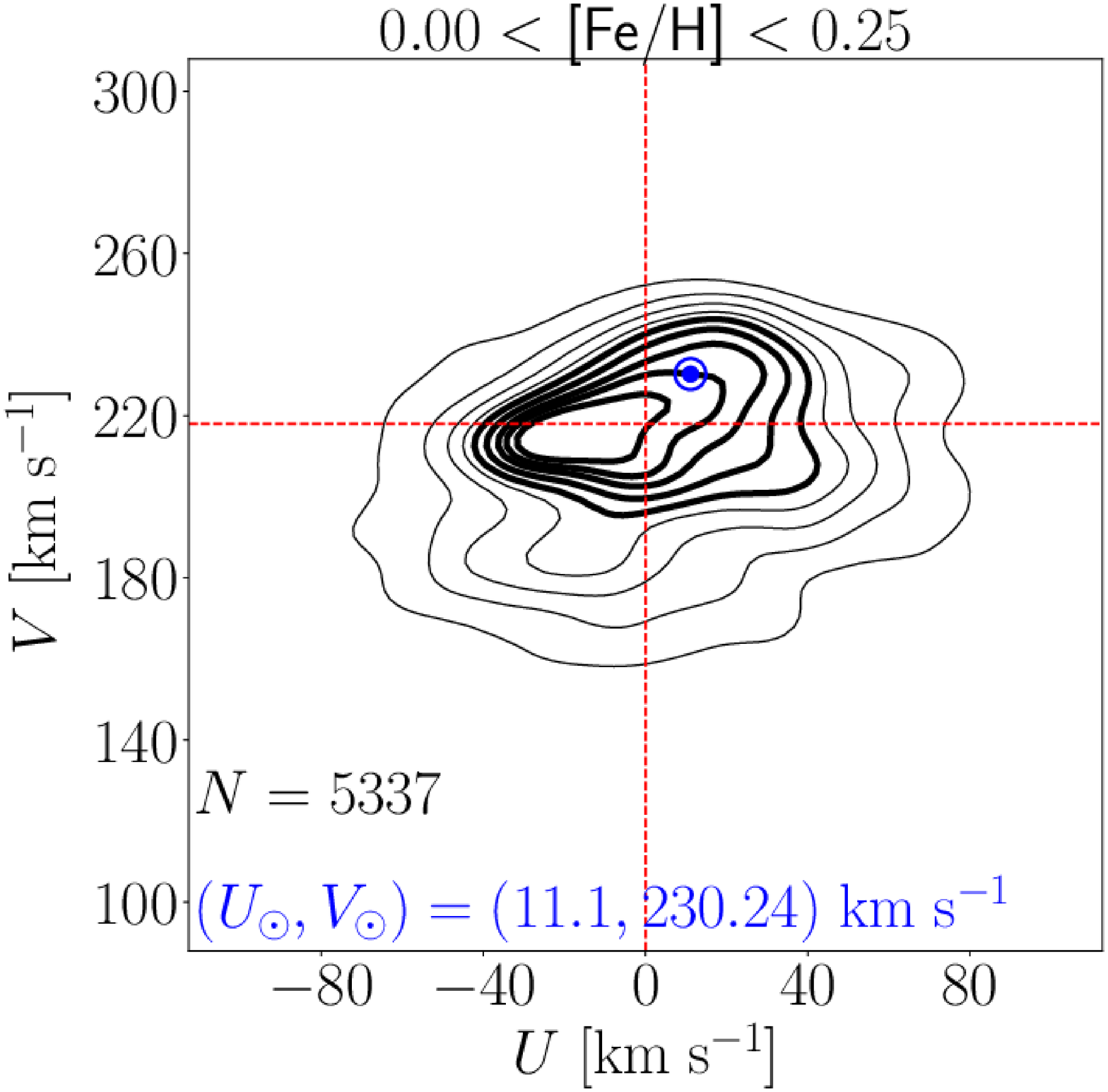} 
 \includegraphics[width=0.65\columnwidth]{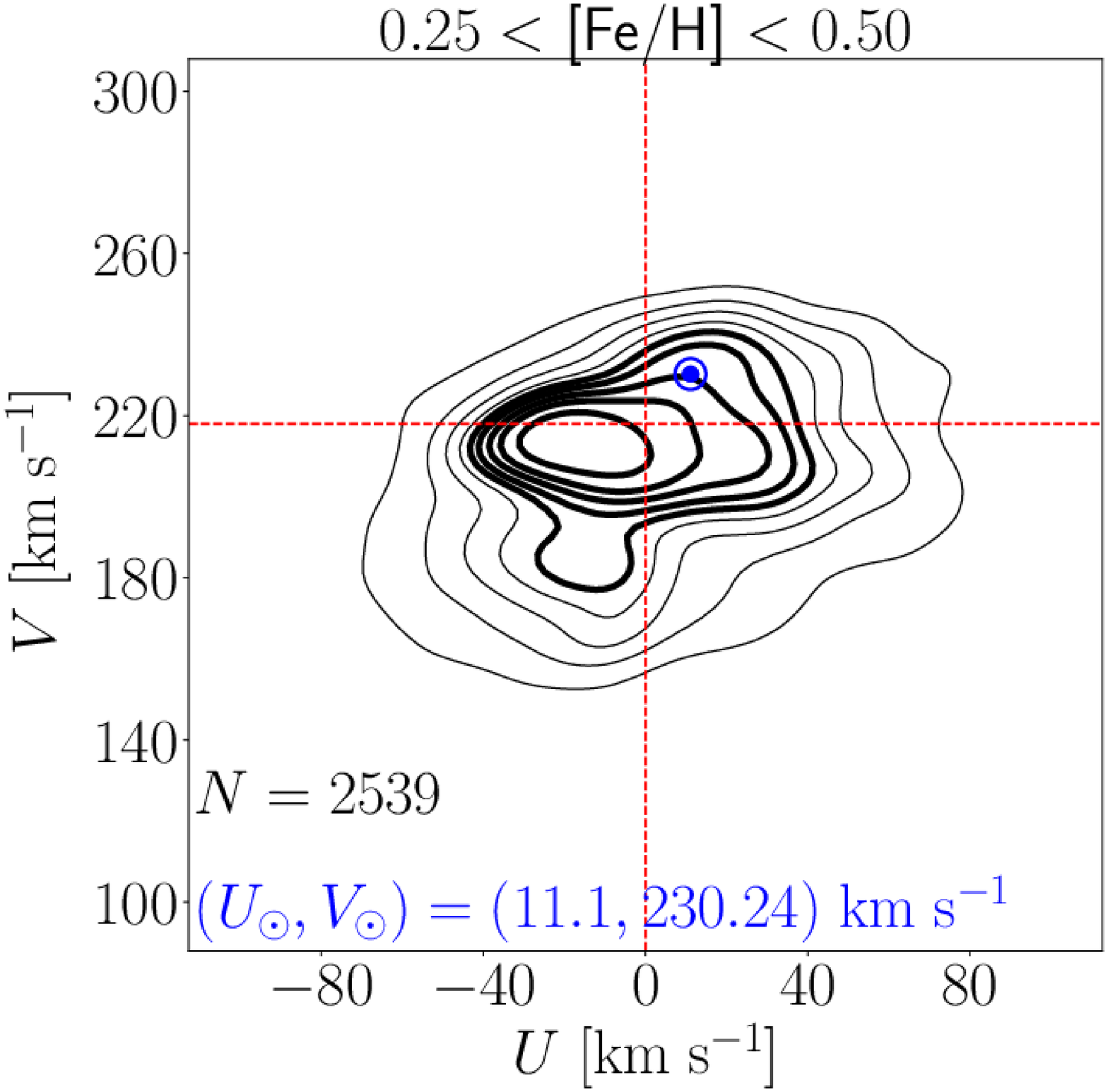} 
\caption{
Distribution of nearby stars in the $(U,V)$ space at various [Fe/H] regions revealed by Gaia (TGAS) and RAVE data. 
The thick and thin contours enclose $10, 20, 30, 40, 50 \%$ and $60, 70, 80, 90 \%$ of the stars, respectively. 
The Solar velocity is shown with a blue point. 
The horizontal and vertical lines correspond to $V=218 \kms$ (the circular velocity at Solar circle; see \citealt{BovyRix2013})
and $U=0 \kms$, respectively. 
The sample size $N$ used in each [Fe/H] region is shown on each panel. 
}
   \label{fig1}
\end{center}
\end{figure*}

Fig. \ref{fig1} shows the [Fe/H] dependence of the velocity distribution. 
In each panel with different [Fe/H],
the thick contours enclose $10, 20, 30, 40$, and $50 \%$ of the stars; 
while the thin contours enclose $60, 70, 80$, and $90 \%$ of the stars. 
In making these plots, 
we use the kernel density estimation method 
with a Gaussian kernel implemented in Python's {\tt scipy} package 
 ({\tt stats.kde.gaussian\_kde}), 
and this method is used throughout this paper for similar plots. 
At the most metal-poor regions 
($-1 \leq$ [Fe/H] $\leq -0.75$ and $-0.75 \leq$ [Fe/H] $\leq -0.5$), 
we see a monomodal distribution, 
which is characterised by a large velocity dispersion. 
As the [Fe/H] increases, the velocity dispersion becomes smaller. 
At the most metal-rich regions 
($0 \leq$ [Fe/H] $\leq 0.25$ and $0.25 \leq$ [Fe/H] $\leq 0.5$), 
we see a clear secondary peak at $(U,V) = (-15, 185) \kms$ (see also \citealt{PerezVillegas2017}), 
which lags behind the azimuthal motion of the Sun by about $45 \kms$.

Following these observational data, in this paper, 
we search for models in which their $(U,V)$ distribution is characterised by the following observational properties: 
\begin{itemize}
\item (P1) The velocity distribution is bimodal at high [Fe/H], but monomodal at low [Fe/H]. 
\item (P2) Hercules-like secondary peak is located at $V \lesssim 0.85 v_0$.
\item (P3) Hercules-like secondary peak is located at $U < 0$.
\end{itemize}
Here, $v_0$ is the Local Standard of Rest velocity in the model (see Section \ref{section:SIS_potential}). 
Throughout this paper, 
we search for models that reproduce the above-mentioned properties of the Hercules stream. 
Hereafter, 
those models that show all of these observational properties (P1)-(P3) are referred to as 
`{\it successful models}'; 
those that satisfy (P1) and (P2) or (P1) and (P3) are referred to as `{\it partially successful models}'; 
while the others are referred to as `{\it unsuccessful models}'.

In the TGAS-RAVE data, the Hercules stream is most prominent at [Fe/H]$>0$. 
However, we do not attempt to find models that show bimodality rigorously at this [Fe/H] range, 
since we use a simple prescription to assign [Fe/H] to each particle in our simulations. 
Also, we do not take into account the relative strength of the main mode and the Hercules stream when we judge the success of our models, 
since it depends on various model parameters (see fig. 14 of \citealt{Fux2001}) which are beyond our interest.

\section{Potential models}\label{section:potential}
 
We calculate the velocity distribution of disc stars in the 2D velocity space (ignoring the velocity perpendicular to the Galactic disc plane) 
under the influence of the rigidly rotating bar and/or spiral potentials.  
To this end, we perform chemo-dynamical 2D test-particle simulations. 
Here we describe the potential models in our simulations.

\subsection{Coordinate systems}\label{section:coordinate_system}

\subsubsection{Non-rotating coordinate system}\label{section:non_rotating_corrdinate_systems}

We first describe our rest frame (non-rotating) coordinate system. 
We use a Galactocentric right-handed cartesian coordinate system $(x,y,z)$, 
in which $(x,y)$ plane corresponds to the Galactic disc plane and the $z$ axis is directed toward the North Galactic Pole. 
We also use a polar coordinate system $(R,\phi, z)$ such that $(x,y)=(R \cos \phi, R \sin \phi)$, 
where $\phi$ increases in the counterclockwise direction when viewed from $z>0$ region. 
Following the convention, we define the $(U,V)$ velocity components such that $(U,V)=(-v_R,-v_\phi)$. 
We note that $V$ is measured with respect to the Galactic rest frame, 
and not with respect to the local standard of rest.
With this definition, $U$ is positive when a star moves towards the Galactic centre 
and $V$ is positive 
when a star is on a prograde 
orbit.

The bar and spiral arms in our models show a {\it prograde} rotation. 
For convenience, 
we want the pattern speeds of the bar $\Omegab$ and spiral arms $\Omegas$ to be expressed by positive numbers. 
Thus we set the spin vector of the bar and spiral arms to be 
$\vector{\Omegab} = - \Omegab \vector{e}_z$ and 
$\vector{\Omegas} = - \Omegas \vector{e}_z$, respectively, 
where $\vector{e}_z$ is the unit vector in the $z$-direction. 
In this paper, 
the unit for the pattern speed is always $\kmskpc$, 
so we use $\Omegab$ and $\Omegabkmskpc$ (or $\Omegas$ and $\Omegaskmskpc$) interchangeably for brevity.

\subsubsection{Rotating coordinate system}\label{section:rotating_corrdinate_systems}

In the presence of a rotating bar or spiral arms, 
it is useful to define the rotating coordinate systems that rotates with the bar or spiral arms. 
Here, 
we define the rotating cartesian coordinate system, $(X_{\rm rot}, Y_{\rm rot}, z)$, 
such that it coincides with $(x,y,z)$ at the current epoch ($t=0$). 
Similarly, we define the rotating cylindrical coordinate system, $(R, \phi_{\rm rot}, z)$,  
such that it coincides with $(R, \phi, z)$ at $t=0$.

\subsubsection{Positions of the Sun, bar, and spiral arms}\label{section:position_of_Sun}

We assume that the Sun is currently located at $(R, \phi ,z) = (R, \phi_{\rm rot}, z) = (R_0, 180^\circ, 0)$ with $R_0 = 8 \kpc$. 
The Galactic bar is currently oriented along $\phi = \phi_{\rm bar} = -25^\circ$, 
and it is always oriented along $\phi_{\rm rot} = \phi_{\rm bar}$ in the bar's rotating frame. 
One of the spiral arms (mimicking the Perseus Arm) and the Solar circle ($R=R_0$) 
intersect at $\phi = \gamma_0 = 135^\circ$ at the current epoch, 
and they always intersect at $\phi_{\rm rot} = \gamma_0$ in the spirals' rotating frame.

\subsection{Total potential}

We assume that the total gravitational potential of the Milky Way is given by
\eq{ \label{eq:total_potential}
\Phi(R,\phi,t) = \Phi_0(R) + G(t) [ \Phi_{\rm b}(R,\phi,t) + \Phi_{\rm s}(R,\phi,t) ] ,
}
where $\Phi_0$, $\Phi_{\rm b}$, and $\Phi_{\rm s}$ are 
the unperturbed, bar, and spiral potentials, respectively. 
Here, $t$ is time and the current epoch is assumed to be $t=0$. 
The function $G(t)$ is introduced so that the motions of stars 
are influenced by a slowly growing perturbation (see Section \ref{section:integration_orbits} for detail).

\subsection{Logarithmic potential} \label{section:SIS_potential}

We assume that the unperturbed potential of the Milky Way is axisymmetric and has a functional form of 
\eq{ \label{eq:log_potential}
\Phi_0(R) = v_0^2 \ln ( R/R_0 ) , 
}
with $R_0=8 \kpc$ and $v_0=220 \kms$. 
Under this potential, the circular velocity is $v_0$ at any radius $R$. 
Note that the adopted value of $v_0$ is close to the circular velocity at the Solar circle, $(218 \pm 10)\kms$, estimated by \cite{BovyRix2013}.

\subsection{Bar potential}

Following \cite{Dehnen2000}, we adopt a rigidly rotating bar potential of the form 
\eq{
\Phi_{\rm b}(R, \phi, t) 
= \frac{\alpha}{3} \left( \frac{R_0}{R_{\rm b}}\right )^{3} f_{\rm b}(R) v_0^2
\cos \left[ 2 (\phi + \Omegab t - \phi_{\rm bar} )\right] . 
}
We fix the current orientation of the bar to be $\phi_{\rm bar}=-25^\circ$ 
and the bar strength parameter to be $\alpha=0.01$. 
The factor $f_{\rm b}(R)$ that governs the radial dependence of the potential is given by 
\begin{align}
f_{\rm b}(R) =
\begin{cases}
     - (R/R_{\rm b})^{-3}, \;\; (R \geq R_{\rm b} ) \\
    (R/R_{\rm b})^{3} - 2, \;\; (R< R_{\rm b} ). 
  \end{cases}
\end{align}
We use 6 values of the pattern speed of the bar, 
$\Omegab = 36.11, 39.12, 42.68, 46.95, 49.42$, and $52.16$ (see Table \ref{table:bar_model_parameters}). 
Note that these values of $\Omegab$ respectively 
have the OLR radius of $R_{\rm OLR}/R_0 = 1.3, 1.2, 1.1, 1.0, 0.95$, and $0.9$, 
in the limit of weak bar ($\alpha \to 0$). 
Also, we set the half-length of the bar to be 
$R_{\rm b}= 0.8 v_0/\Omegab$ ($80\%$ of the co-rotating radius of the bar in the limit of weak bar; 
see \citealt{Aguerri2003,Aguerri2015}). 
For example, 
a slow bar with $\Omegab=36.11$ is as long as $R_{\rm b}=4.87 \kpc$ (which is consistent with the recent claim of $\sim 5\kpc$ by \citealt{Wegg2015});
while a fast bar with $\Omegab=49.42$ has a half-length of $R_{\rm b}=3.56 \kpc$. 
(As mentioned in Section \ref{section:non_rotating_corrdinate_systems}, 
we use $\Omegab$ and $\Omegabkmskpc$ interchangeably throughout this paper.)

\subsection{Spiral potential}
We adopt a rigidly rotating logarithmic spiral potential model (see \citealt{Seigar1998} section 3.4; \citealt{BT2008} equation 8.124)  given by
\eq{
&\Phi_{\rm s}(R, \phi, t) = - v_{\rm s}^2(R)  \nonumber\\
&\times \cos \left[ m (\phi + \Omega_{\rm s}t - \gamma_0) - \frac{m}{\tan \left( \theta_{\rm pitch} \right)} \ln \left( \frac{R}{R_0} \right) \right]. 
}
Detailed description for the parameters are presented below.

\subsubsection{Pattern speed $\Omega_{\rm s}$ of the spiral potential}

The pattern speed $\Omega_{\rm s}$ of spiral arms in the literature 
are distributed at around $17 < \Omegas < 28$ \citep{Gerhard2011}. 
In order to cover this range, 
we adopt 7 values of $\Omegas = 17, 19, 20, 21, 23, 25$, and $28$ (see Table \ref{table:spiral_model_parameters}). 
(As mentioned in Section \ref{section:non_rotating_corrdinate_systems}, 
we use $\Omegas$ and $\Omegaskmskpc$ interchangeably throughout this paper.)

\subsubsection{Number and geometry of the spiral arms}

We assume that the number of spiral arms is either $m=2$ or $m=4$. 
The choice of $m$ is motivated by the observational result that 
there seem to be four gas-rich spiral arms \citep{Steiman2010}
but only two of them (Perseus Arm and Scutum-Centaurus Arm) may be dynamically important  
since the other two (Sagittarius Arm and Norma/Outer Arm) do not show an over-density of stars along them \citep{Churchwell2009}.

Also, 
we fix the pitch angle of the spiral $\theta_{\rm pitch}=15^\circ$, 
and the current position of the spiral arm at the Solar circle $\gamma_0=135^\circ$ 
to roughly match the observed position of the Perseus Arm \citep{QuillenMinchev2005, Churchwell2009}.
To put it differently, 
we locate a Perseus-like logarithmic spiral arm such that it passes through a position $(R,\phi)=(R_0, \gamma_0)$ at the current epoch, 
and symmetrically locate the remaining $(m-1)$ spiral arm(s).

\subsubsection{Radial dependence of the spiral amplitude}

The radial dependence of the spiral amplitude is governed by $v_{\rm s}(R)$. 
Here we adopt a functional form of 
\eq{
v_{\rm s}^2(R) &= v_{{\rm s},0}^2 \times \left(\frac{R}{R_0} \right) \nonumber \\
&\times 
\begin{cases}
    \exp \left[ \frac{R_0-R}{\sigma_2} \right] , \;\;\ (R > R_3)\\
    \exp \left[ \frac{R_0-R_3}{\sigma_2} + \frac{R-R_3}{\sigma_1} \right] , \;\;(R<R_3). 
  \end{cases}
}
This functional form is inspired by \cite{Steiman2010}, 
and thus we adopt $(R_3, \sigma_1, \sigma_2) = (2.9, 0.7, 3.1) \kpc$ following their study.
Throughout this paper, we adopt two spiral morphologies,
namely the SS20m2 model with $(v_{\rm s,0}, m)=(20 \kms, 2)$ 
and the SS15m4 model with$(v_{\rm s,0}, m)=(15 \kms, 4)$.


\section{2D test-particle simulation} \label{section:simulation}

We run a large number of chemo-dynamical 2D test-particle simulations 
under the influence of the rigidly rotating bar and/or spiral arms.

Each simulation is performed in three steps: 
\begin{itemize}
\item Generating initial condition of stars by using axisymmetric distribution function models. 
\item Integrating stellar orbits under our model potential. 
\item Assigning [Fe/H] to the stars. 
\end{itemize}
In the following, we provide the prescription for these steps. 

\subsection{Initial condition} \label{section:IC}

In our simulations, we assume that 
the stellar disc is formed at $t=-12 \Gyr$ and the star formation rate is constant as a function of $t$.

For each simulation, we use $1.2\times10^7$ stars. 
These stars consist of 120 cohorts of coeval stars, 
such that $i$th cohort includes $10^5$ stars with an age of $\tau_i = (0.1 i) \Gyr$ ($i= 1,2, \cdots, 120$) 
at current epoch. 
We assign the initial conditions for each cohort of stars depending on their age  
in order to realistically model the evolution of the stellar disc. 
We note that our initial condition 
does not depend on the properties of the bar or spirals, $(\Omegab, m, \Omegas)$.

The way we assign the initial condition for $i$th cohort depends on 
whether or not $\tau_i$ is older than $-\tform$.

If $\tau_i \leq -\tform$, 
i.e., if $i$th cohort of stars are born after the formation of the bar and spirals, 
we simply assign the position and velocity of the stars at $t=-\tau_i$ according to the 
2D stellar disc distribution function model introduced by \cite{Dehnen1999a}. 
In this model, the stellar disc surface density profile and radial velocity dispersion profile 
are described as 
\eq{
\Sigma (R) &= \Sigma_0 \exp \left[ -\frac{R-R_0}{R_d} \right] ,\\
\sigma_R^2 (R) &= \sigma_{R0}^2 \exp \left[ -\frac{R-R_0}{R_\sigma} \right] ,
}
and their action distribution is set such that the distribution function model 
is in a dynamical equilibrium under the axisymmetric potential $\Phi = \Phi_0$ [equation (\ref{eq:log_potential})]. 
Here, we set $\Sigma_0={\rm const.}$, $R_d = R_0/3$, $R_\sigma = R_0$, and $\sigma_{R0}=22 \kms = 0.1 v_0$ 
independent of $i$ as long as $\tau_i \leq -\tform$.

If $\tau_i >  -\tform$, i.e., if $i$th cohort of stars are born before the formation of the bar and spirals, 
we generate the position and velocity of stars at the epoch of $t=\tform$ (instead of $t = -\tau_i$)
and regard them as the `initial' condition of the $i$th cohort of stars. 
Specifically, we assign the position and velocity of $i$th cohort of stars at $t=\tform$ 
by using the same 2D distribution function model with $\Sigma_0={\rm const.}$, $R_d = R_0/3$ and $R_\sigma = R_0$ as before, 
but by adopting 
\eq{ \label{equation:oldIC}
\sigma_{R0} = (50 \kms) \left(\frac{ ( \tau_i + \tform ) + 0.5 \Gyr} { 12.5 \Gyr } \right)^{0.25} . 
}
This functional form is motivated by numerical simulations 
\citep{Jenkins1990, Kokubo1992, Hanninen2002, AumerBinney2009, Aumer2016, Kumamoto2017}.\footnote{
If $\tau_i >  -\tform$, 
$i$th cohort has been evolved in the Milky Way 
for a period of $(\tau_i + \tform)$ at the epoch of $t=\tform$. 
In our prescription, 
we adopt hotter initial condition for larger value of $(\tau_i + \tform)$ (older cohort), 
in order to mimic the effect of the internal heating 
(such as that induced by Giant Molecular Clouds, which is not explicitly modelled in our simulations). 
} 
Our prescription 
makes it easier for us to run realistic test-particle simulations 
that roughly satisfy the age-velocity-dispersion relationship 
(AVR)
\citep{Nordstrom2004}. 
The AVR of the Solar-neighbour disc stars in our simulation is 
discussed in Appendix \ref{appendix:AVR}.

In order to minimise the computational cost, 
we generate the initial conditions of test particles only at $3 \kpc < R < 15 \kpc$ for each cohort. 
We have confirmed that this radial range is enough to explore the velocity distribution within a few $\kpc$ from the 
Solar Circle.

\subsection{Integration of the orbits} \label{section:integration_orbits}

After assigning the initial condition, we integrate the orbit from the initial condition until the current epoch.

In order to make the results of our simulation smooth, 
we add a noise in the stellar age, such that 
the age of
a star that belongs to the $i$th cohort 
follows a distribution 
$\tau \sim \tau_i + \mathcal{U}(0, 0.1 \Gyr)$, 
where $\mathcal{U}(a,b)$ is 
a uniform distribution
between $a$ and $b$.
Also, in order to slowly introduce the perturbation, 
we adopt a certain functional form for $G(t)$ in equation (\ref{eq:total_potential}):
\eq{\label {eq:G(t)}
G(t)  = 
\begin{cases}
0, \; &(t \leq T), \\
\frac{3}{16} \xi^5 - \frac{5}{8} \xi^3 + \frac{15}{16} \xi + \frac{1}{2}, \; &(T < t < T + \delta T), \\
1, \; &(T + \delta T <t), 
\end{cases}
}
with $\xi=\frac{2 (t - T)}{\delta T} - 1$ and $\delta T = 0.25 \Gyr$, following \cite{Dehnen2000}. 
Here, we adopt $T = \tform$ for old cohorts of stars that are born before $t=\tform$; 
while we adopt $T = -\tau$ for younger cohorts of stars (that are born at $t=-\tau > \tform$). 
The prescription for younger cohorts of stars is adopted 
so that the orbits of young stars (that are designed to be in dynamical equilibrium at birth with the axisymmetric potential $\Phi_0$) 
are not drastically perturbed just after $t=-\tau$ due to the pre-existing perturbations from bar and spirals.\footnote{
The gas clouds from which disc stars are formed 
have been exposed to the perturbation from the bar/spirals for some time 
and thus the orbits of recently-born stars are partially phase-mixed. 
Our prescription can mimic this situation by slowly perturbing the orbits of recently-born stars, 
so it is not too unrealistic. 
}
We have confirmed with some representative simulations that the resultant velocity distribution is almost identical 
when we set $\delta T \to 0$. 
However, throughout this paper we adopt $\delta T = 0.25 \Gyr$ 
so that we can avoid possible velocity substructure arising from the sudden introduction of the bar/spirals 
and so that the resultant velocity distribution becomes smoother.

After adopting $G(t)$ for each star, the stellar orbit is integrated from $t=T$ to $t=0$. 
Throughout this paper, orbit integration is always performed 
with the explicit embedded Runge-Kutta Prince-Dormand (8,9) method (with an adaptive time step; and with the relative tolerance error of $10^{-5}$) 
implemented in GNU Scientific Library \citep{Galassi2009}. 


\subsection{Chemical evolution model} \label{section:chemistry}

In order to compare the observed data with our simulation,
we assign [Fe/H] to our test particles by using the prescription in \cite{SandersBinney2015}, 
in which inside-out formation of the Galactic disc is assumed. 

Their prescription is useful for our study in two ways. 
First, they provide an observationally motivated semi-analytic model 
of the gas-phase metallicity ${\rm [Fe/H]_{\rm gas}}(\tau, R)$ in the disc 
as a function of look-back time $\tau$ and Galactocentric radius $R$ (see their equation 1).
Secondly, they provide a probabilistic formulation that connects the initial azimuthal action $J_\phi'$ 
and the current 3D action $\vector{J}=(J_R, J_\phi, J_z)$ of a star due to internal heating with some plausible assumptions. 
Here, $J_R$, $J_\phi$, and $J_z$ ($\equiv 0$ in our 2D simulations) are the (current) radial, azimuthal, and vertical action, respectively. 
By using their formulation, 
the probability that a star's initial azimuthal action is $J_\phi'$ given its current action $\vector{J}$ and age $\tau$ 
can be expressed as 
\eq{ \label{eq:P_Jprime_J_tau}
P(J^{\prime}_\phi | \vector{J}, \tau ) 
= \frac
{\mathcal{G}(J_\phi, J^{\prime}_\phi, \tau) J^{\prime}_\phi \exp{[-J^{\prime}_\phi / (R_d v_0)]}}
{\int {\rm d} J^{\prime}_\phi \; \mathcal{G}(J_\phi, J^{\prime}_\phi, \tau) J^{\prime}_\phi \exp{[-J^{\prime}_\phi / (R_d v_0)]}}
}
when the potential is $\Phi_0(R)$ and the disc scale length is $R_d$ (see Appendix \ref{appendix:P_Jprime_J_tau} for derivation). 
Here, $\mathcal{G}(J_\phi, J^{\prime}_\phi, \tau)$ is a normalised Green's function, 
and is given by equation (23) in \cite{SandersBinney2015}.

In this paper, the metallicity [Fe/H] of a test particle 
follows a distribution 
given by 
\eq{
{\rm [Fe/H]} \sim {\rm [Fe/H]_{\rm gas}}(\tau, R_i) + \mathcal{N}(0,\sigma_{\rm [Fe/H]}) .
}
Here, ${\rm [Fe/H]_{\rm gas}}$ is the gas-phase metallicity adopted from \cite{SandersBinney2015}, 
$\tau$ is the stellar age, 
$R_i$ is the Galactocentric radius at birth, 
and 
$\mathcal{N}(0, \sigma_{\rm [Fe/H]})$ is a Gaussian 
distribution 
with mean $0$ and dispersion $\sigma_{\rm [Fe/H]} \equiv 0.05$. 
The value of $\sigma_{\rm [Fe/H]}$ can be regarded as 
the intrinsic metallicity scatter of the stars 
or the observational error on [Fe/H]. 

For old stars with $-\tau < \tform$, we use the azimuthal action $J_\phi$ at $t=\tform$ 
(the initial condition of the orbit integration) and the stellar age at that moment, $\tau + \tform$, 
and probabilistically assign the azimuthal action $J_\phi'$ at $t=-\tau$. 
Then we simply set $R_i = J_\phi'/v_0$, by approximating the born radius by the guiding centre radius at birth. 

For young stars with $\tform < -\tau$, we set $R_i = J_\phi'/v_0$ (the guiding centre radius at birth), 
where $J_\phi'$ is the value of the azimuthal angular momentum at $t=-\tau$. 
We note that we do not set $R_i$ to be the initial radius at which the star was born, 
in order to make our prescription for young stars ($\tform < -\tau$) more consistent with that for old stars ($-\tau < \tform$).


\begin{table}
  \begin{center}
  \caption{Important parameters in the bar-only models}
  \begin{tabular}{| c |  c | cl c|} \hline
  	Category 		& $\Omegab$	&$R_{\rm b}$ 	&$\tform$ 	\\ 
				&$\kmskpc$	&$\kpc$ 		&$\Gyr$	 \\ 
	\hline
	Slow			&36.11		&$4.87$ 		&$-1, -3, -5$ \\ 
	Slow			&39.12		&$4.50$ 		&$-1, -3, -5$  \\ 
	Medium		&42.68		&$4.12$		&$-5$	\\ 
	Medium		&46.95		&$3.75$		&$-5$ 	\\ 
	Fast			&49.42		&$3.56$		&$-1,-2,-3,-4,-5$ \\ 
	Fast			&52.16		&$3.37$		&$-1,-2,-3,-4,-5$ \\ 
	\hline
\end{tabular}
\label{table:bar_model_parameters}
\end{center} 
\end{table}

\begin{table}
  \begin{center}
  \caption{Important parameters in the spiral-only models}
  \begin{tabular}{| c |  c |l |} \hline
  	$\Omegas$	&$m^\dagger$	&$\tform$ 	\\ 
	$\kmskpc$	&		&$\Gyr$  \\ 
	\hline
	17			&2, 4		&$-1, -3, -5$ \\ 
	19			&2, 4		&$-1, -3, -5$ \\ 
	20			&2, 4		&$-1, -3, -5$ \\ 
	21			&2, 4		&$-1, -3, -5$ \\ 
	23			&2, 4		&$-1, -3, -5$ \\ 
	25			&2, 4		&$-1, -3, -5$ \\ 
	28			&2, 4		&$-1, -3, -5$ \\ 
	\hline
\end{tabular}
\label{table:spiral_model_parameters}
\end{center} 
\begin{flushleft}
$^\dagger${Spiral models with $m=2$ and $m=4$ are referred to as SS20m2 and SS15m4 models, respectively.}\\
\end{flushleft}
\end{table}

\begin{table*}
  \begin{center}
  \caption{Representative bar-only and spiral-only models}
  \begin{tabular}{| c |  c |c |c | c |c| c |  c |} \hline
  	Bar		&$\Omegab$	&$m$	&$\Omegas$	&$\tform$	& Under-dense region orbit & Figure						&Section\\ 
			&$\kmskpc$	& 		&$\kmskpc$	&$\Gyr$	 &					& 							&\\ 
	\hline
	Fast$^\dagger$ 		
			&49.42		&		&			&$-1$ 		& highly non-closed 		& \ref{fig_bar49closed}, \ref{fig_model_UV_FeH_detail}, \ref{fig_origin_of_Hercules}, \ref{fig:map_bimodality}(a), \ref{fig:10kpc}  & \ref{section:fastbar_models}\\ 
	Slow$^\ddagger$		&36.11	&			&			&$-1$ 					& $-$		 		& \ref{fig_bar36closed} & \ref{section:slowbar_models}\\ 
			&			&4		&21			&$-3$		& highly non-closed 		& \ref{fig_spiral21m4closed} & \ref{section:omegas21m4_models}\\ 
			&			&2		&28			&$-1$ 		& highly non-closed 	& \ref{fig_spiral28m2closed} & \ref{section:omegas28m2_models}\\ 
	\hline
\end{tabular}
\label{table:bar_only_spiral_only_model_parameters}
\end{center} 
\begin{flushleft}
$^\dagger${This young, fast-bar-only model is the only successful model in this table.}\\
$^\ddagger${The slow-bar-only model shows only a monomodal velocity distribution.}
\end{flushleft}
\end{table*}

\begin{table*}
  \begin{center}
  \caption{Representative bar$+$spiral models that satisfy observational properties (P1), (P2), and (P3) listed in Section \ref{section:data}}
  \begin{tabular}{| c |  c |c |c | c |c| c |  c |c |c|} \hline
  	Bar		&$\Omegab$	&$m$	&$\Omegas$	&$\tform$	 & Under-dense region orbit & Figure						&Section\\ 
			&$\kmskpc$	& 		&$\kmskpc$	&$\Gyr$	 &					& 							&\\ 
	\hline
	Fast 		&49.42		&4		&21			&$-5$ 		& highly non-closed 		& \ref{fig_bar49_spiral21m4_closed} & \ref{section:bar49_spiral21m4_closed}\\ 
	Fast 		&49.42		&4		&23			&$-5$ 		& highly non-closed 		\\
	\hline
	Slow		&36.11		&4		&23			&$-3$ 		& highly non-closed 		& \ref{fig_bar36_spiral23m4_closed}, \ref{fig:map_bimodality}(b), \ref{fig:12kpc} & \ref{section:bar36_spiral23m4_closed}\\ 
	Slow		&36.11		&4		&20			&$-3$ 		& hot CR orbits of bar 	& \ref{fig_bar36_spiral20m4_closed} 	& \ref{section:bar36_spiral20m4_closed}\\ 
	Slow		&39.12		&4		&23			&$-1$ 		& hot CR orbits of bar 	\\
	Slow		&39.12		&4		&20			&$-5$ 		& highly non-closed		\\
	Slow		&36.11		&2		&25			&$-3$ 		& hot CR orbits of spiral	\\

	\hline
\end{tabular}
\label{table:good_model_parameters}
\end{center} 
\end{table*}

\section{Result of simulations}\label{section:result}

In this Section, we describe the simulated velocity distribution of stars in the Solar neighbourhood. 
Here, the Solar neighbourhood is defined by $7.75<R/(\kpc)<8.25$ and $175^\circ < \phi <185^\circ$. 
This simulated Solar neighbourhood is larger than the spatial distribution of our TGAS-RAVE data (within $200 \pc$ of the Sun), 
but we do not see significant change in the velocity distribution across this simulated area. 
Some representative models to be discussed are listed in Tables \ref{table:bar_only_spiral_only_model_parameters} and \ref{table:good_model_parameters}.

\subsection{Bar-only models}\label{section:barOnly}

First, we ran 18 bar-only simulations as shown in Table \ref{table:bar_model_parameters}.
We find that 
bar-only models are successful 
(recovering all the observational properties (P1)-(P3) of the Hercules stream listed in Section \ref{section:data}) 
if the bar is fast-rotating ($\Omegab = 49.42, 52.16$) as well as dynamically young ($\tform = -1 \Gyr$). 
One of these successful models is shown on the top row of Fig. \ref{fig_bar49closed}. 
In order to make a fair comparison with the observed data in Fig. \ref{fig1}, 
the thick contours enclose $10, 20, 30, 40$, and $50 \%$ of the stars; 
while the thin contours enclose $60, 70, 80$, and $90 \%$ of the stars
in our models.

Previously, many authors have attributed the origin of the Hercules stream 
to the resonance of the fast-rotating bar whose 2:1 OLR radius is slightly inside the Solar circle \citep{Dehnen2000}. 
Indeed, both of our successful fast-bar-only models 
locate the 2:1 OLR radius at around $R \simeq 0.8$-$0.95 R_0$. 
Intriguingly, our bar-only simulations suggest that the fast-rotating bar has to be as young as $\sim 1 \Gyr$ 
in order to reproduce the Hercules stream. 
The success of young, fast-rotating bar models is consistent with many previous models, 
as epitomised by the pioneering work of \cite{Dehnen2000} in which the dynamical age of the bar is $\sim0.5 \Gyr$.

Our result that old bar-only models can not reproduce the Hercules stream 
is to some extent consistent with previous study by \cite{Monari2013}, 
in which 3D test-particle simulations were performed with various dynamical age of the bar. 
\cite{Monari2013} found that the Hercules-like signature is weakened 
depending on the age of the bar (see their fig. 5).

Here we note that  
\cite{Monari2013} found that their model with $\tform \simeq -3 \Gyr$ 
was able to reproduce a Hercules-like structure (and thus they used this model for further investigation), 
while our fast-bar-only simulation with $\tform = -3 \Gyr$ is unsuccessful in reproducing the Hercules stream. 
This apparent difference may be partially attributed to the fact that 
\cite{Monari2013} use more realistic 3D potential models while we use simple 2D models. 
However, we emphasise a more fundamental difference between their simulations and ours. 
In each of their simulations, all the particles have a common age, 
and they have been influenced by the bar for a common duration of time, $|\tform|$. 
In other words, they conducted simulations of mono-age populations. 
In contrast, 
in each of our simulations, the particles have a range of stellar age $0 \leq \tau/\Gyr \leq 12$ 
and the duration of time that a star has been influenced by the bar depends on $\tau$. 
In this regard, our results can be considered as a superposition 
of simulations by \cite{Monari2013} with various age of the bar, $|\tform|$. 
Since the strength (and even the existence) of the Hercules-like stream in their simulations depends on $|\tform|$, 
a superposition of their simulations with varying $|\tform|$ 
may weaken or erase the Hercules-like stream. 
Thus, 
the fact that their mono-age simulation with $\tform \simeq -3 \Gyr$ shows a Hercules-like stream 
and 
the fact that our simulation with $\tform = -3 \Gyr$ is unsuccessful in reproducing a Hercules-like stream 
are not as contradictory as they may sound. 
In any case, our simulations suggest that the observed properties of the Hercules stream favour 
a young fast-rotating bar, if the rotating bar is the only source of perturbation in the stellar disc. 
(In Section \ref{section:fastbar_models}, we speculate the reason why a young bar is favoured.)

It is interesting to note that our slow-bar-only models 
can not reproduce the Hercules stream independent of the dynamical age of the bar (see top row of Fig. \ref{fig_bar36closed}). 
Our results seem to disagree with the study by \cite{PerezVillegas2017}, 
in which the co-rotation (CR) resonance of the bar is claimed to cause the Hercules stream. 
This disagreement might be due to our simpler bar model, 
or the fact that
they derive a time-averaged velocity distribution 
albeit the fact that simulations by us or \cite{Monari2013} indicate a time-dependence of the velocity distribution. 
In any case, it is intriguing to note that (as we shall see in Sections \ref{section:barSpiral} and \ref{section:various_bar+spiral_models})
our slow-bar$+$spiral models can reproduce the Hercules stream.

\subsection{Spiral-only models} \label{section:spiralOnly}

Next, we ran 42 spiral-only simulations as shown in Table \ref{table:spiral_model_parameters}.
We found that 
six 4-armed spiral models (SS15m4 models) with $\Omegas = 20, 21$ and $\tform / \Gyr = -1,-3$, and $-5$; 
and a 2-armed spiral model with $\Omegas = 28$ and $\tform / \Gyr = -1$ 
show a bimodal velocity distribution and reproduce the observed properties (P1) and (P2) in Section \ref{section:data}. 
As seen in the top row of Figs. \ref{fig_spiral21m4closed} and \ref{fig_spiral28m2closed}, 
for these (partially successful) models, the Hercules-like secondary peak is more prominent in the metal-rich region and is located at $V/v_0 \simeq 0.8$, 
but it is located at $U \simeq 0$, which is inconsistent with the observed property (P3). 
This result of near-zero radial velocity for the secondary peak is consistent with the spiral-only models in \cite{Antoja2009}. 
In any case, these partially successful models with velocity bimodality are still intriguing, 
since none of our long-standing bar models with $|\tform| \geq 2 \Gyr$ can reproduce a bimodal structure.

We note that 4-armed spiral models with $\Omegas = 19$ also show bimodal structure, 
but in these models the bimodality is seen not only in the metal-rich region but also in the metal-poor region.

\subsection{Bar$+$spiral models} \label{section:barSpiral}

Lastly, we ran 168 bar$+$spiral models.  
We used 4 values for $\Omegab = 36.11, 39.12, 49.42$, and $52.16$; 
3 values of $\tform$; 
and for each combination of $(\Omegab, \tform)$ 
we used 14
spiral models (7 values for $\Omegas$; 2 values for $m$) 
as in Table \ref{table:spiral_model_parameters}. 
The parameter $\Omegab$ 
is not as densely sampled as $\Omegas$, 
since one of our motivations in this study is 
to understand whether or not 
a slow-bar model can ever reproduce the Hercules stream by adequately choosing the spiral model.

The velocity distribution for some successful bar$+$spiral models that recover 
the observed properties (P1)-(P3) 
are shown on the top row of Figs. \ref{fig_bar49_spiral21m4_closed}, \ref{fig_bar36_spiral23m4_closed}, and \ref{fig_bar36_spiral20m4_closed}. 
Also, some other successful models are summarised in Table \ref{table:good_model_parameters}. 
We can immediately see from Table \ref{table:good_model_parameters} that the Hercules-like stream can be reproduced 
not only by fast-bar$+$spiral models but also by slow-bar$+$spiral models. 
The fact that the Hercules-like stream can be reproduced with slow-bar$+$spiral models has a significant implication, 
since a slowly rotating bar is favoured by recent claims that the Galactic bar is as long as $\sim 5 \kpc$ in radius (see \citealt{Wegg2015}). 
Also, our result is intriguing from a theoretical aspect as well, 
since the slow-bar-only or spiral-only models in our simulations can not reproduce all of the observed properties (P1)-(P3).

We note that our judgment of `successful models' 
depends solely on the adopted criteria of (P1)-(P3). 
Although some of the slow-bar$+$spiral models satisfy (P1)-(P3), 
in general 
successful slow-bar$+$spiral models 
seem to result in `less successful' similarity to the observed velocity distribution 
than successful fast-bar($+$spirals) models.

\begin{figure*}
\begin{center}
 \includegraphics[width=0.6\columnwidth]{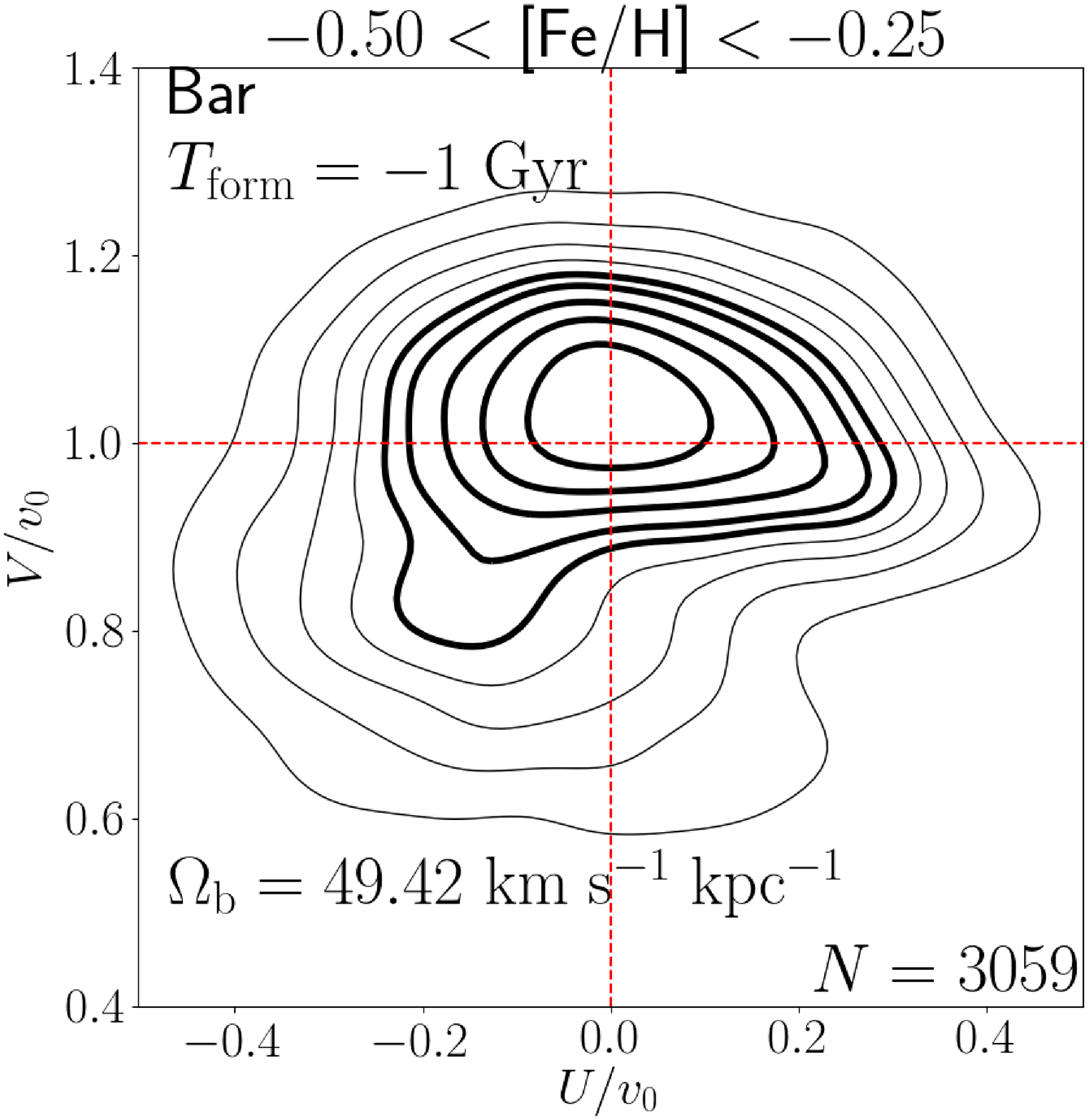} 
 \includegraphics[width=0.6\columnwidth]{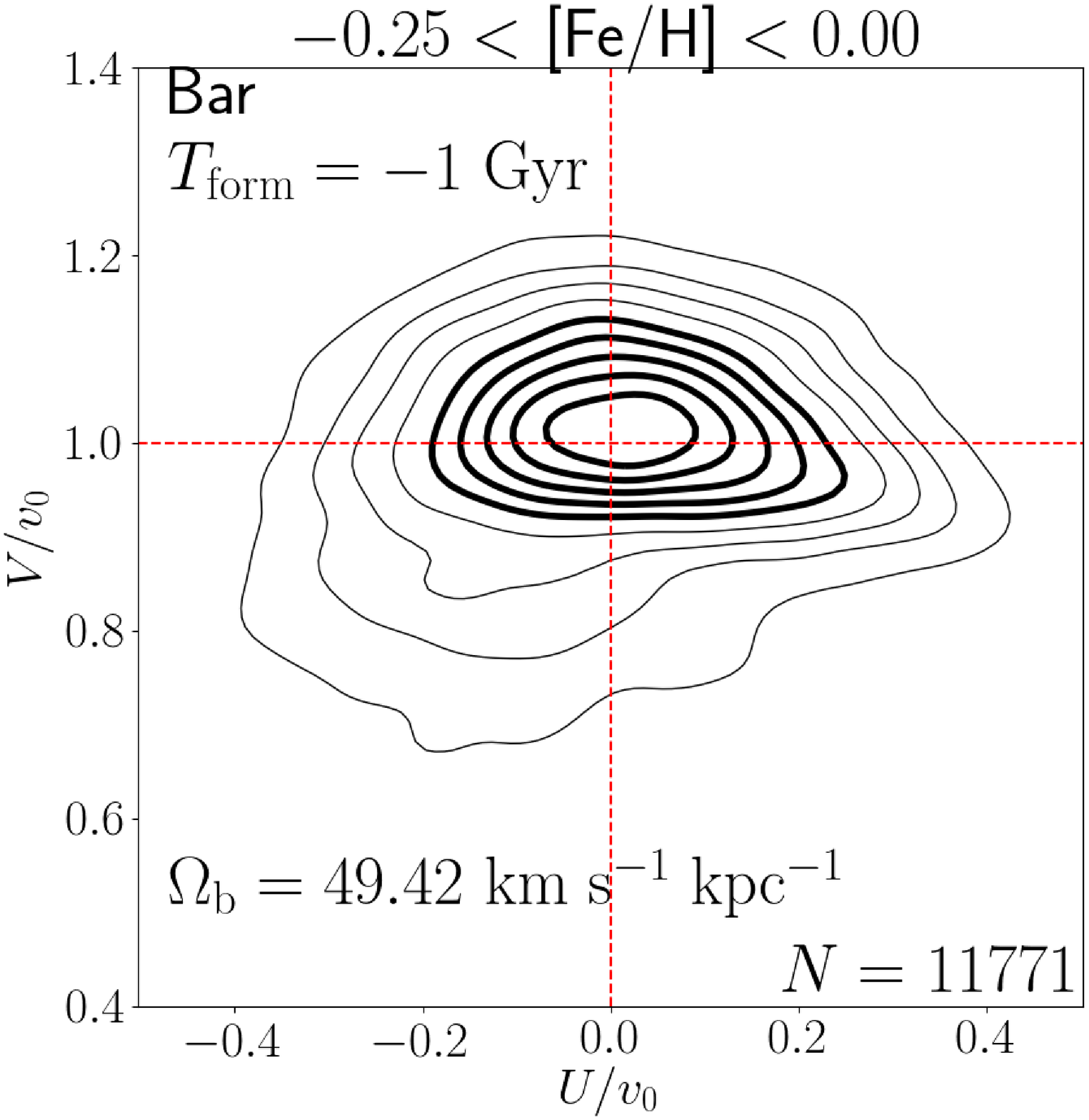} 
 \includegraphics[width=0.6\columnwidth]{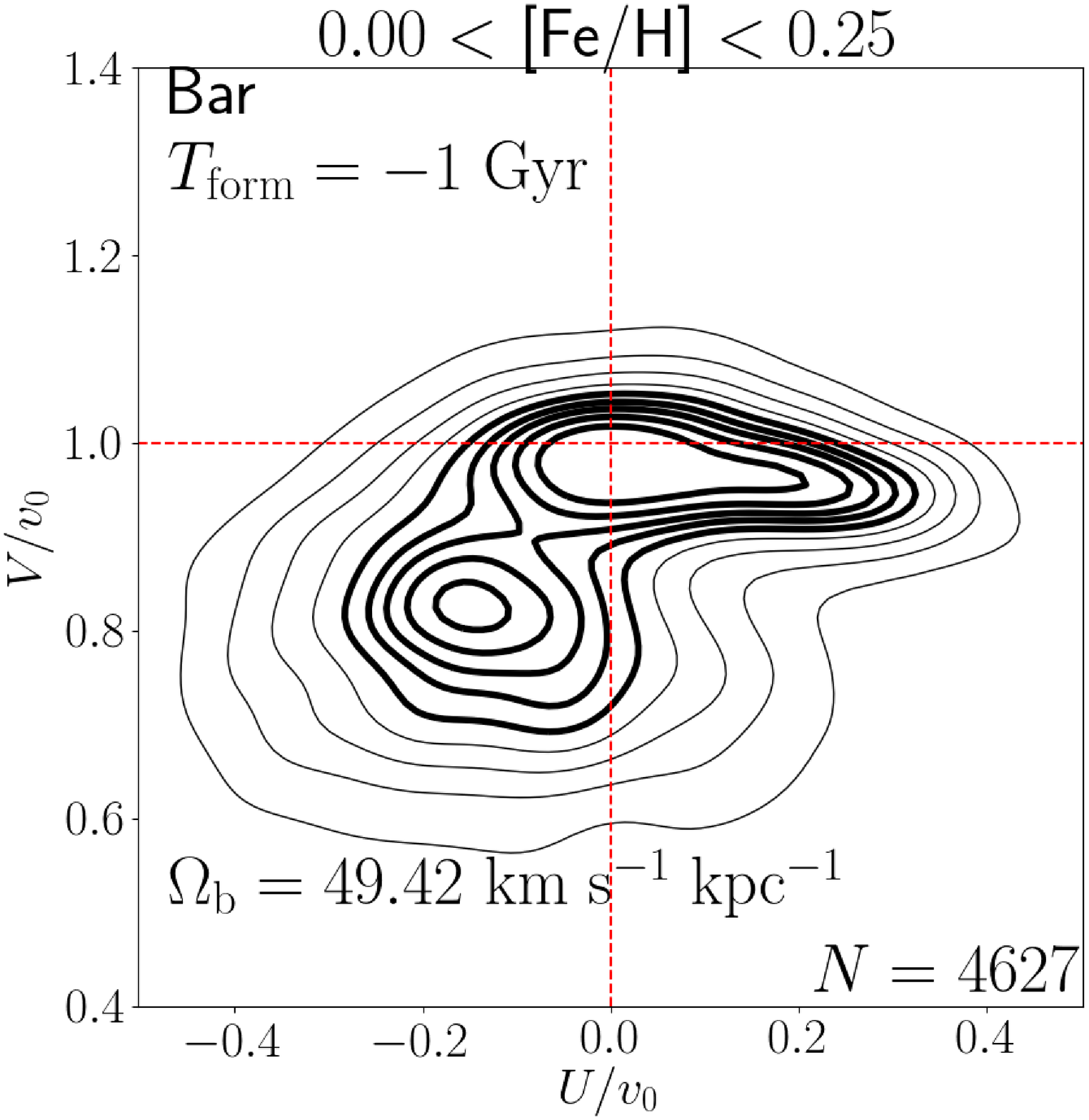} \\
 \includegraphics[width=0.8\columnwidth]{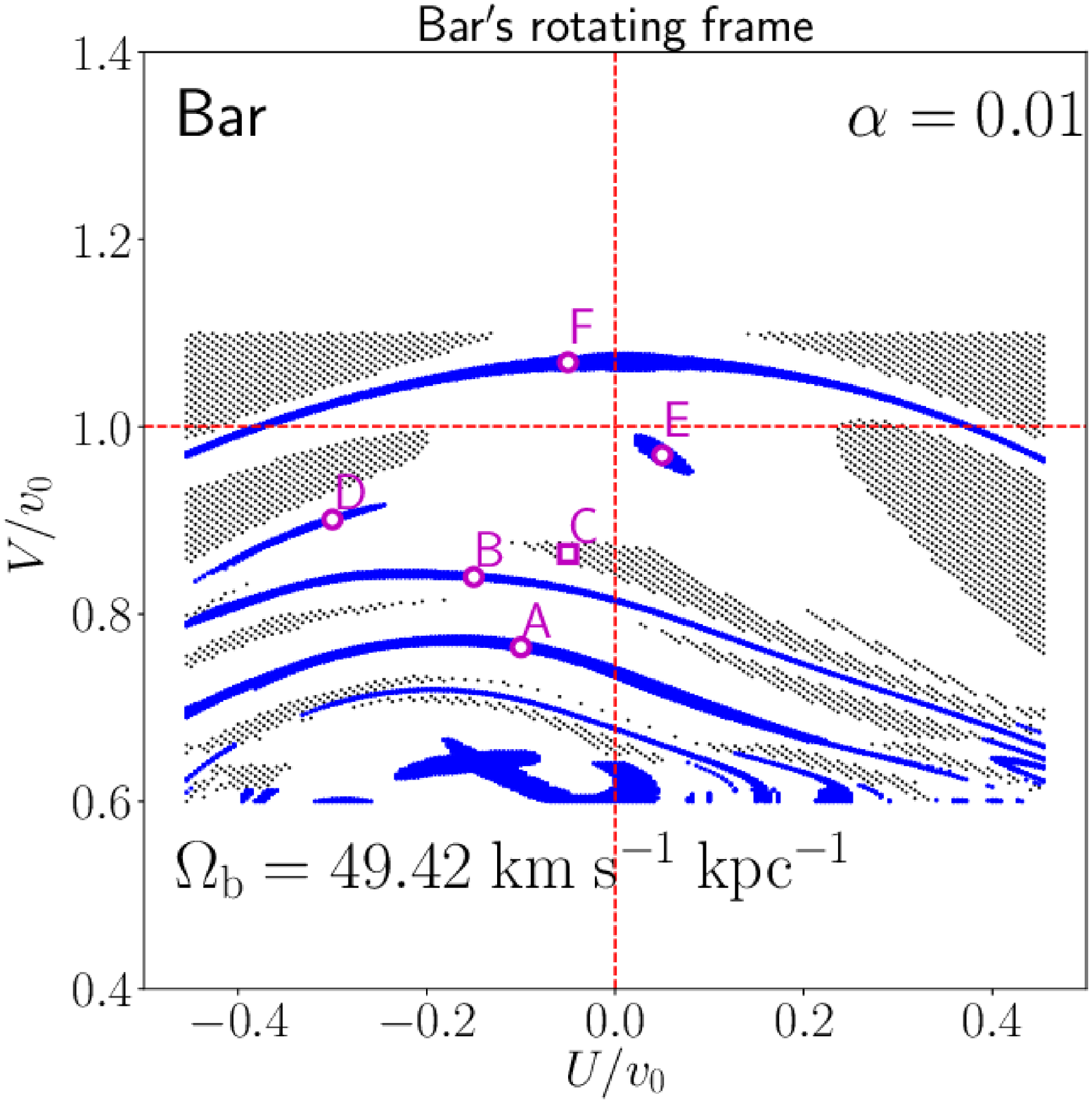} 
 \includegraphics[width=0.8\columnwidth]{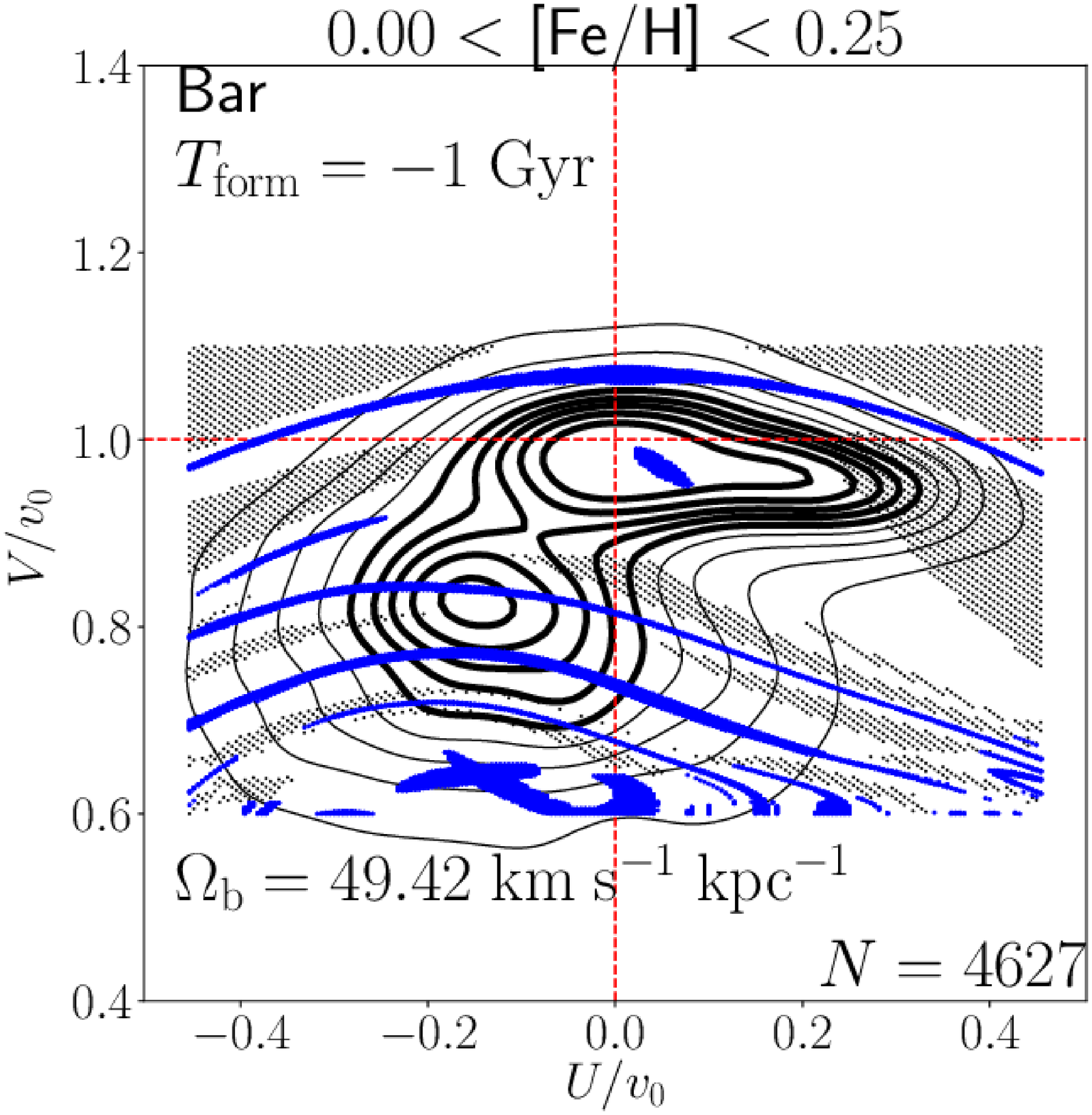} \\
 \includegraphics[width=0.5\columnwidth]{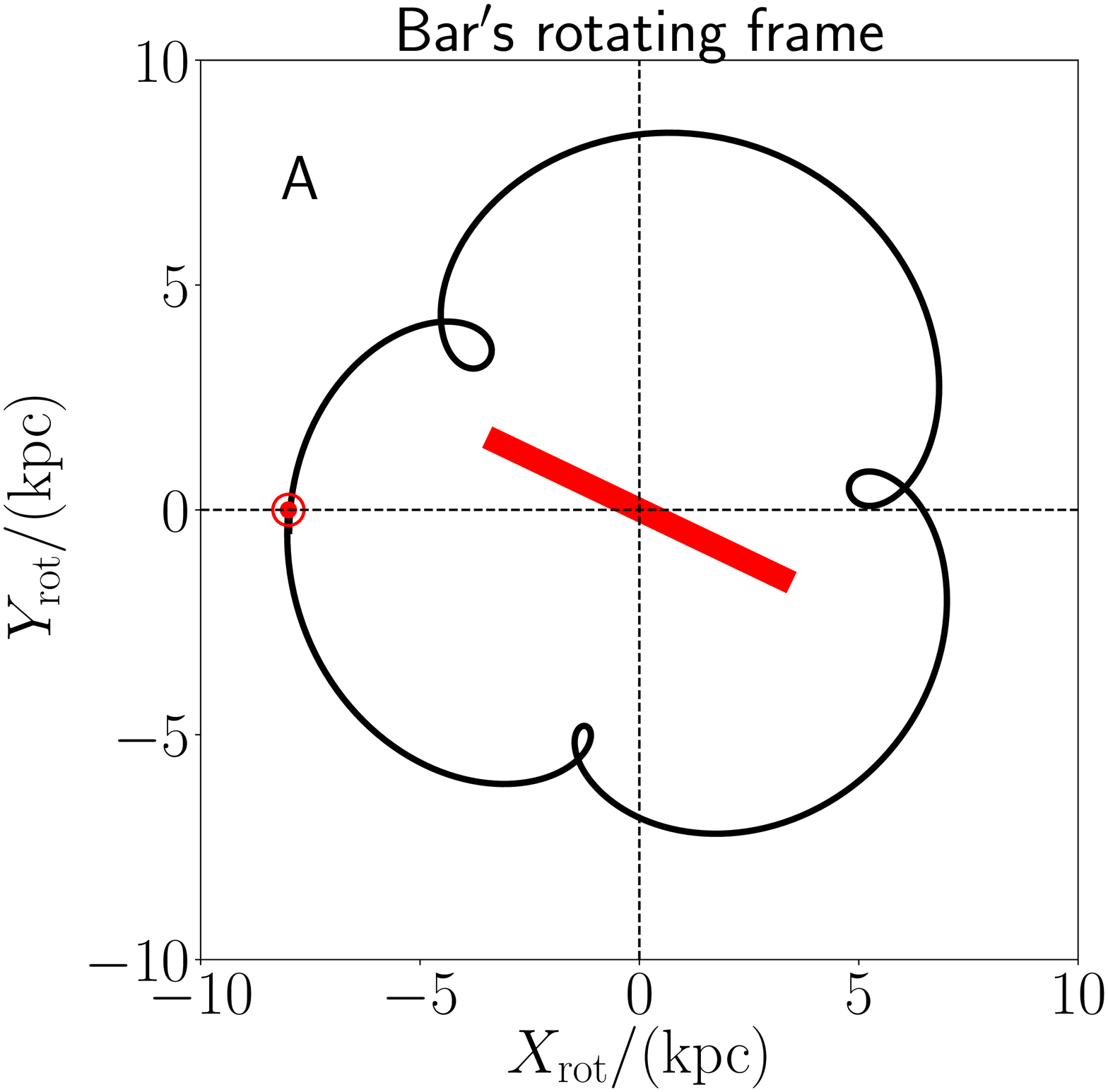} 
 \includegraphics[width=0.5\columnwidth]{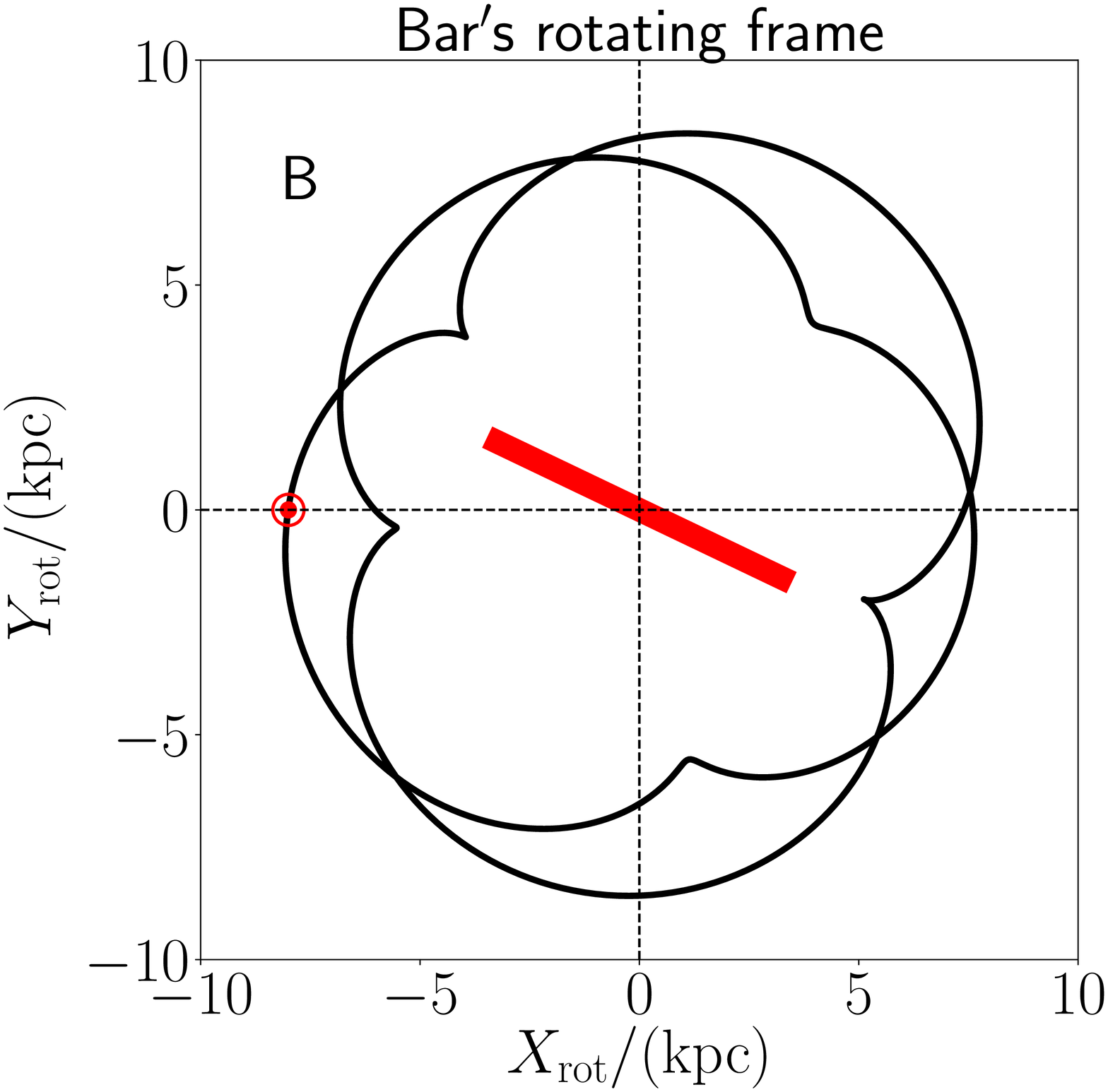} 
 \includegraphics[width=0.5\columnwidth]{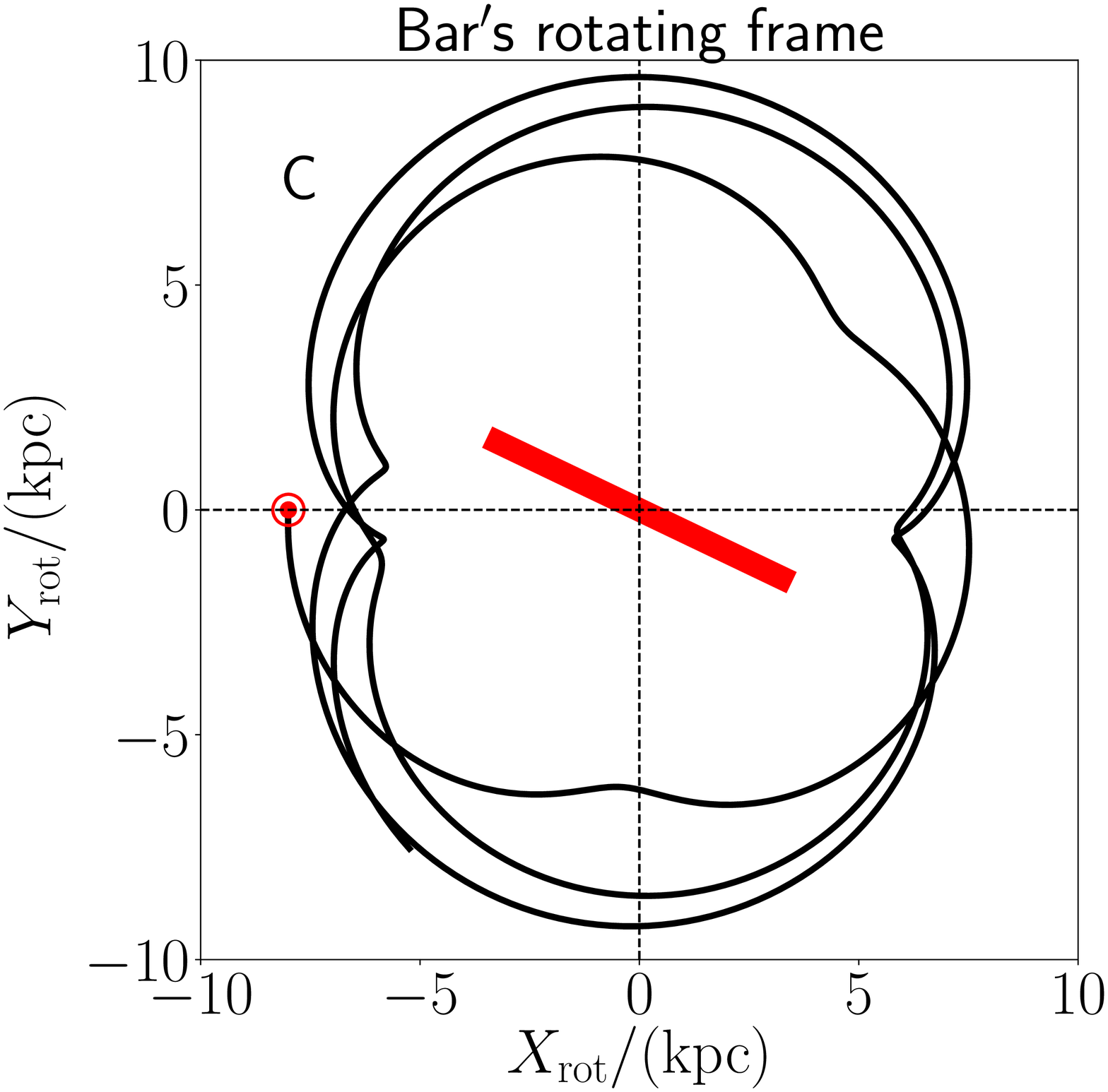} \\
 \includegraphics[width=0.5\columnwidth]{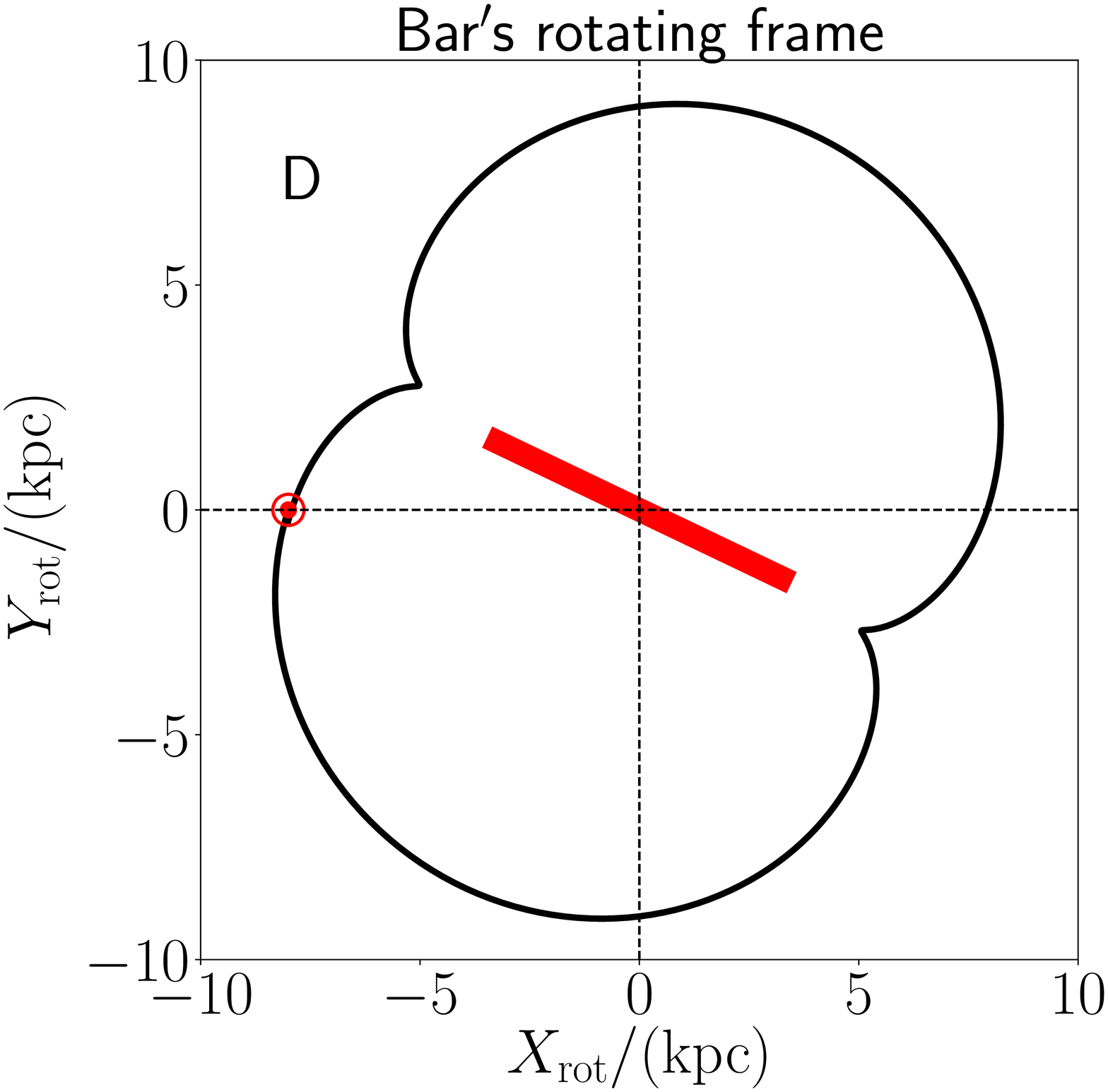} 
 \includegraphics[width=0.5\columnwidth]{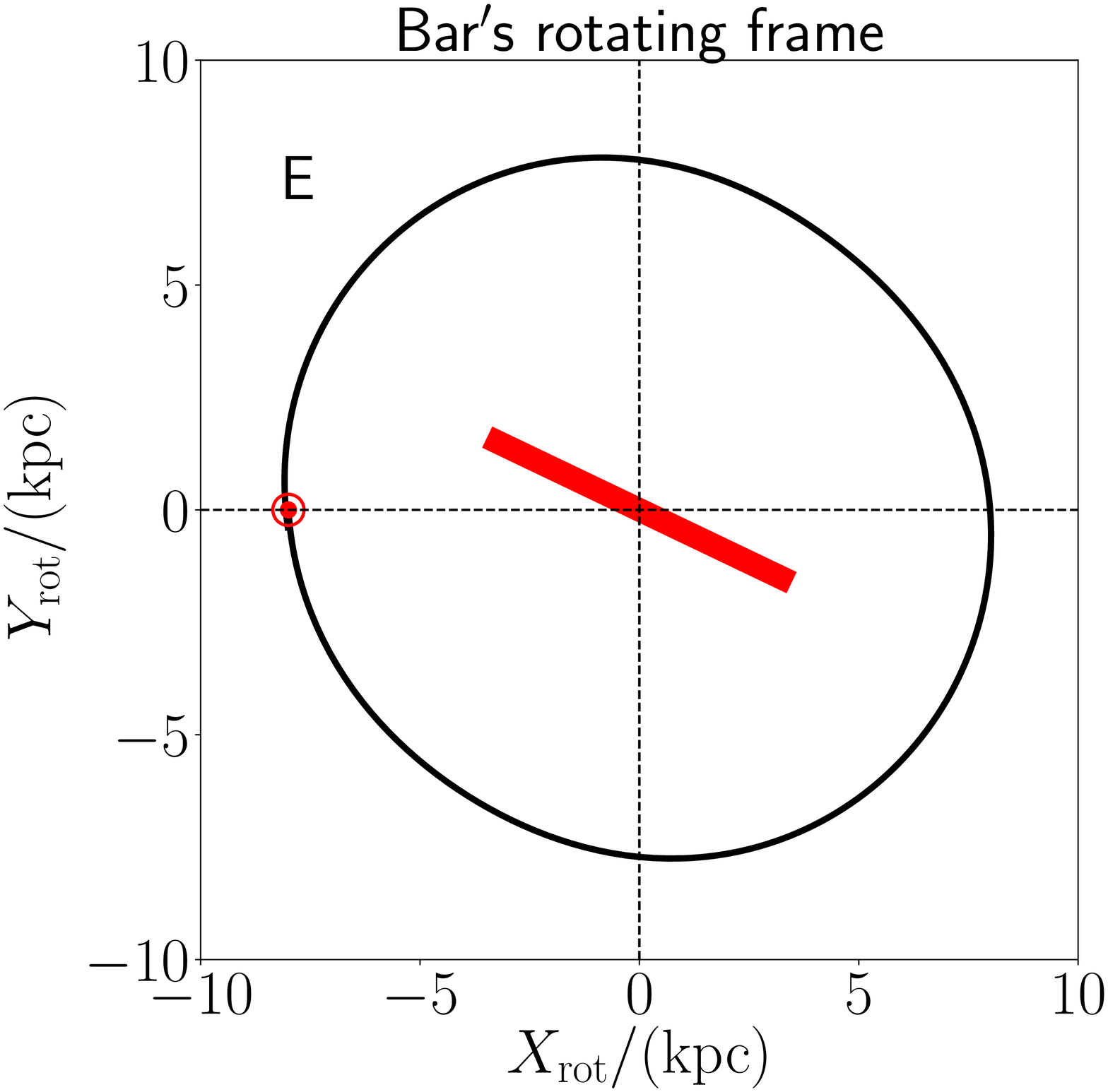} 
 \includegraphics[width=0.5\columnwidth]{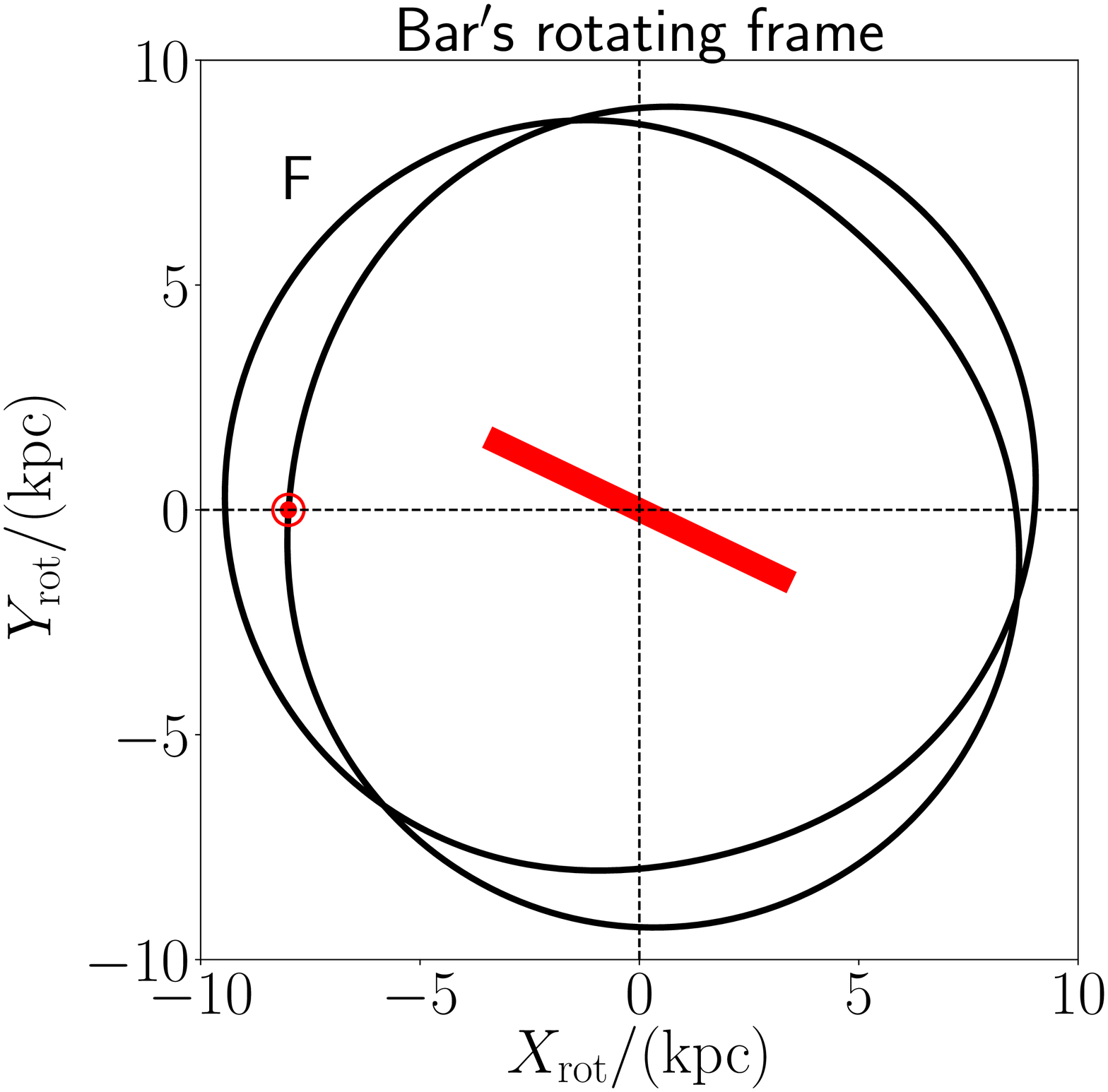} 
\caption{
Orbital analyses of a successful, fast-bar-only model with $\Omegab=49.42 \kmskpc$. 
We note that this model reproduces observational properties (P1)-(P3) in Section \ref{section:data}. 
(Top row)
The velocity distribution of Solar-neighbour stars in our test-particle simulation at different [Fe/H] regions. 
The sample size $N$ is shown at the bottom of each panel. 
(Second row, left) 
The locations in $(U,V)$-plane of Solar neighbour disc stars with closed and non-closed orbits 
mapped by non-closed-ness parameter $\eta$ defined in equation (\ref{eq:eta}). 
The blue regions correspond to nearly closed orbits with $\eta < 0.03$, 
while the black shaded regions correspond to highly non-closed orbits with $\eta>0.3$. 
Unmarked (white) regions correspond to intermediate orbits with $0.03<\eta<0.3$. 
(Second row, right) 
The same as the left-hand panel on the second row, but the simulated velocity distribution of metal-rich stars is plotted as well. 
Comparison of the panels on the second row enables us to characterise the orbits for each substructure in the velocity space. 
(Third and fourth rows)
Some representative orbits in the bar's rotating frame $(X_\mathrm{rot}, Y_\mathrm{rot})$. 
The corresponding locations in the $(U,V)$ plane are marked in the left-hand panel in the second row. 
The red dot at $X_\mathrm{rot} = -8 \kpc$ is the current Solar location. 
The tilted red line segment at the centre represents the length and the current orientation of the bar. 
}
\label{fig_bar49closed}
\end{center}
\end{figure*}

\begin{figure*}
\begin{center}
 \includegraphics[width=0.6\columnwidth]{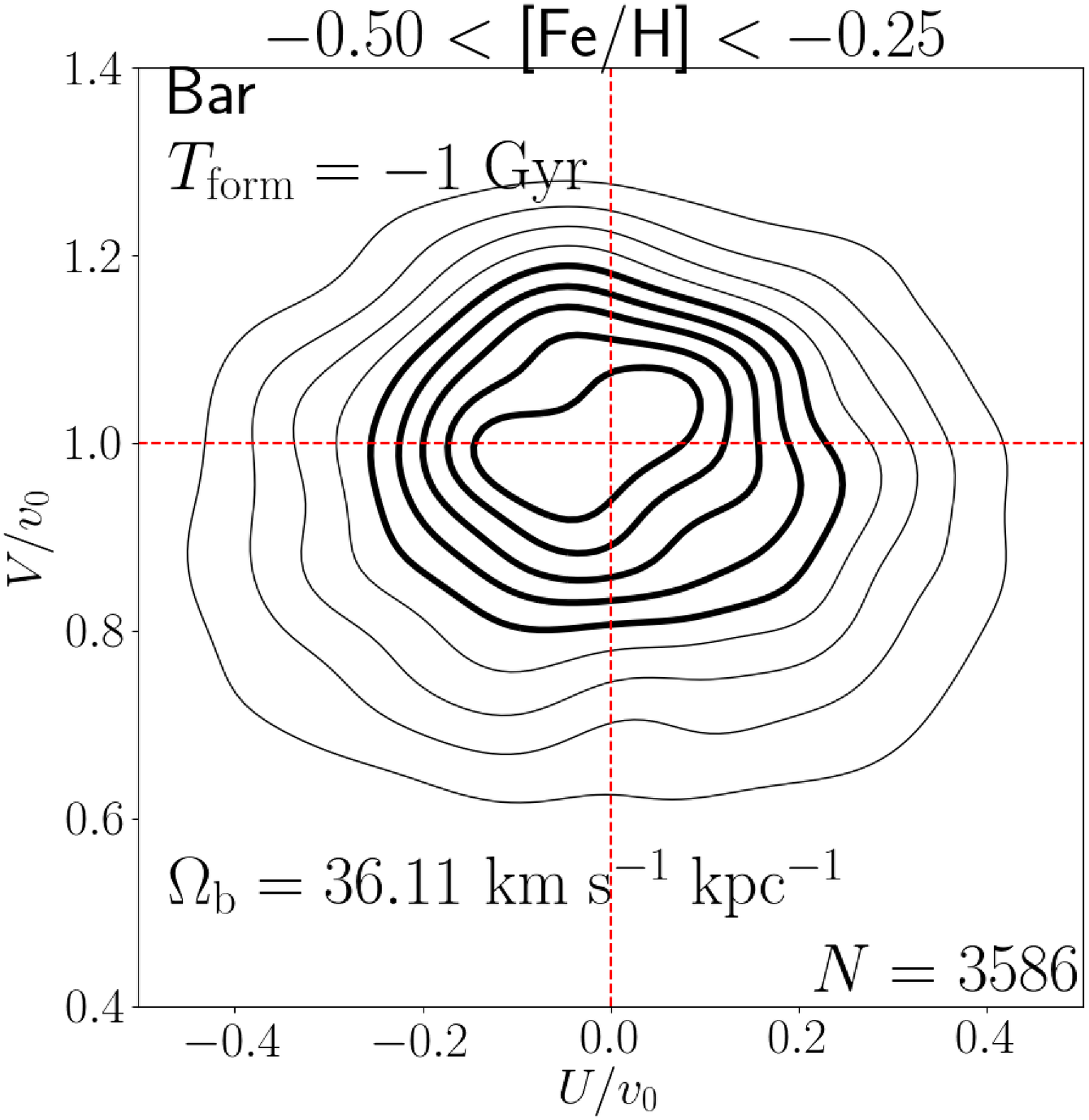} 
 \includegraphics[width=0.6\columnwidth]{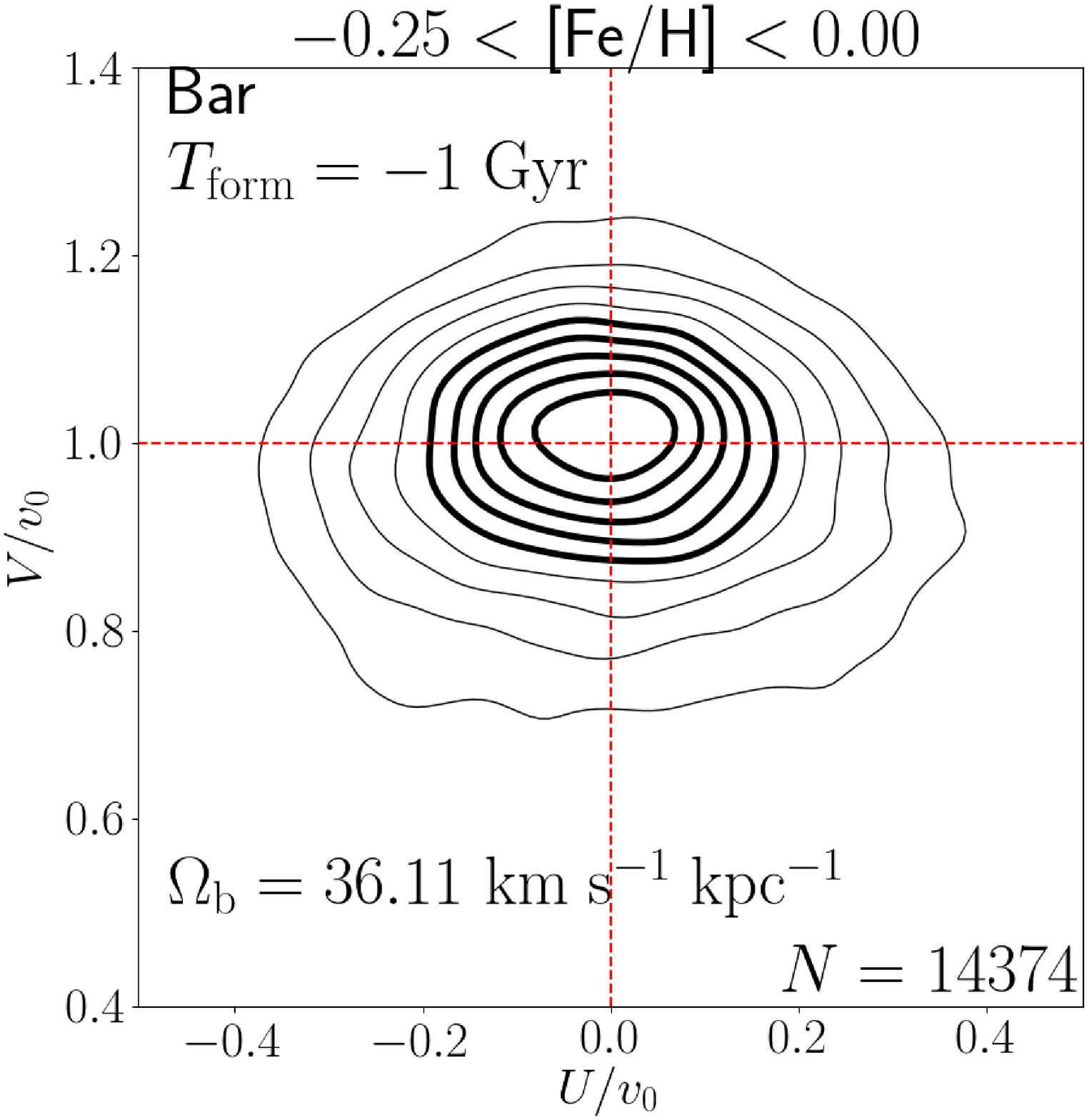} 
 \includegraphics[width=0.6\columnwidth]{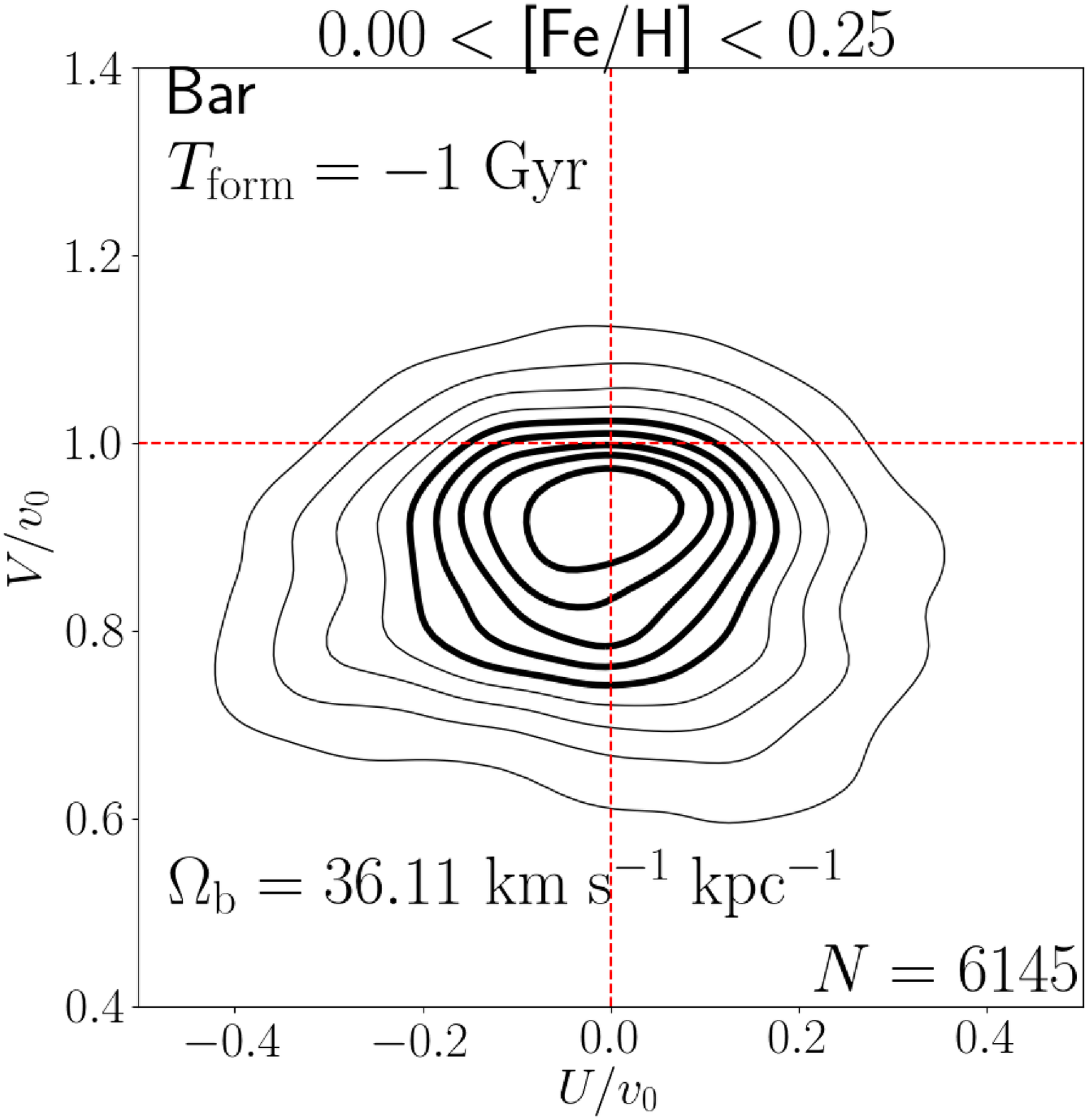} \\
 \includegraphics[width=0.8\columnwidth]{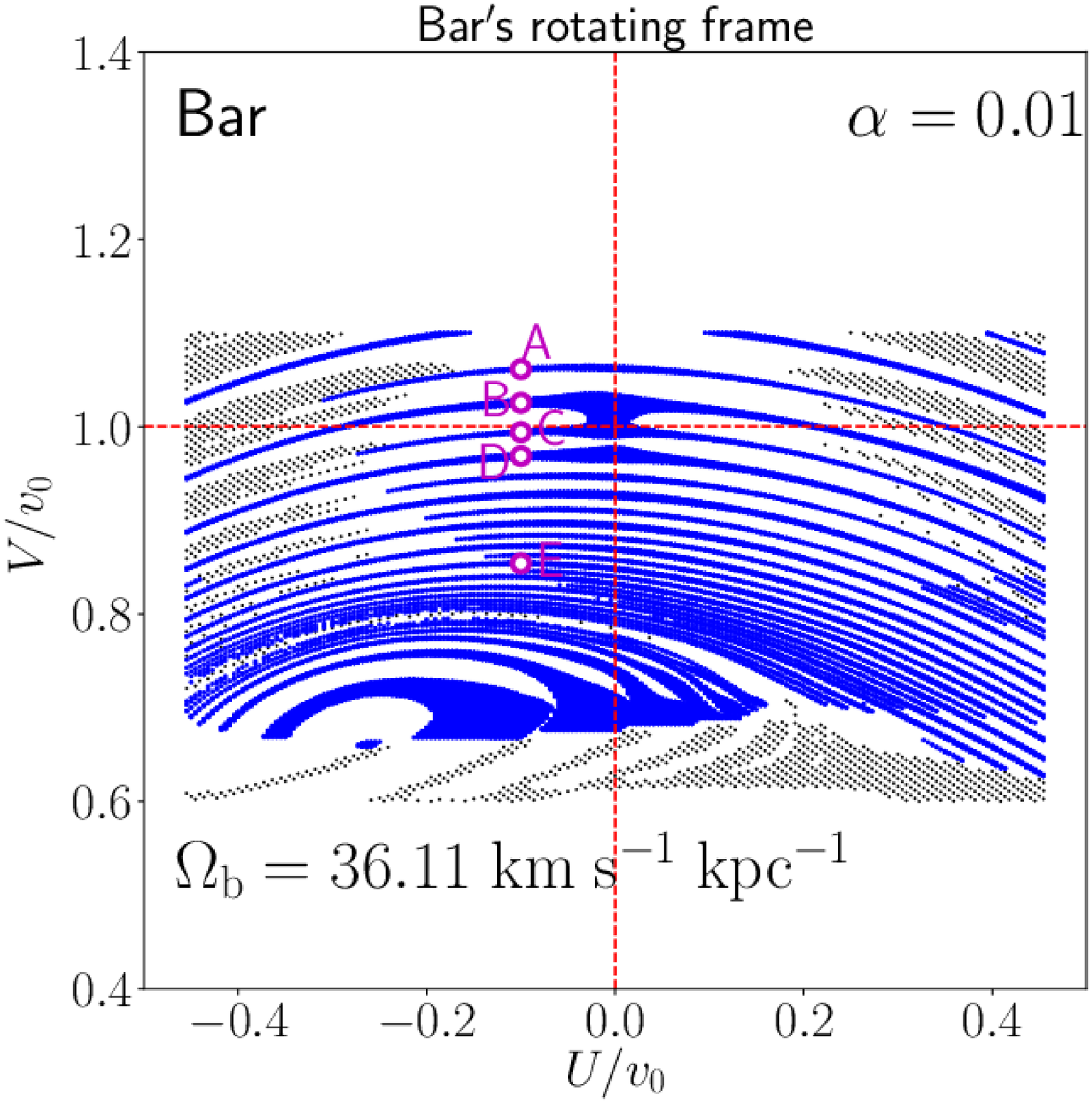} 
 \includegraphics[width=0.8\columnwidth]{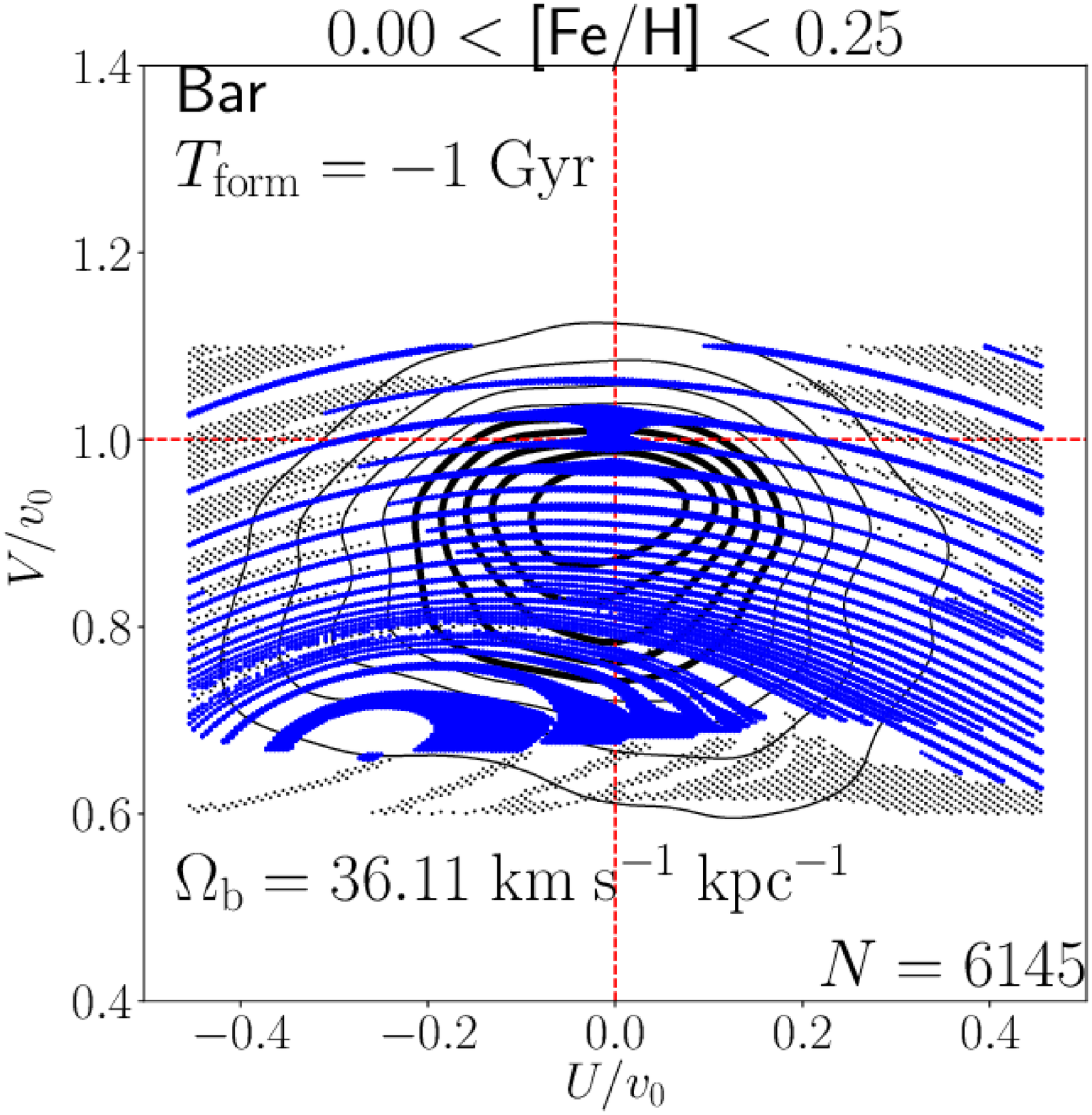} \\
 \includegraphics[width=0.5\columnwidth]{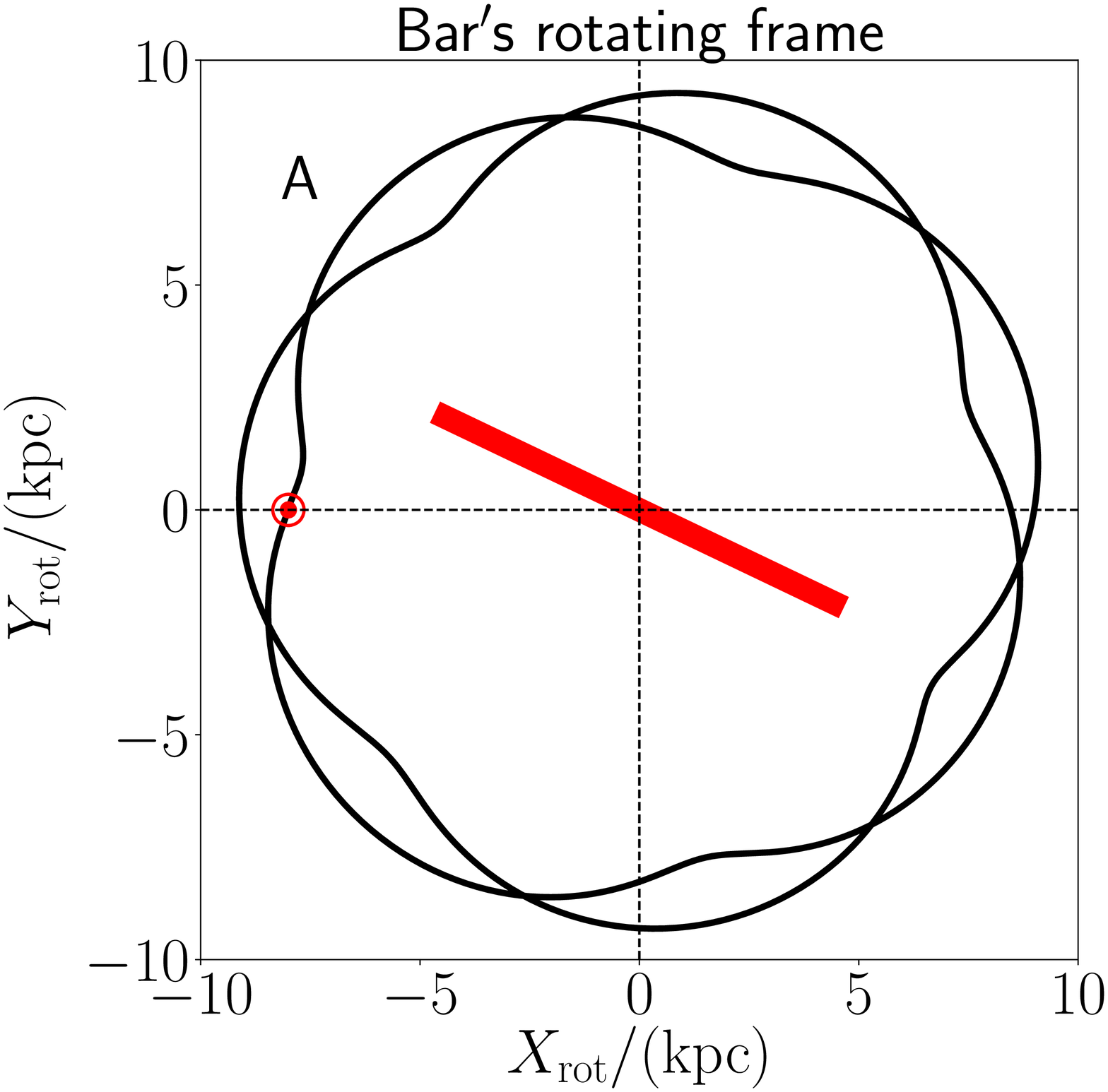} 
 \includegraphics[width=0.5\columnwidth]{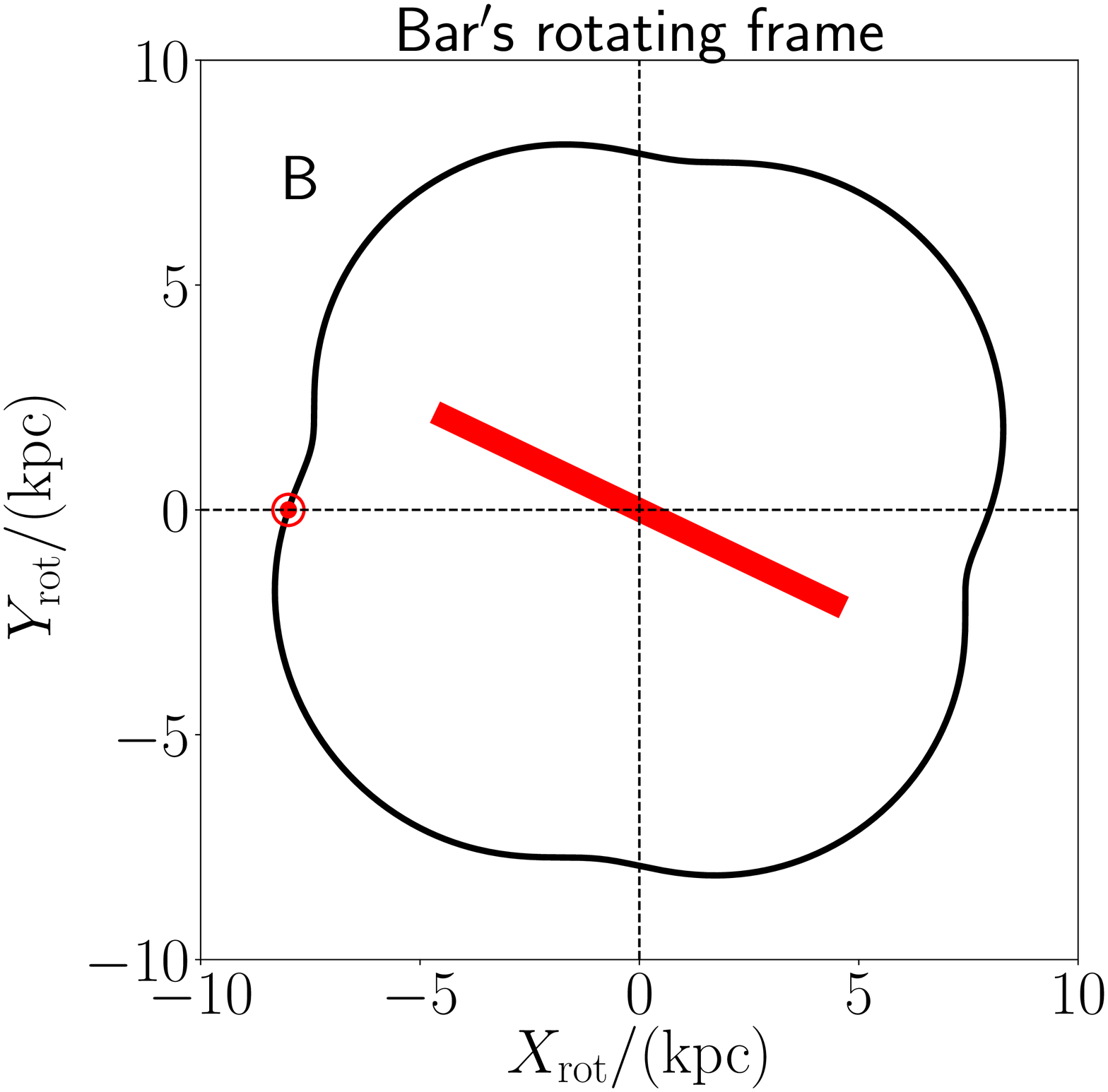} 
 \includegraphics[width=0.5\columnwidth]{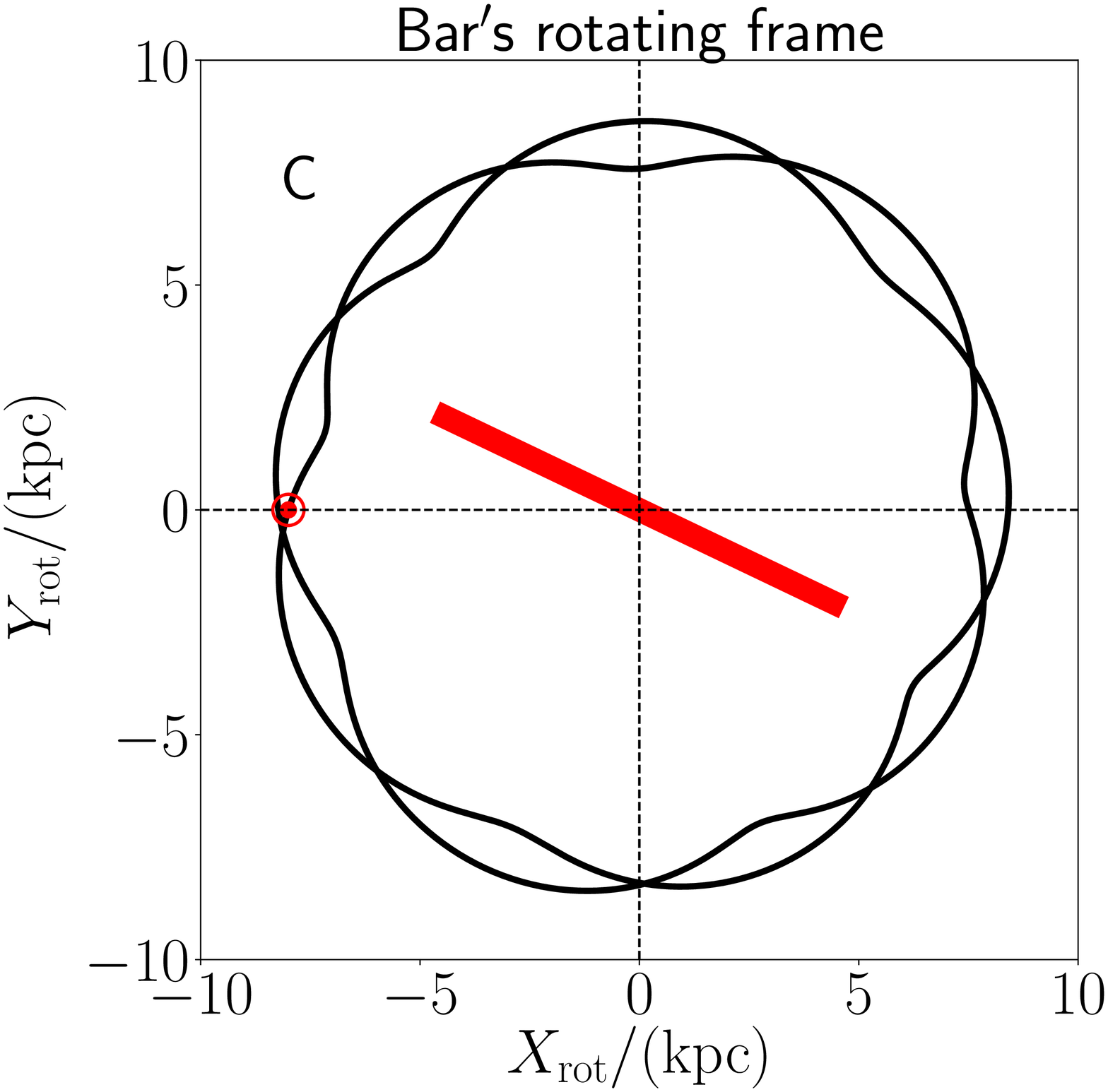} \\
 \includegraphics[width=0.5\columnwidth]{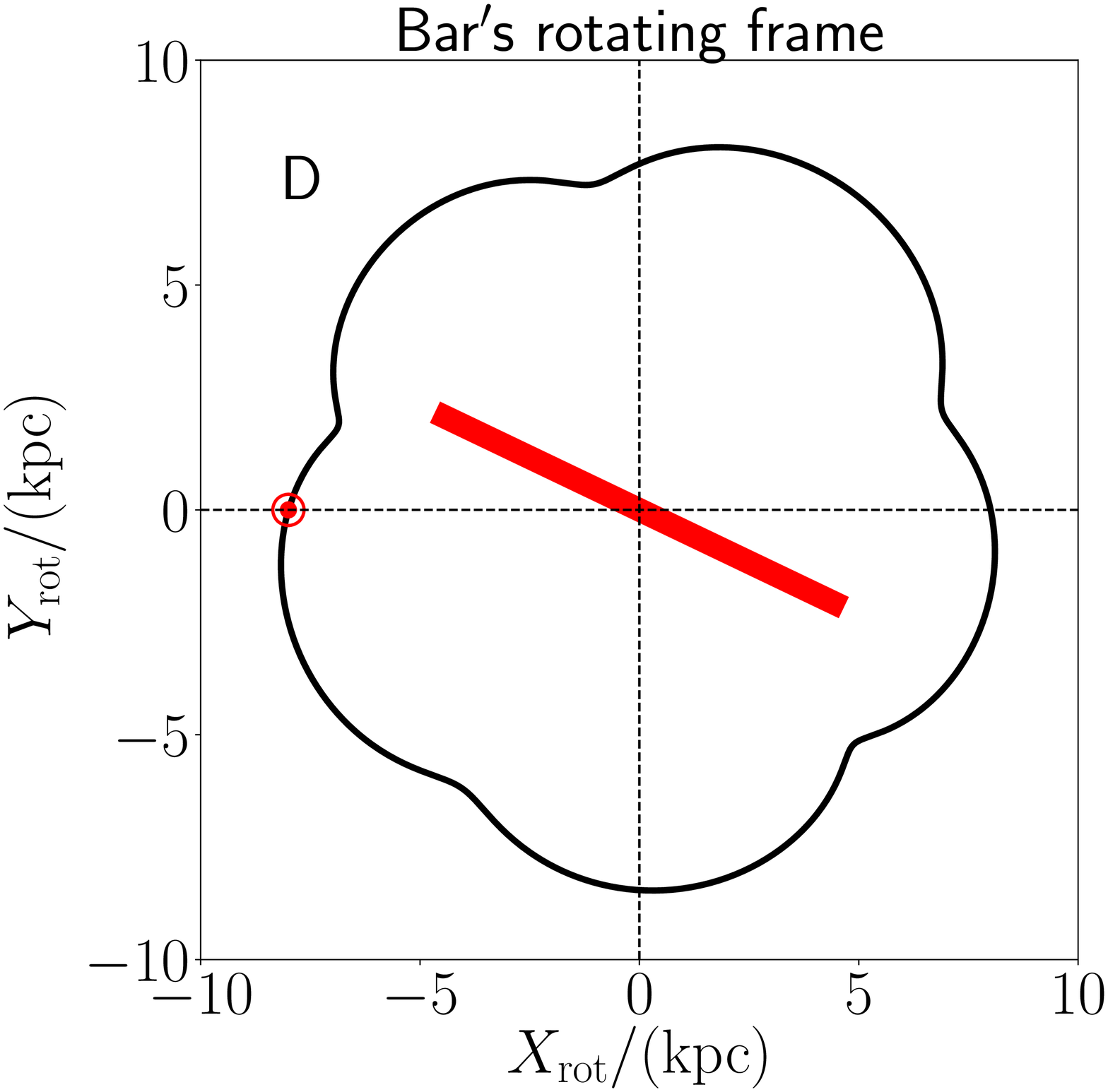} 
 \includegraphics[width=0.5\columnwidth]{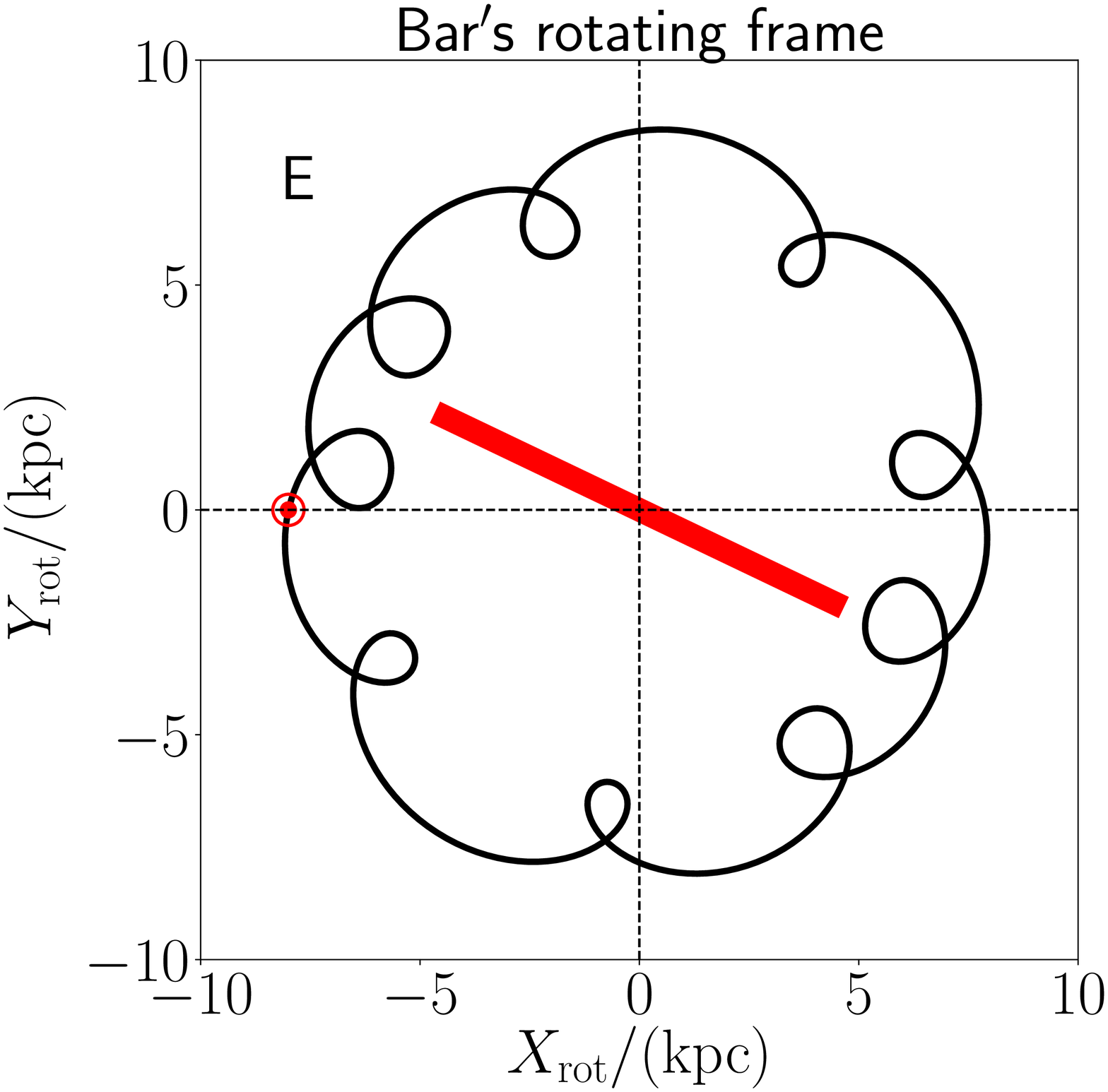} 
\caption{
Similar to Fig. \ref{fig_bar49closed}, but in the case of an unsuccessful slow-bar-only model with $\Omegab=36.11 \kmskpc$. 
We note that this model does not show bimodal velocity distribution at any [Fe/H] range (see the observational property (P1) in Section \ref{section:data}). 
On the left-hand panel of the second row, 
the five magenta markers at $U/v_0=-0.1$ correspond to the orbit A, B, C, D, and E (in descending order of $V$). 
On the third and fourth row, 
the bar's rotating frame $(X_\mathrm{rot}, Y_\mathrm{rot})$ is used. 
This slow bar is assumed to be longer than the fast bar in Fig. \ref{fig_bar49closed}. 
}
\label{fig_bar36closed}
\end{center}
\end{figure*}

\begin{figure*}
\begin{center}
 \includegraphics[width=0.6\columnwidth]{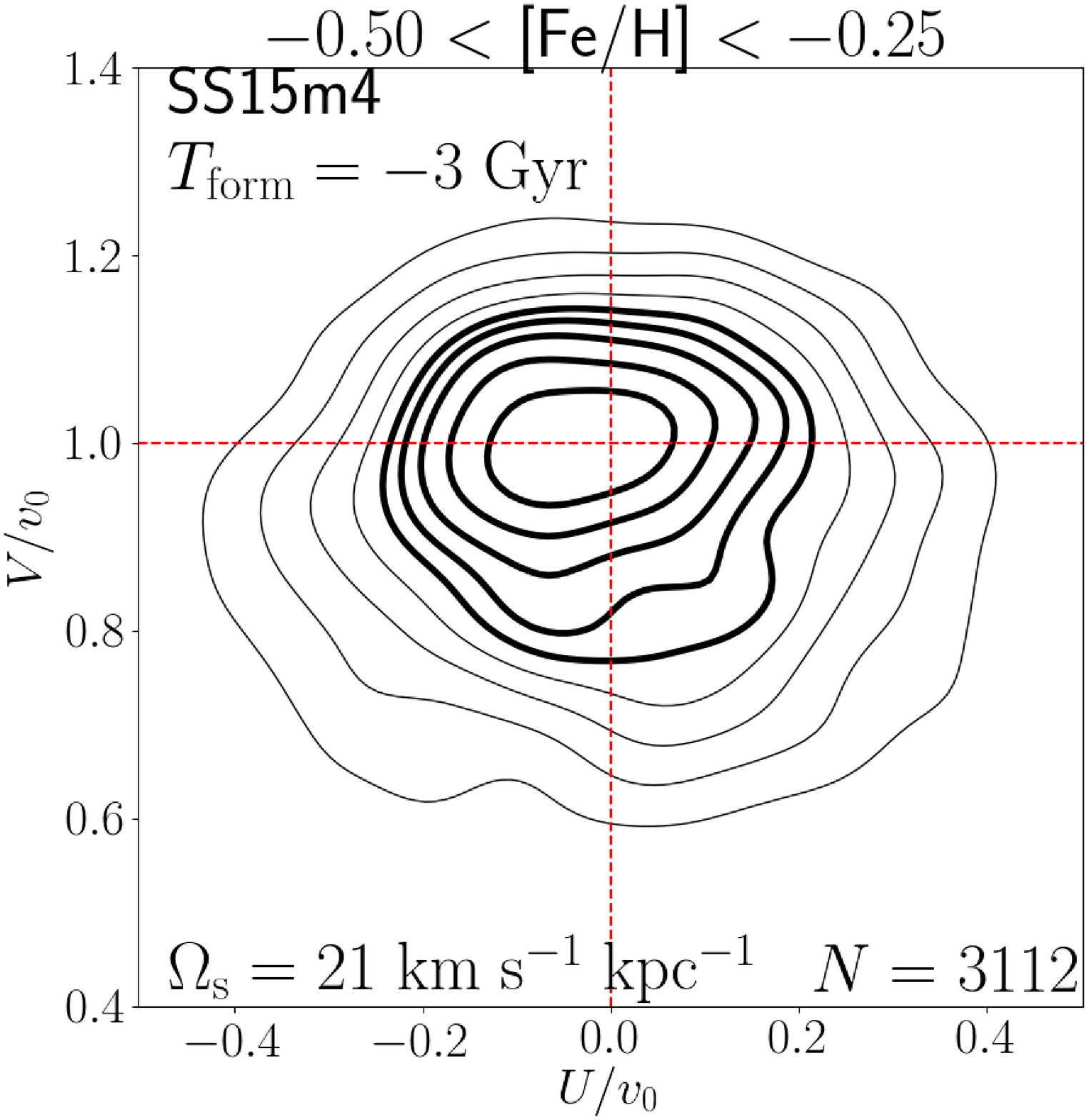} 
 \includegraphics[width=0.6\columnwidth]{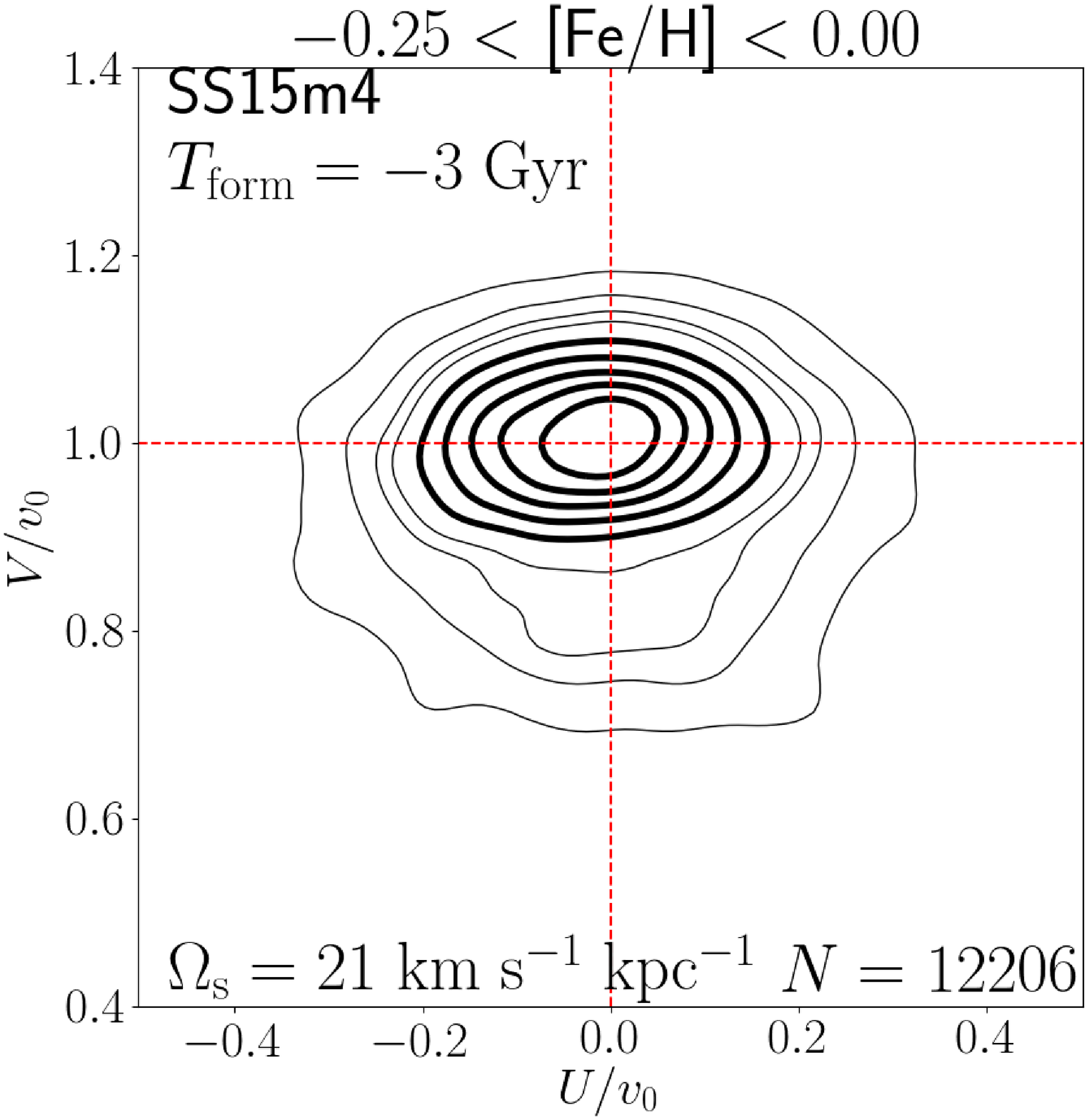} 
 \includegraphics[width=0.6\columnwidth]{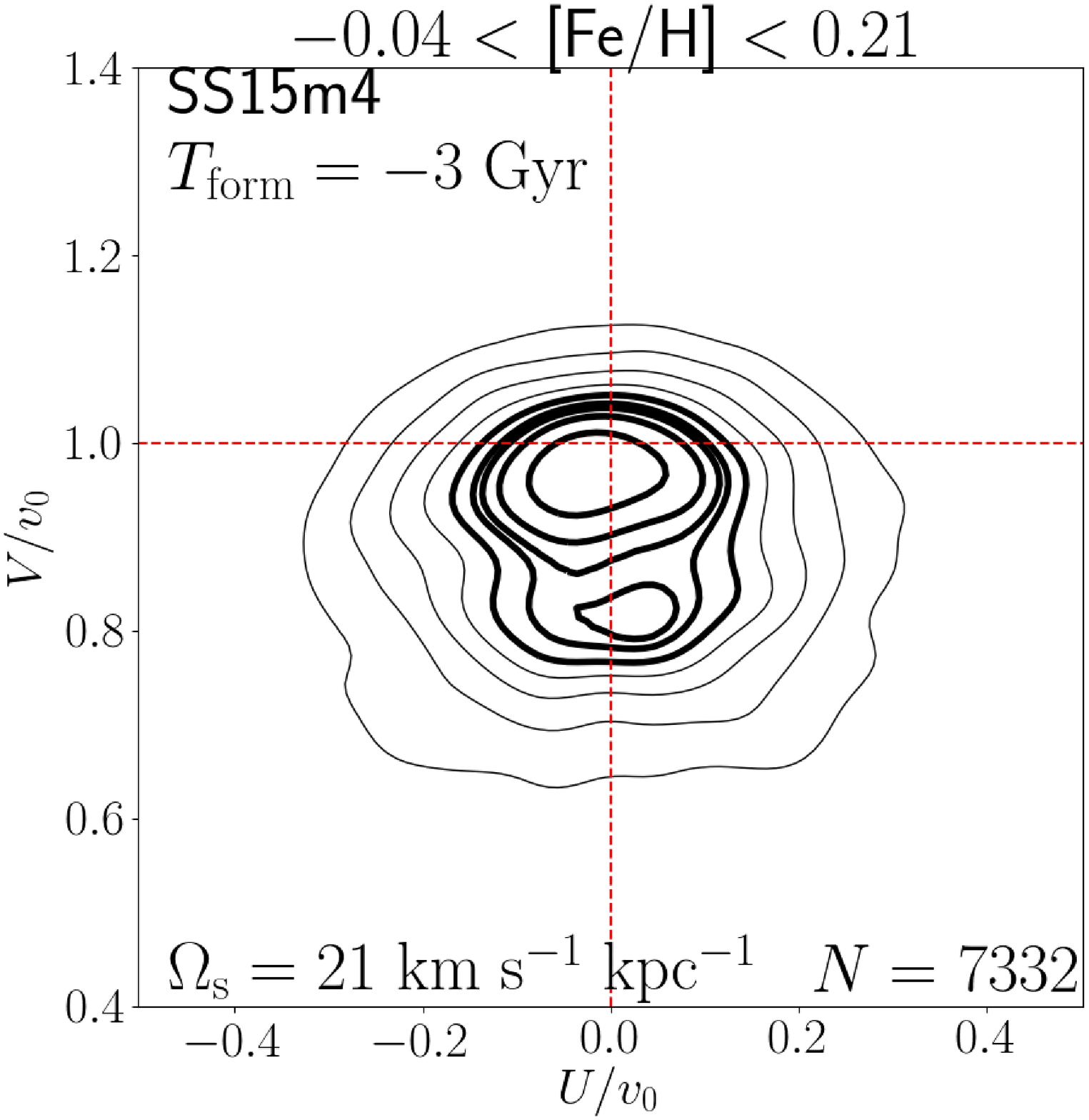} \\
 \includegraphics[width=0.8\columnwidth]{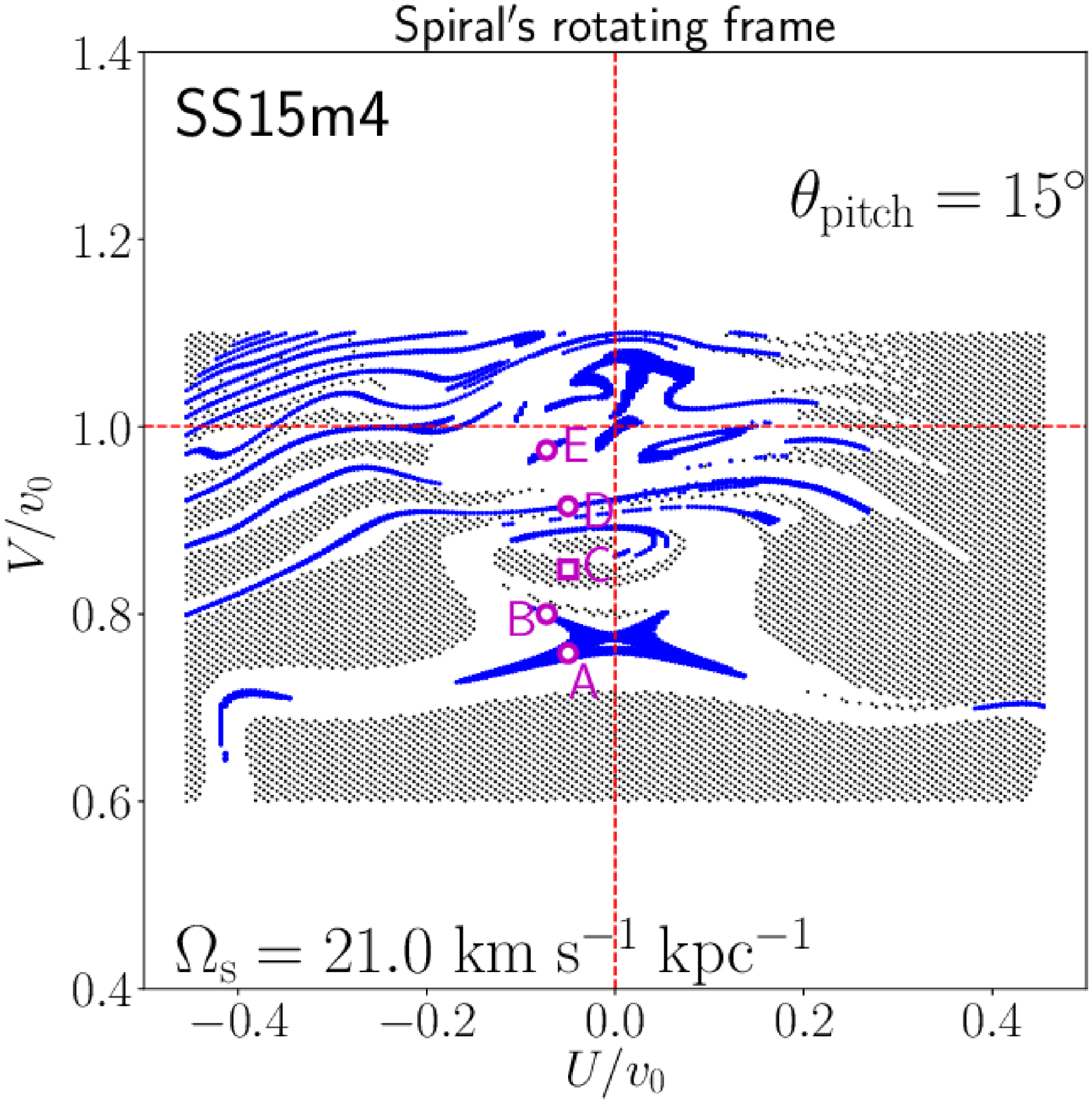} 
 \includegraphics[width=0.8\columnwidth]{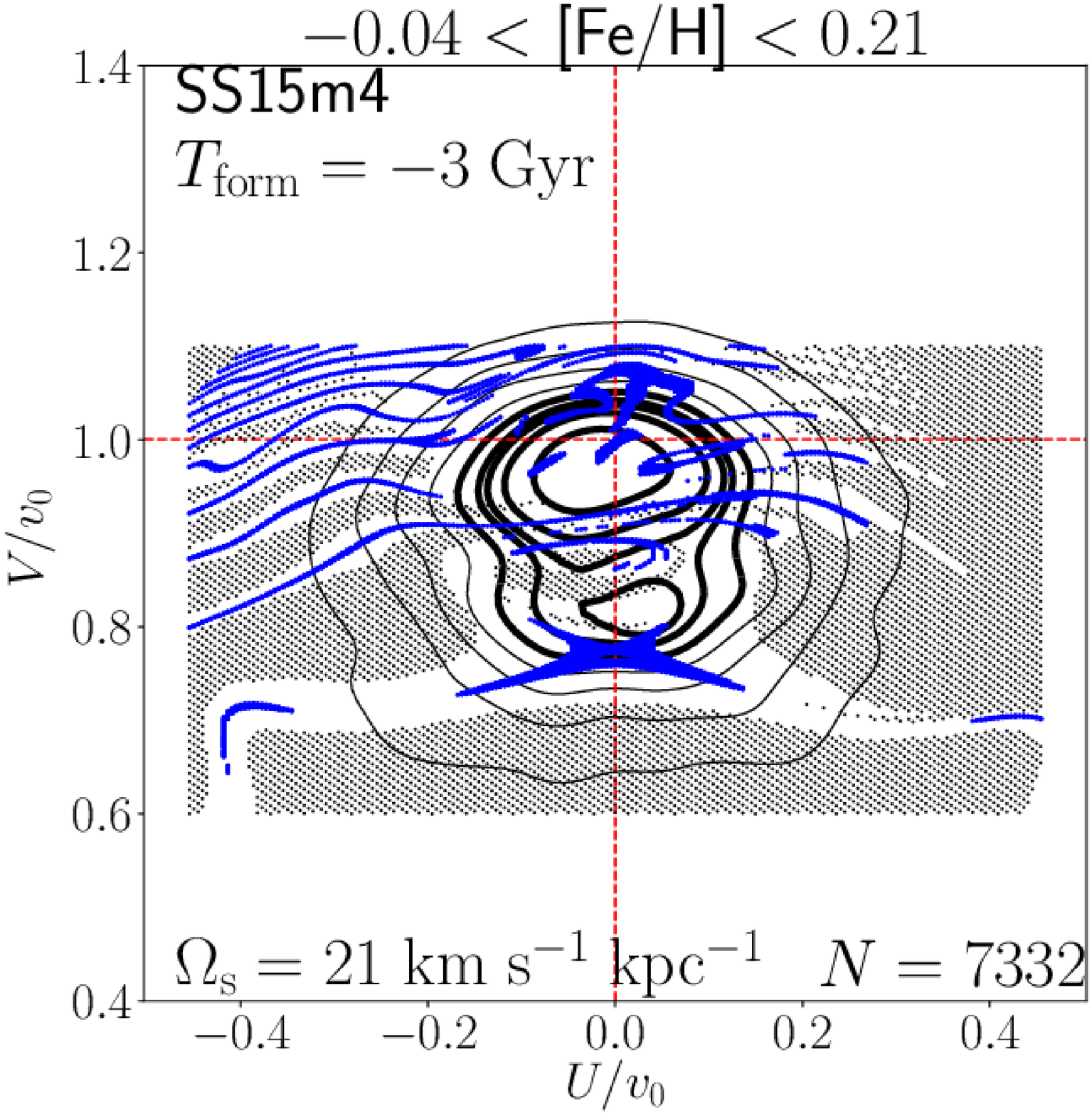} \\
 \includegraphics[width=0.5\columnwidth]{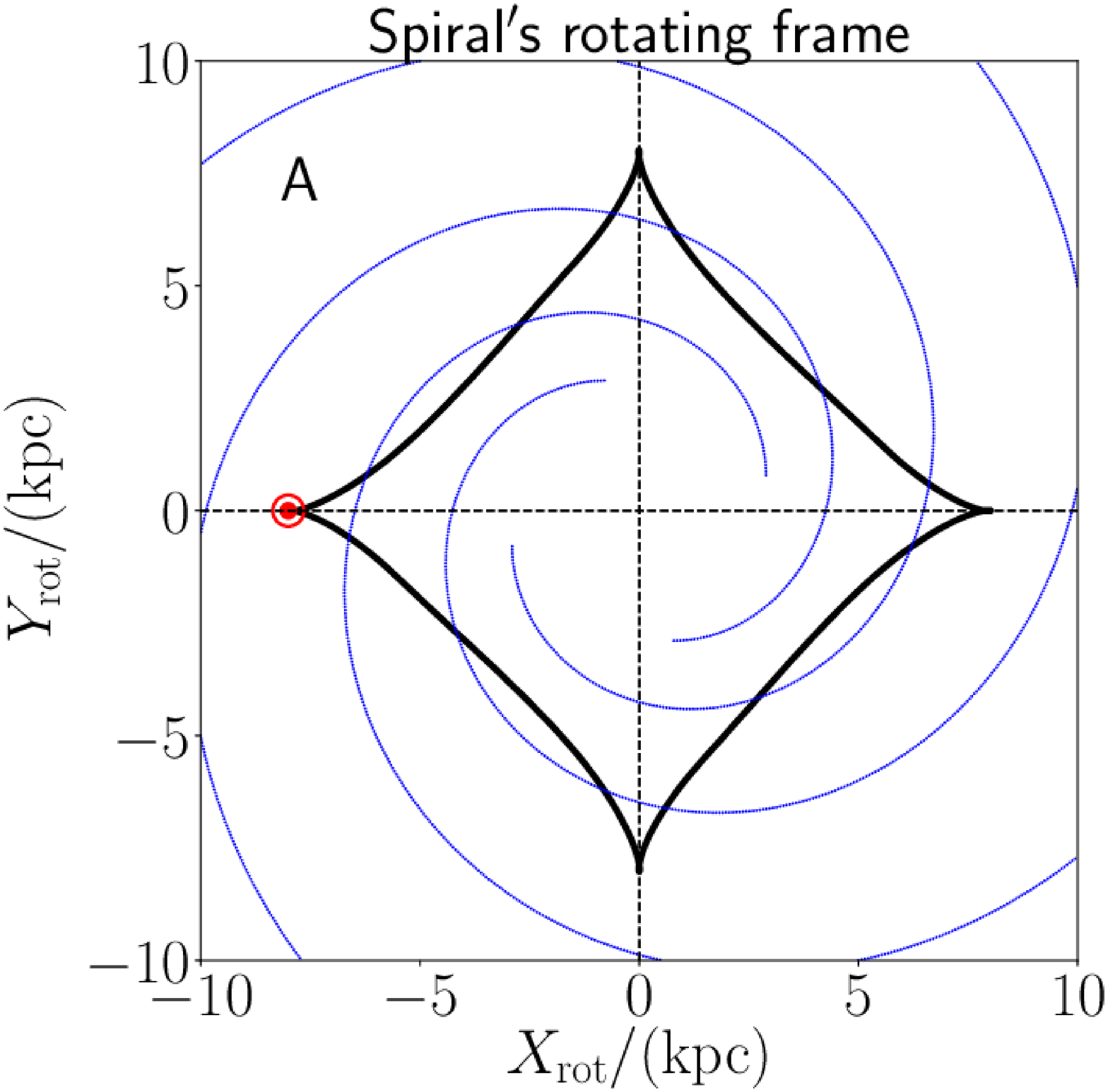} 
 \includegraphics[width=0.5\columnwidth]{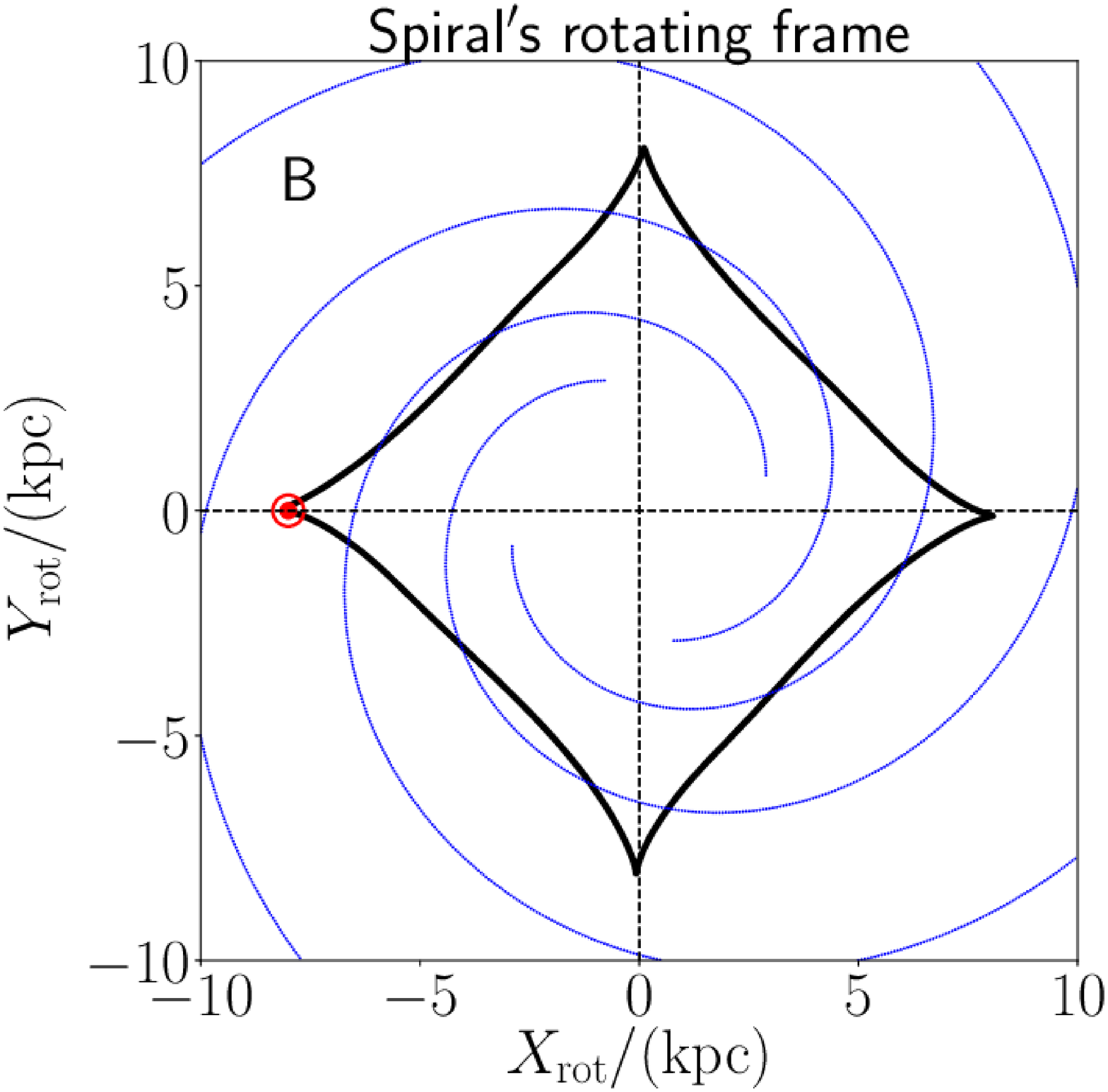} 
 \includegraphics[width=0.5\columnwidth]{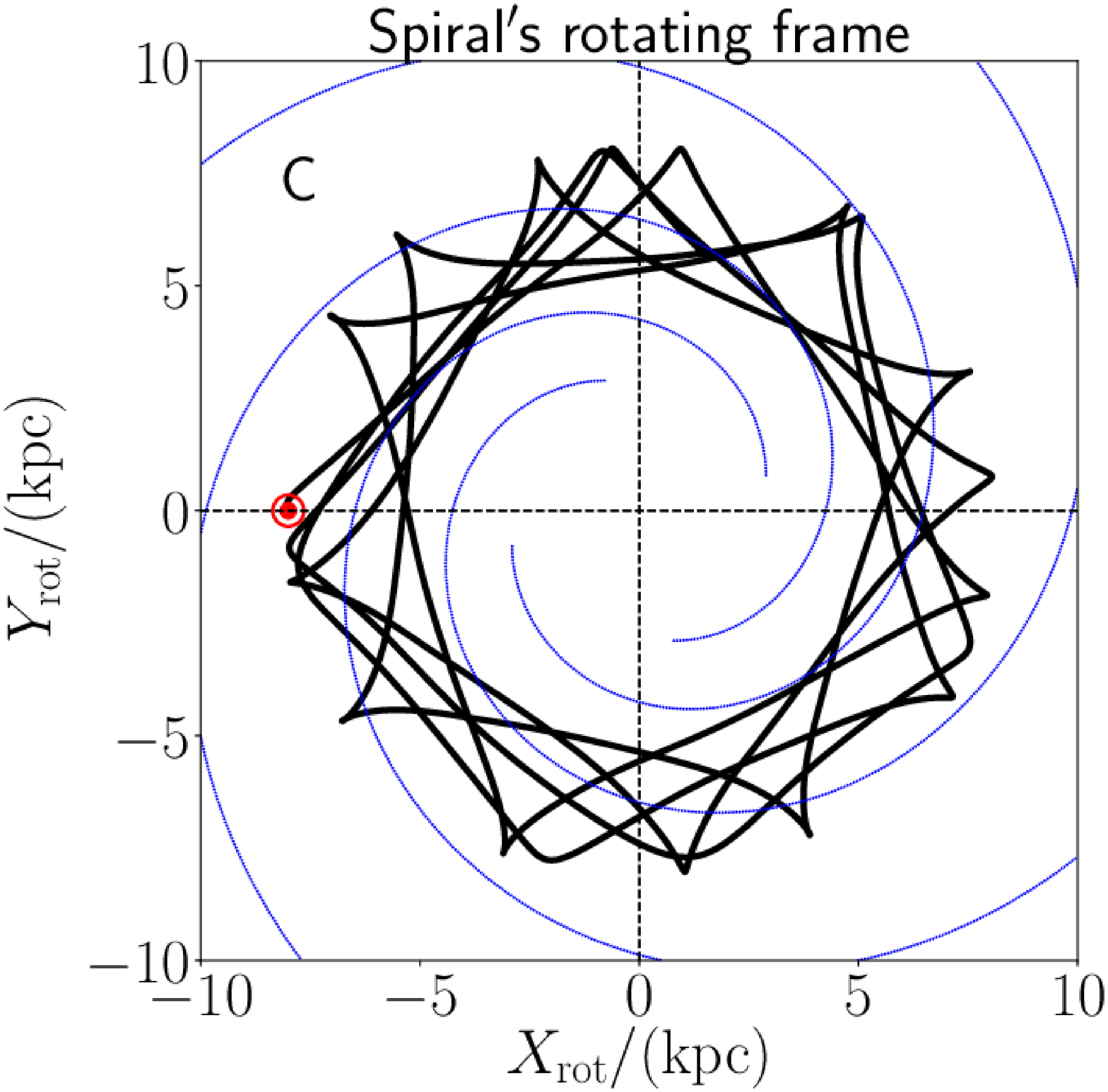} \\
 \includegraphics[width=0.5\columnwidth]{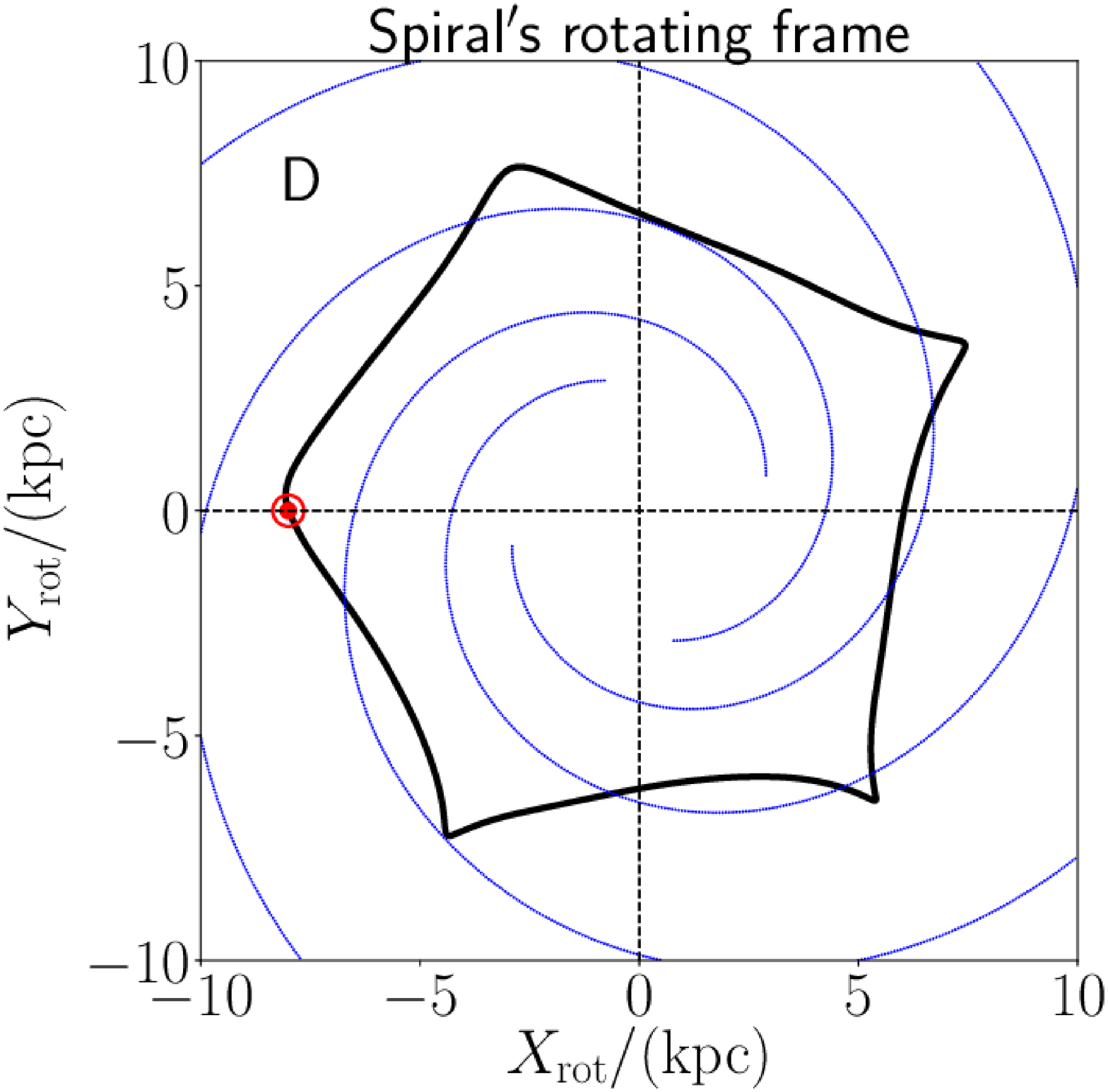} 
 \includegraphics[width=0.5\columnwidth]{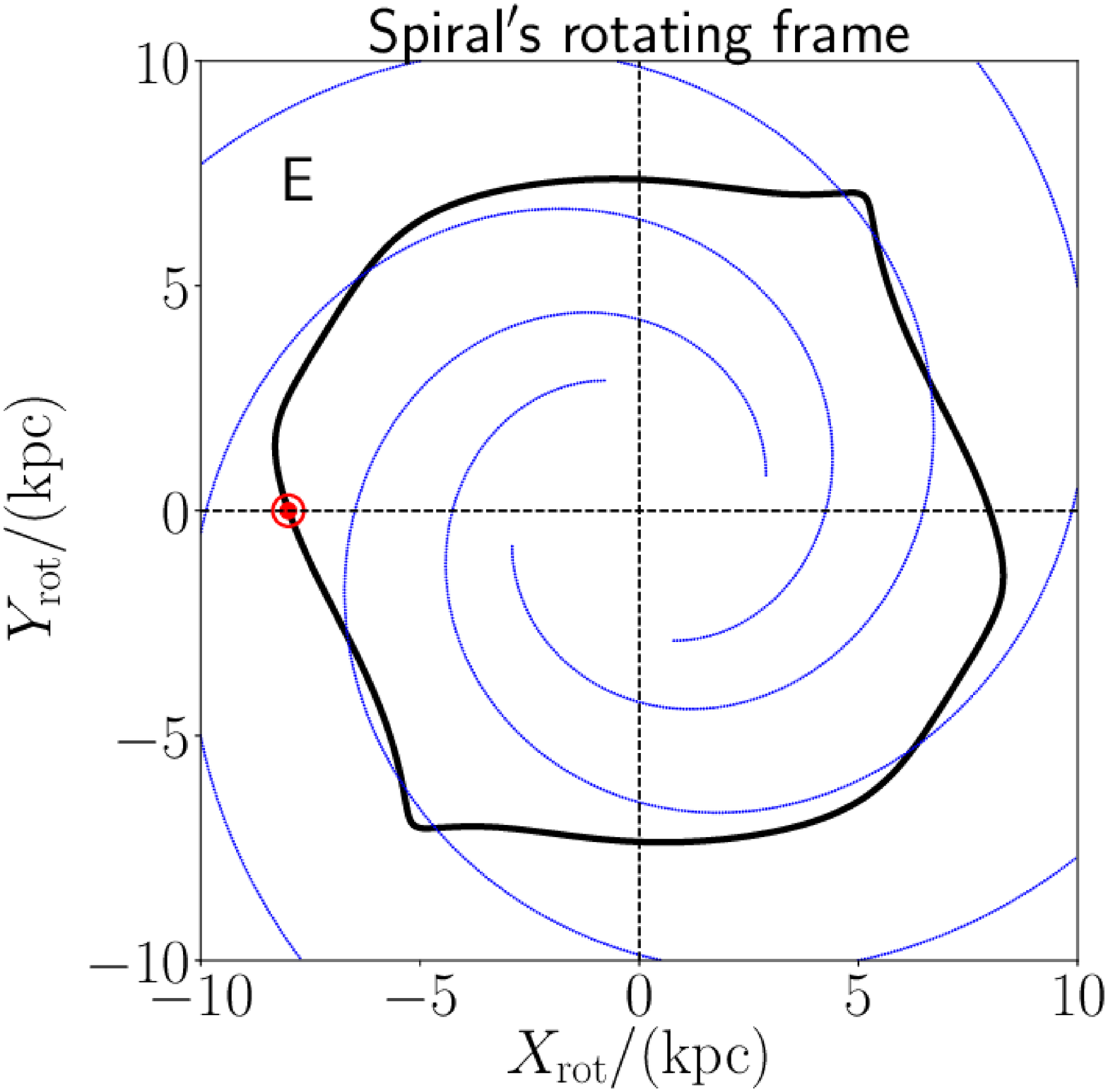} 
\caption{
Similar to Fig. \ref{fig_bar49closed}, but in the case of a partially successful, spiral-only model with $(\Omegas, m) = (21, 4)$. 
We note that this model reproduces observational properties (P1) and (P2) in Section \ref{section:data} but not (P3). 
On the third and fourth row, 
the spirals' rotating frame $(X_\mathrm{rot}, Y_\mathrm{rot})$ is used. 
The blue curved lines represents the current location of the spiral arms. 
}
\label{fig_spiral21m4closed}
\end{center}
\end{figure*}

\begin{figure*}
\begin{center}
 \includegraphics[width=0.6\columnwidth]{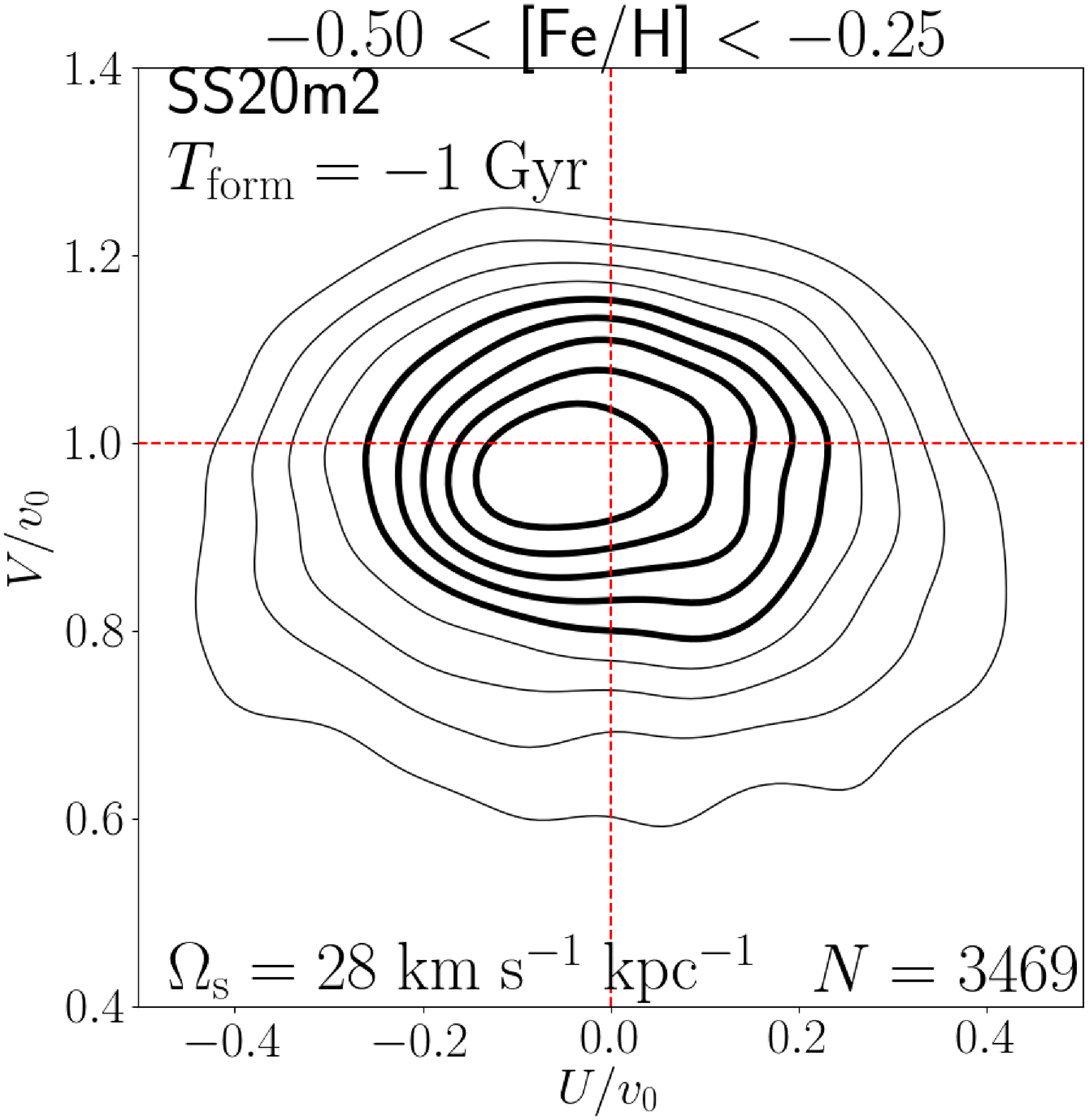} 
 \includegraphics[width=0.6\columnwidth]{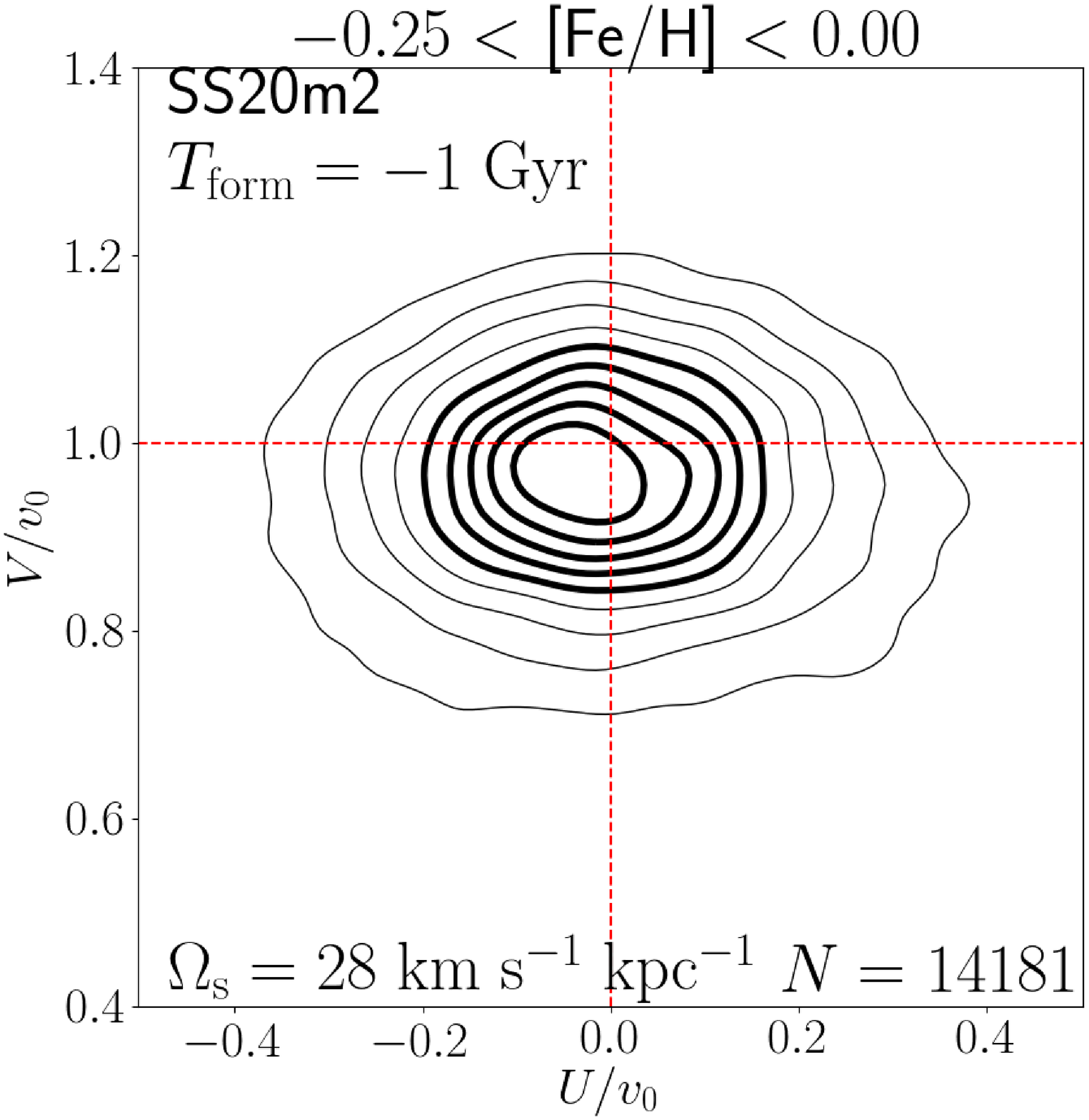} 
 \includegraphics[width=0.6\columnwidth]{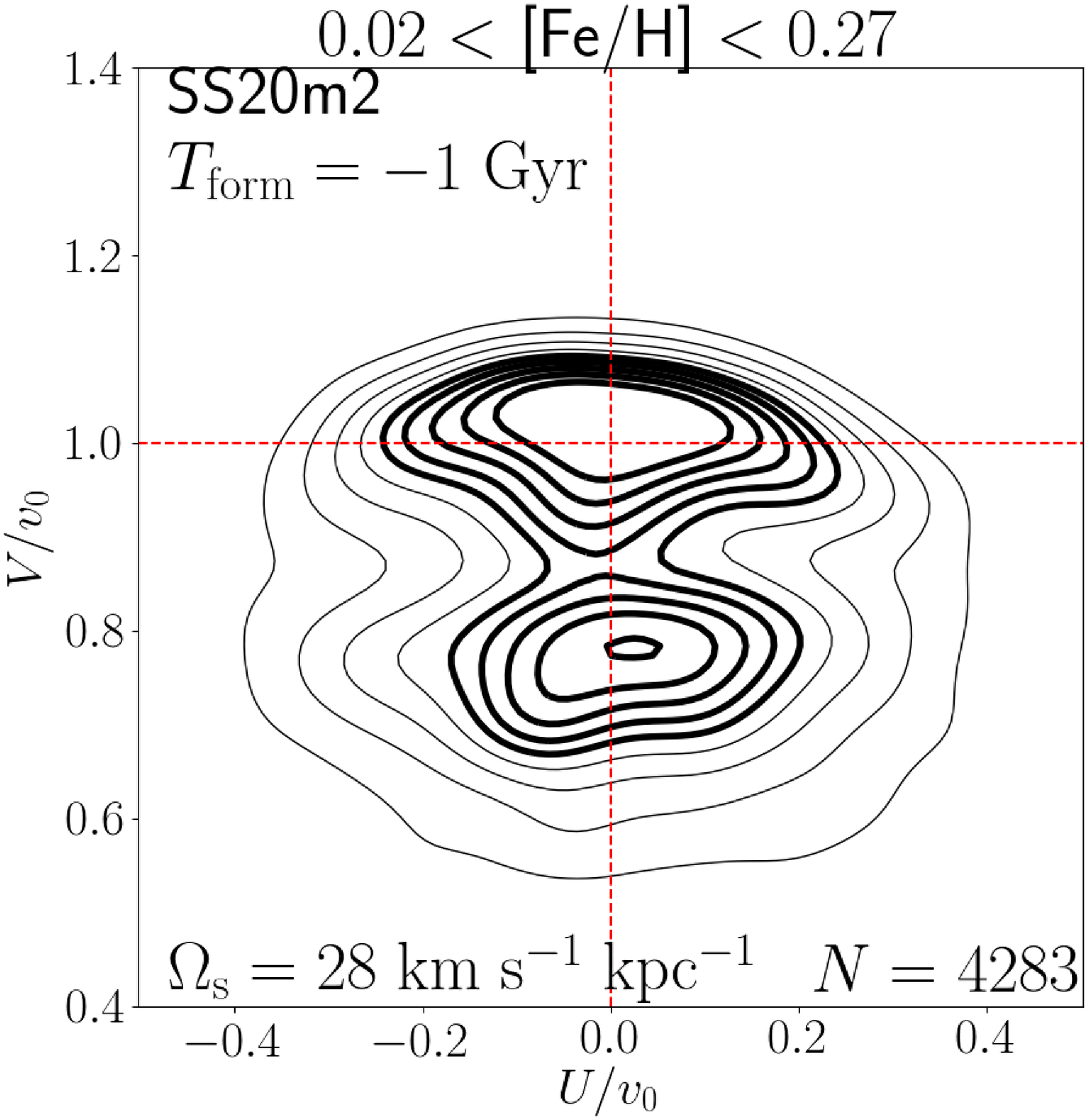} \\
 \includegraphics[width=0.8\columnwidth]{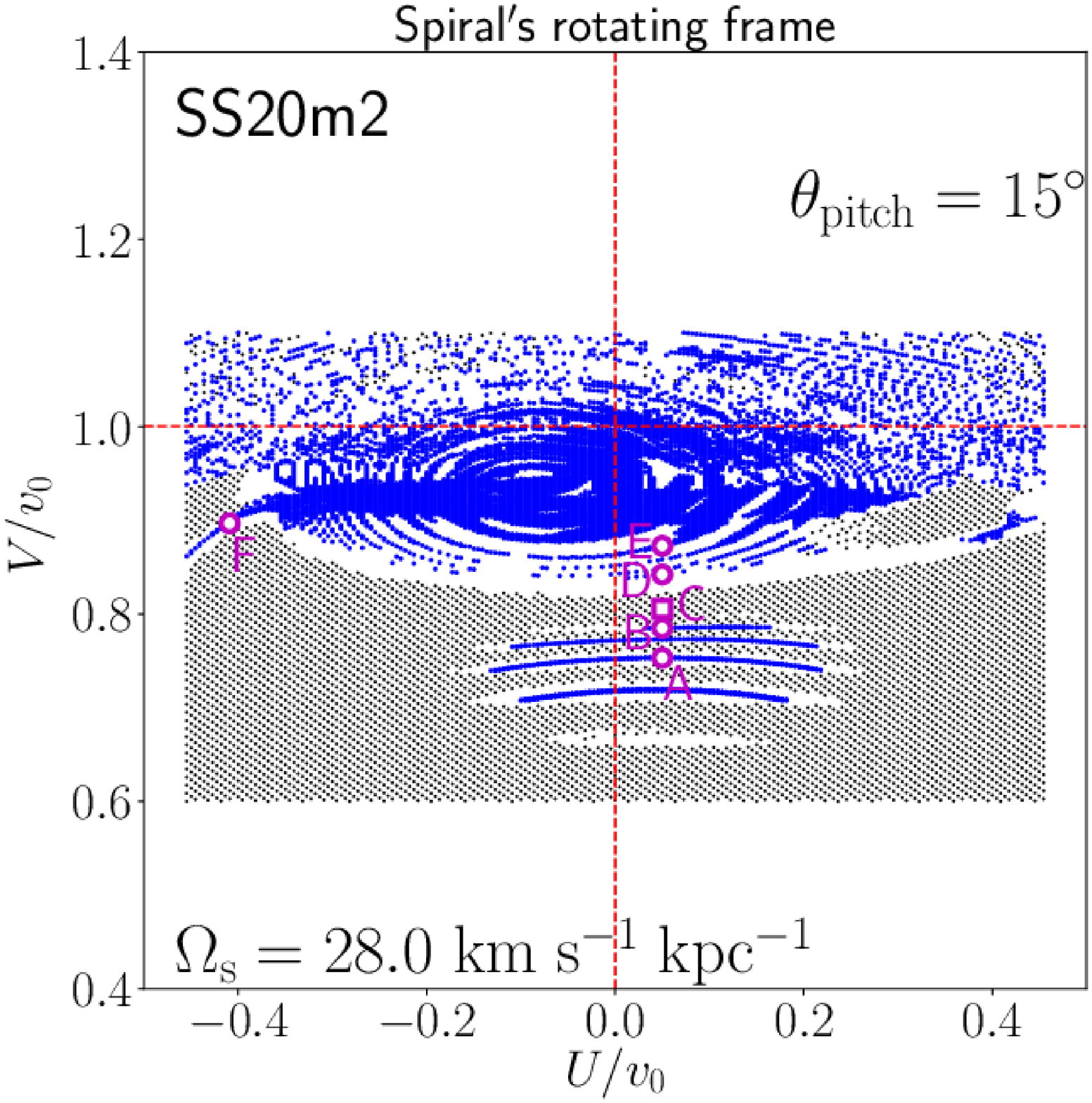} 
 \includegraphics[width=0.8\columnwidth]{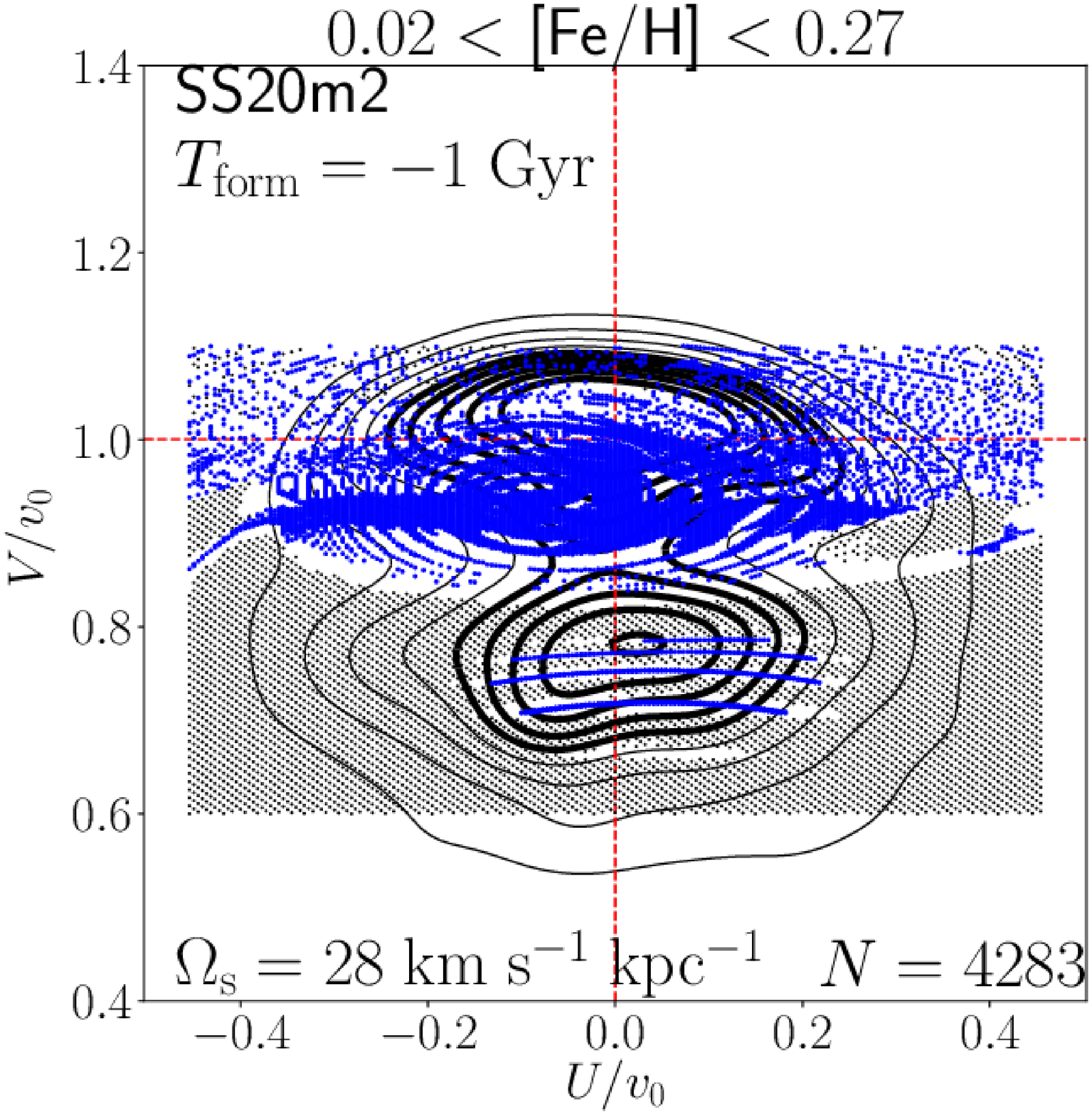} \\
 \includegraphics[width=0.5\columnwidth]{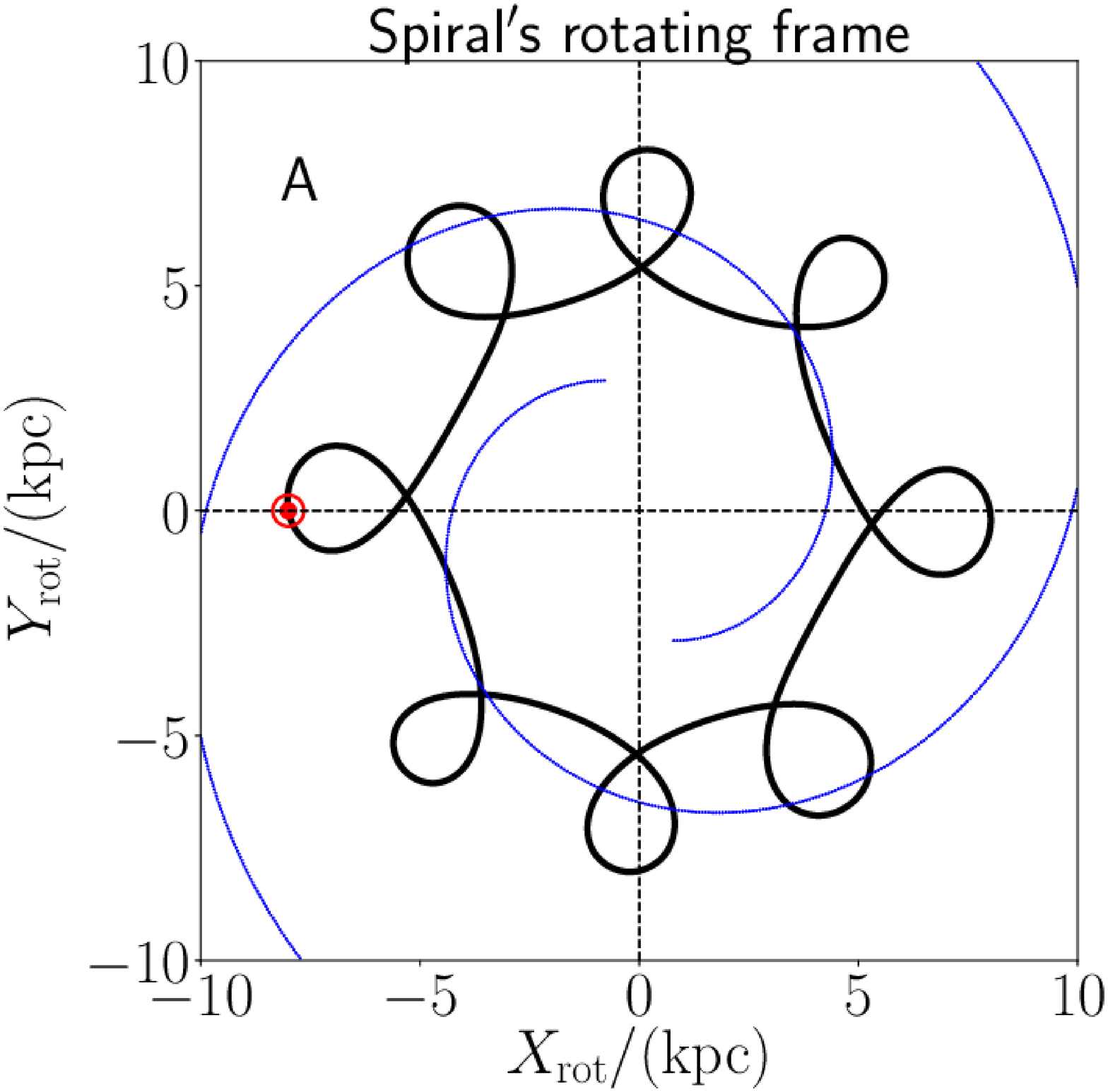} 
 \includegraphics[width=0.5\columnwidth]{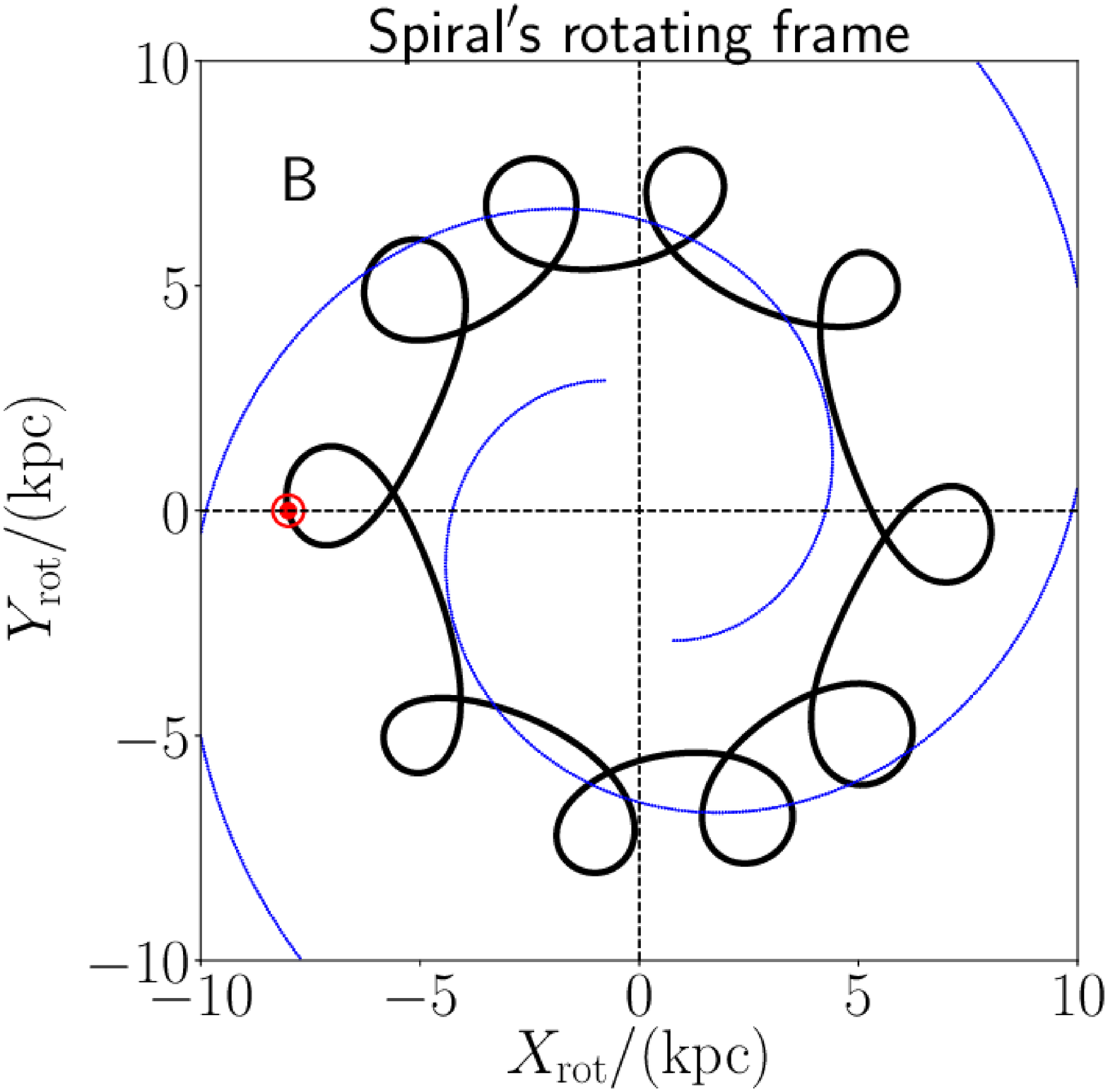} 
 \includegraphics[width=0.5\columnwidth]{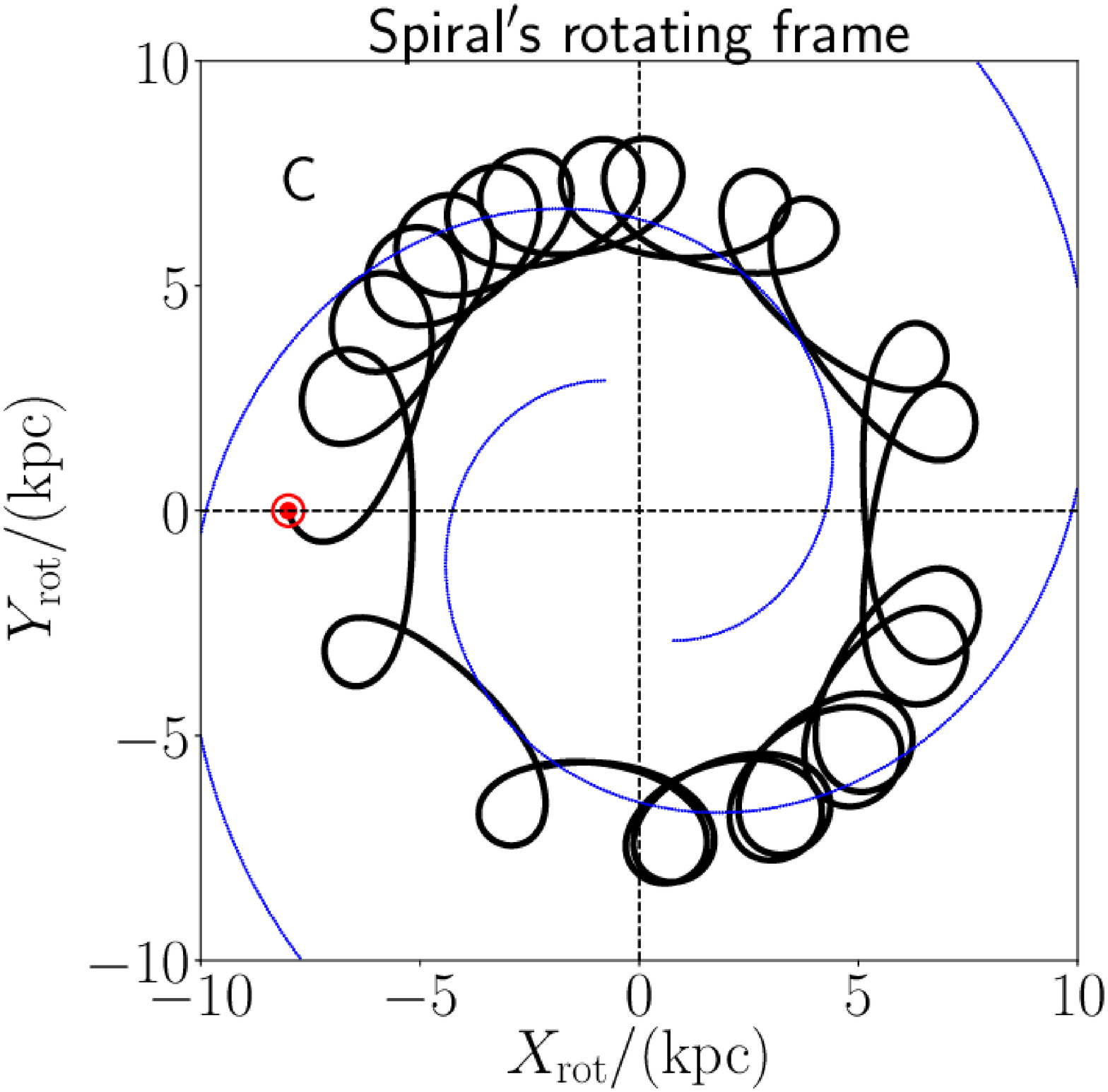} \\
 \includegraphics[width=0.5\columnwidth]{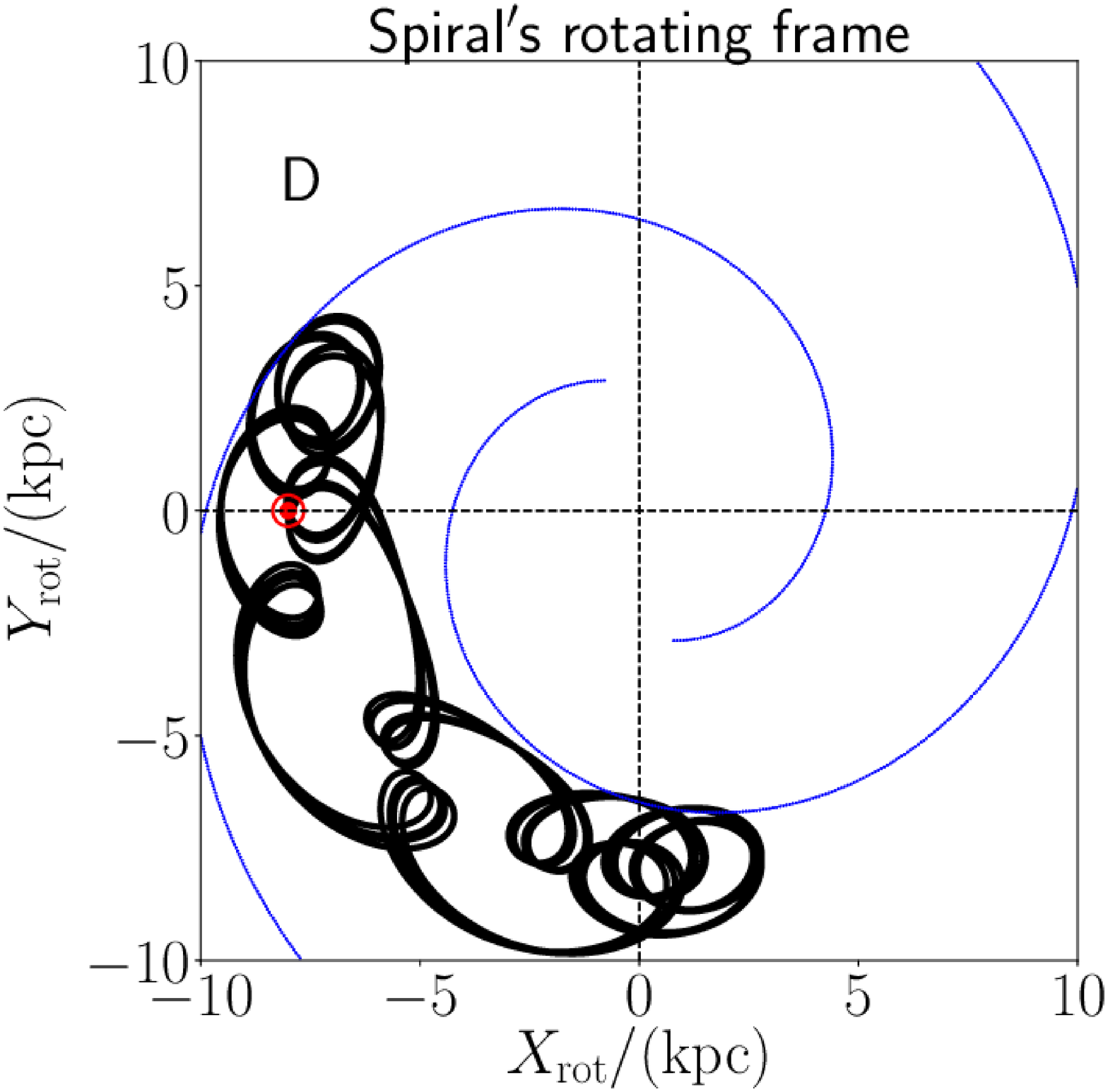} 
 \includegraphics[width=0.5\columnwidth]{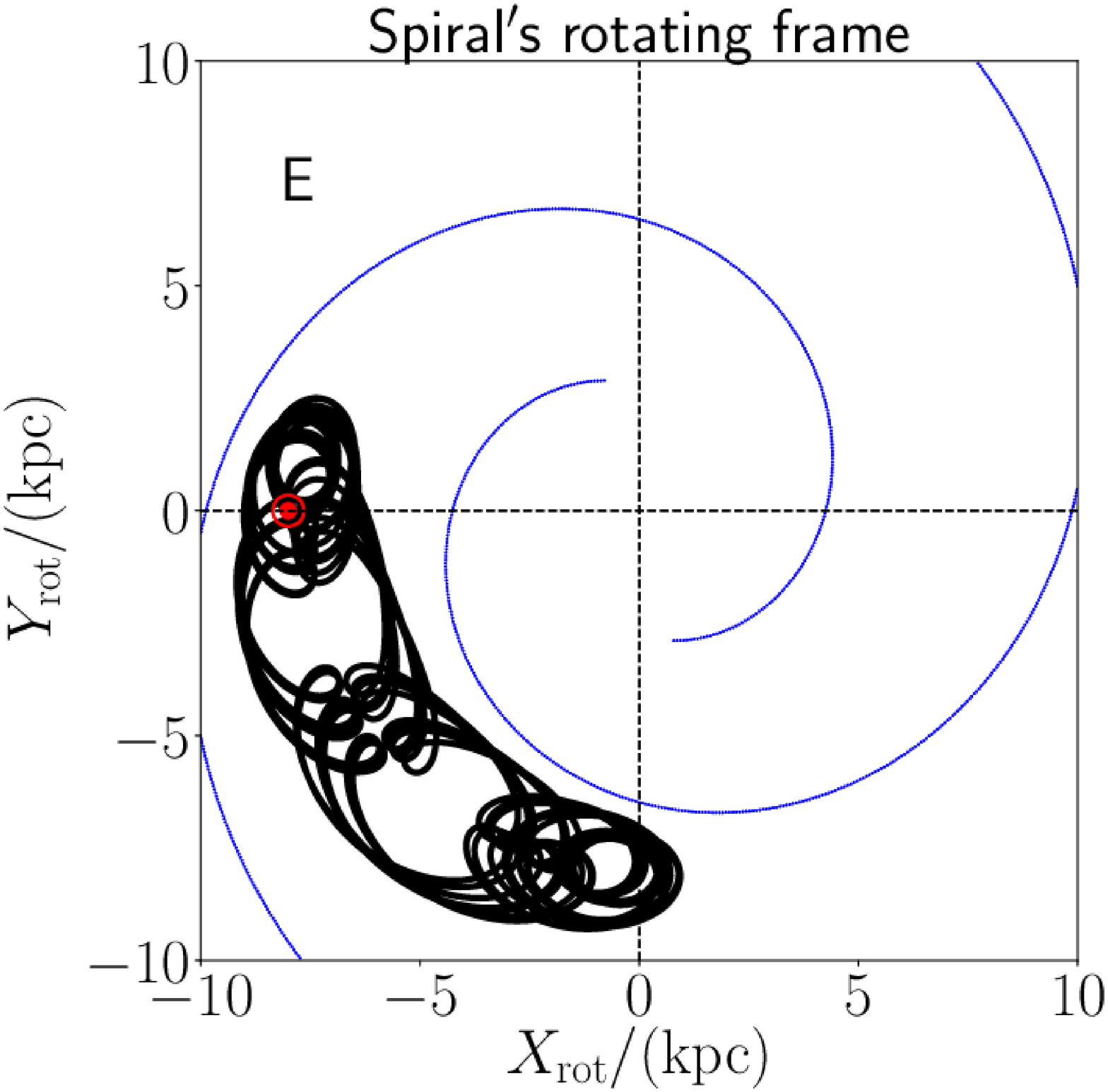} 
 \includegraphics[width=0.5\columnwidth]{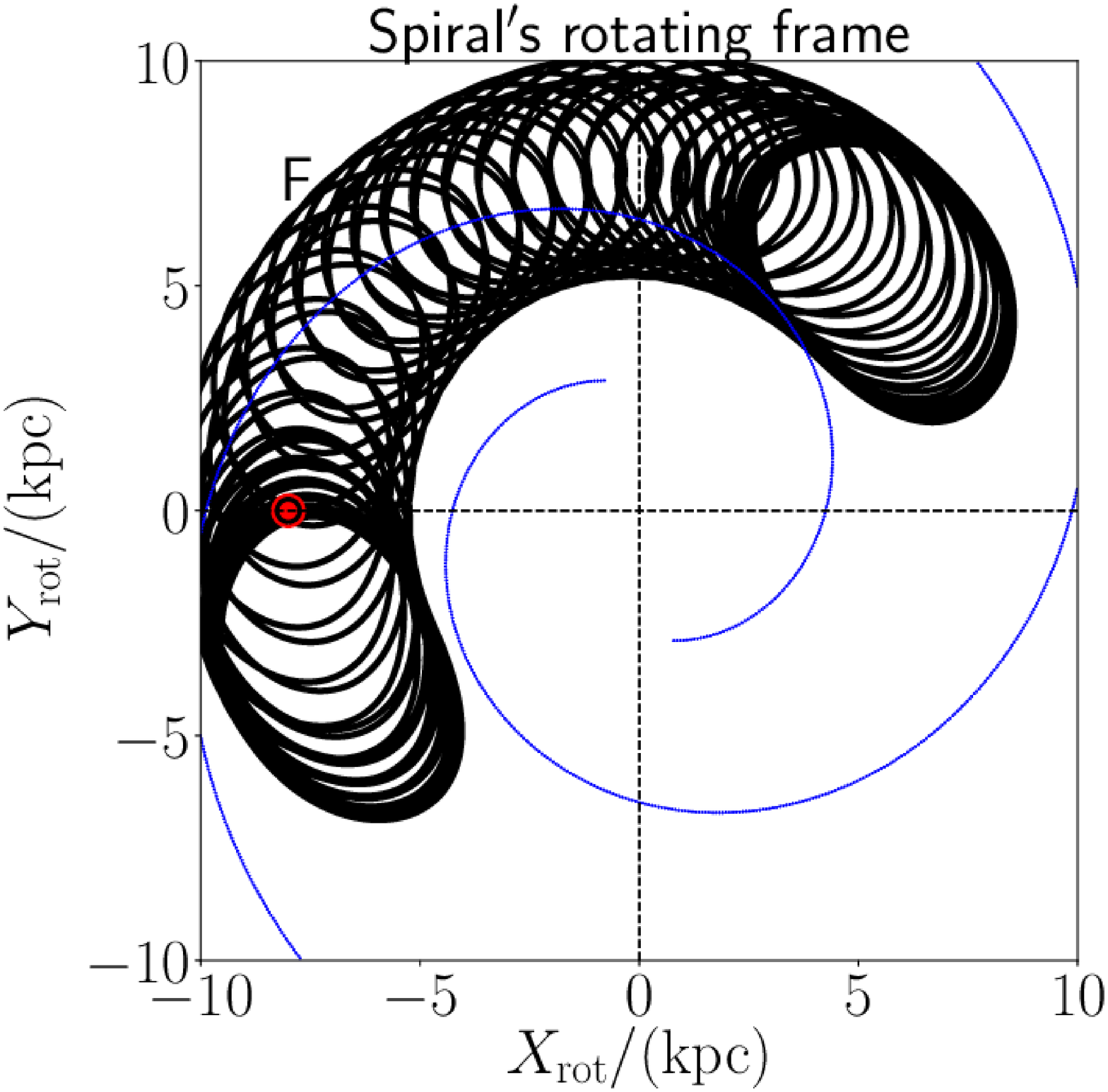}  
\caption{
Similar to Fig. \ref{fig_bar49closed}, but in the case of a partially successful, spiral-only model with $(\Omegas, m) = (28, 2)$. 
We note that this model reproduces observational properties (P1) and (P2) in Section \ref{section:data} but not (P3). 
On the left-hand panel of the second row, 
the five magenta markers at $U/v_0=0.05$ correspond to the orbit A, B, C, D, and E (in ascending order of $V$), while the marker at $U/v_0=-0.41$ corresponds to the orbit F. 
On the third and fourth row, 
the spirals' rotating frame $(X_\mathrm{rot}, Y_\mathrm{rot})$ is used. 
The blue curved lines represents the current location of the spiral arms. 
}
\label{fig_spiral28m2closed}
\end{center}
\end{figure*}

\begin{figure*}
\begin{center}
 \includegraphics[width=0.6\columnwidth]{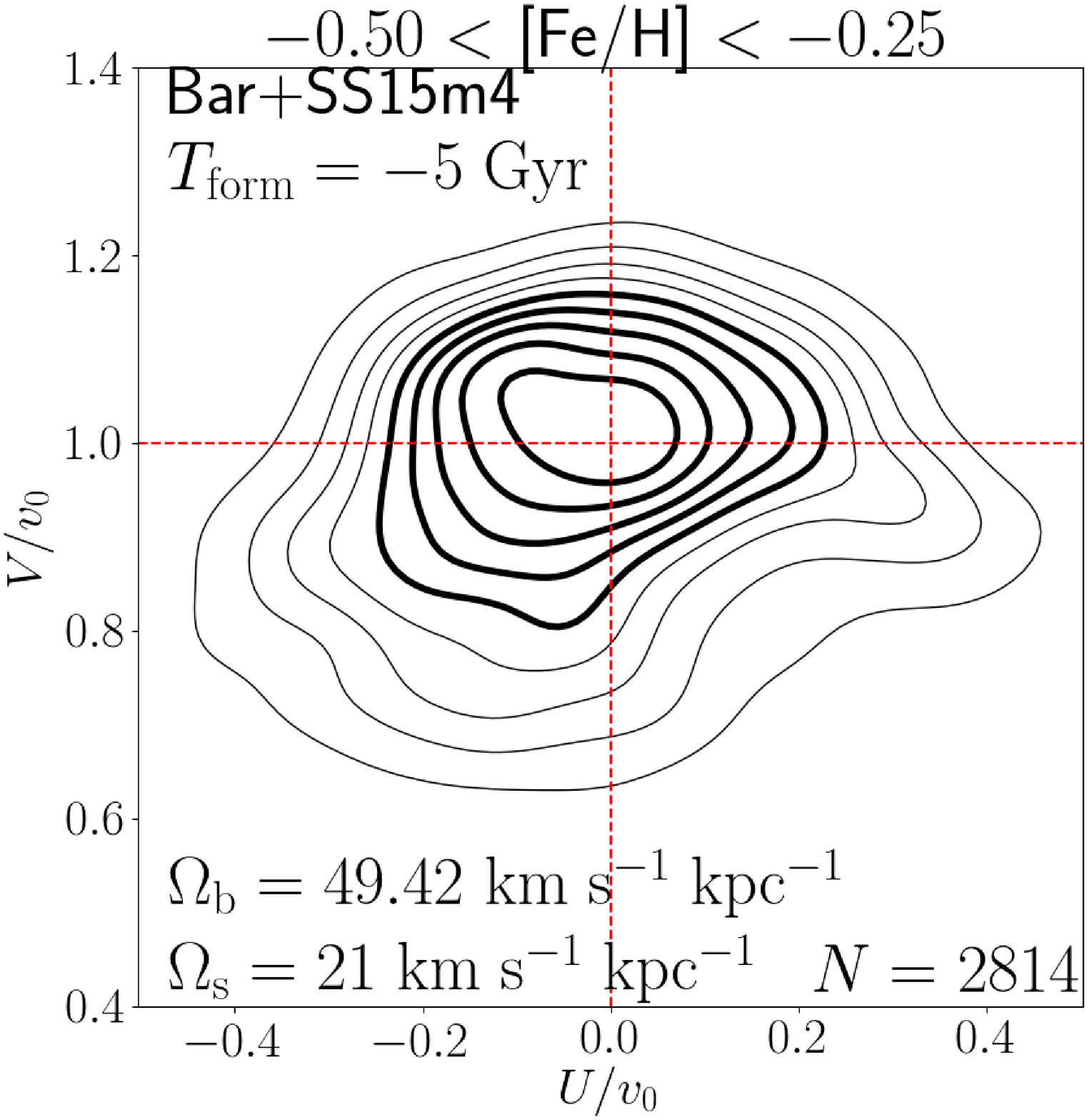} 
 \includegraphics[width=0.6\columnwidth]{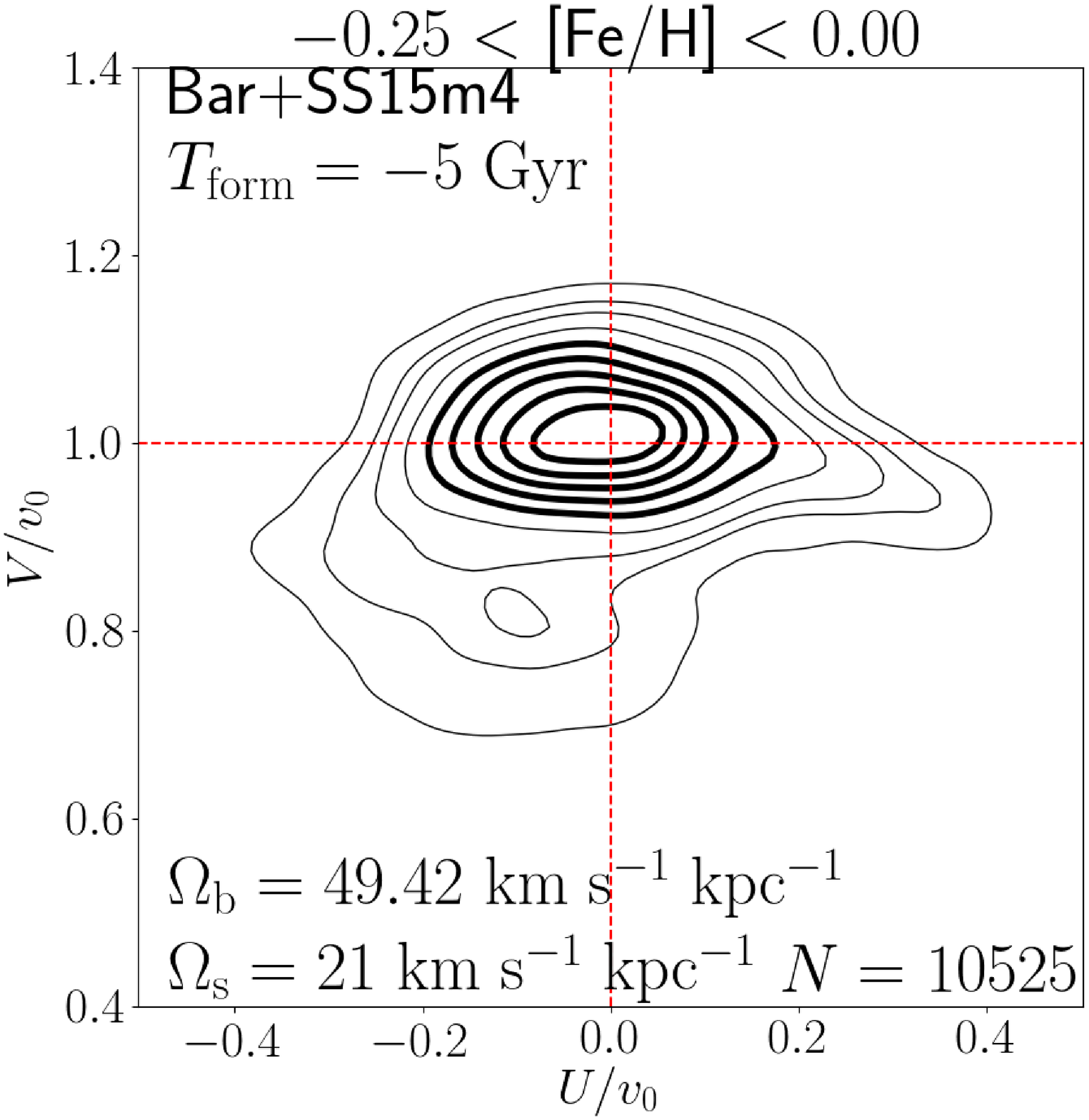} 
 \includegraphics[width=0.6\columnwidth]{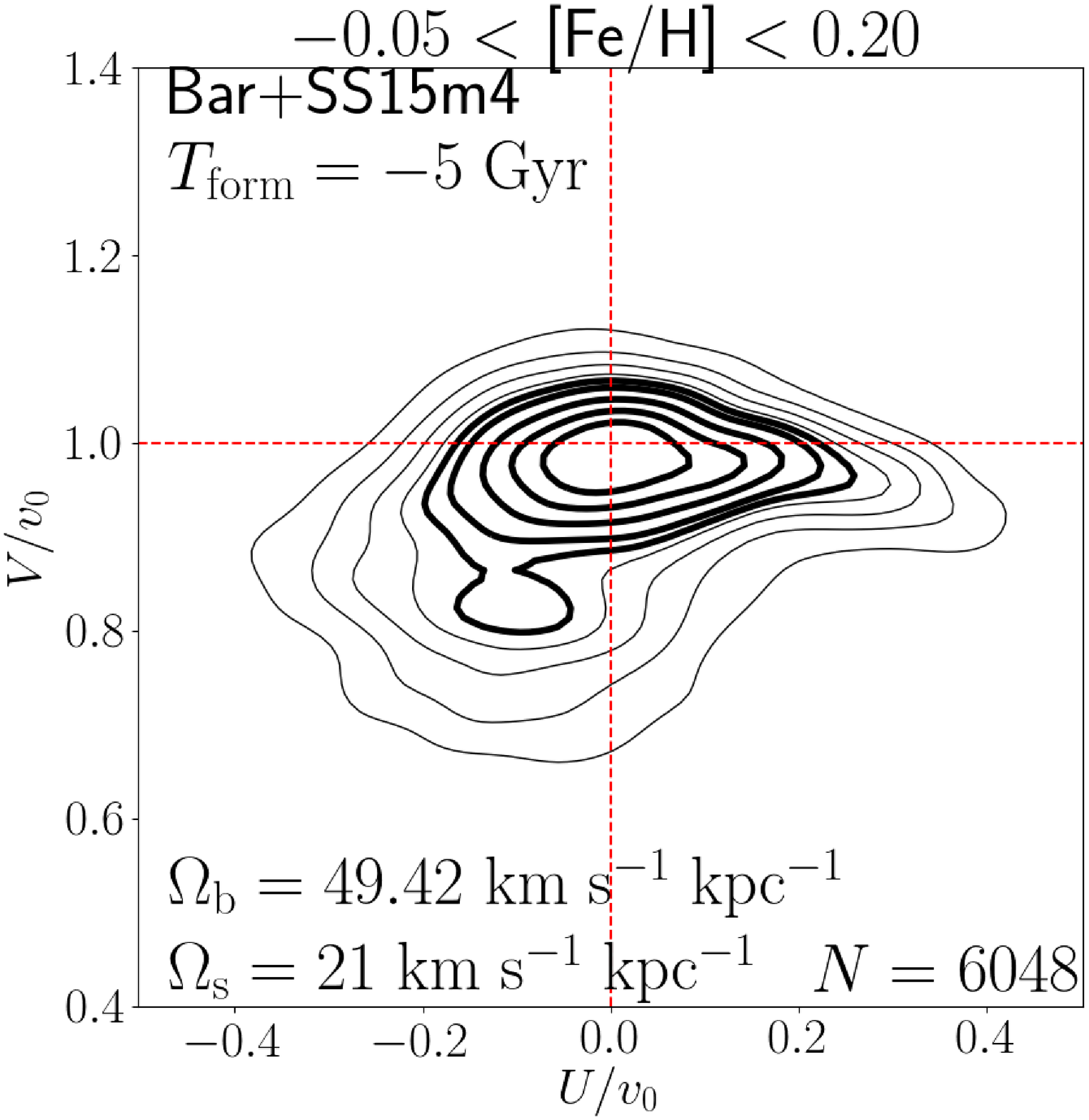} \\
 \includegraphics[width=0.8\columnwidth]{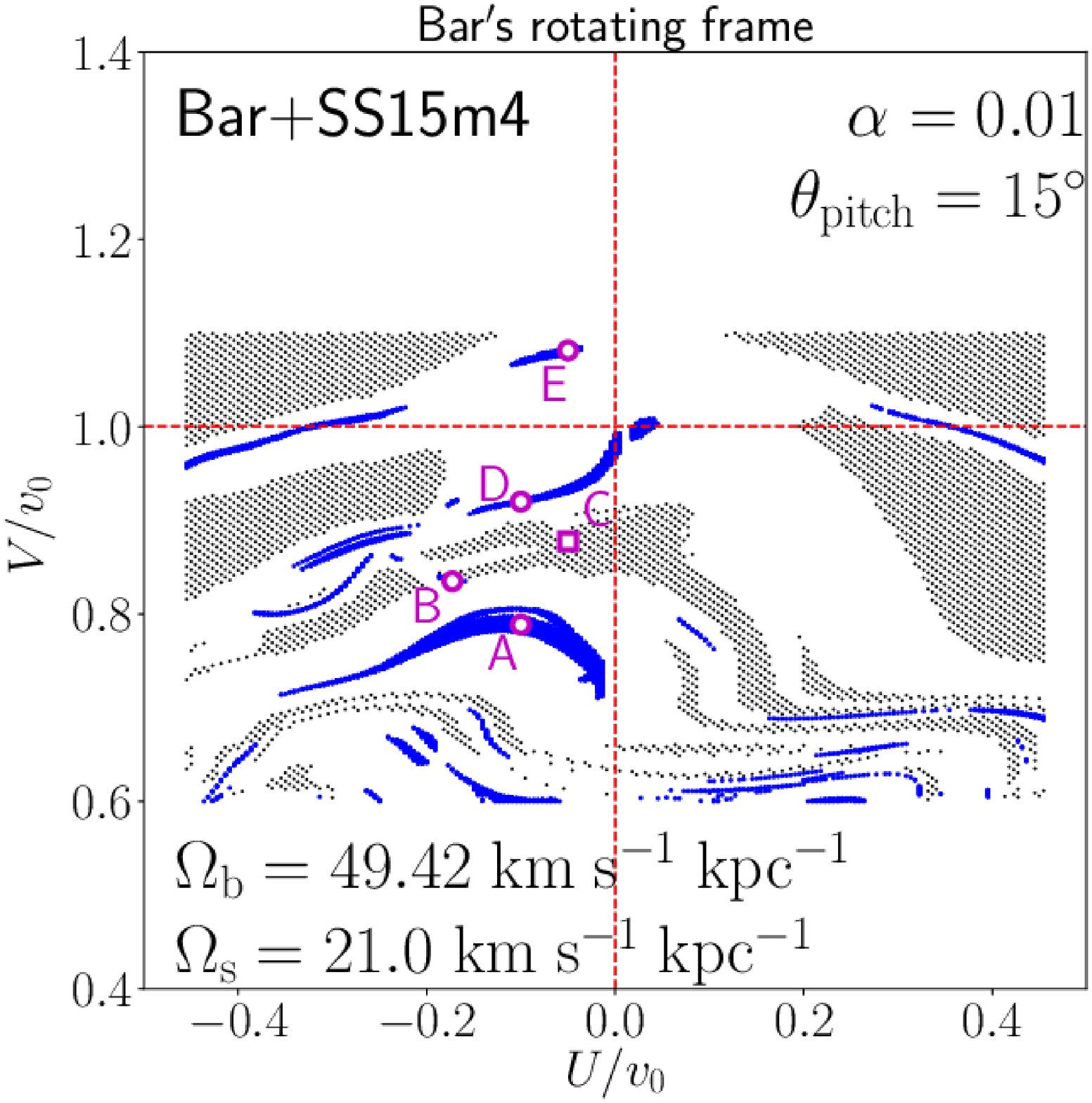} 
 \includegraphics[width=0.8\columnwidth]{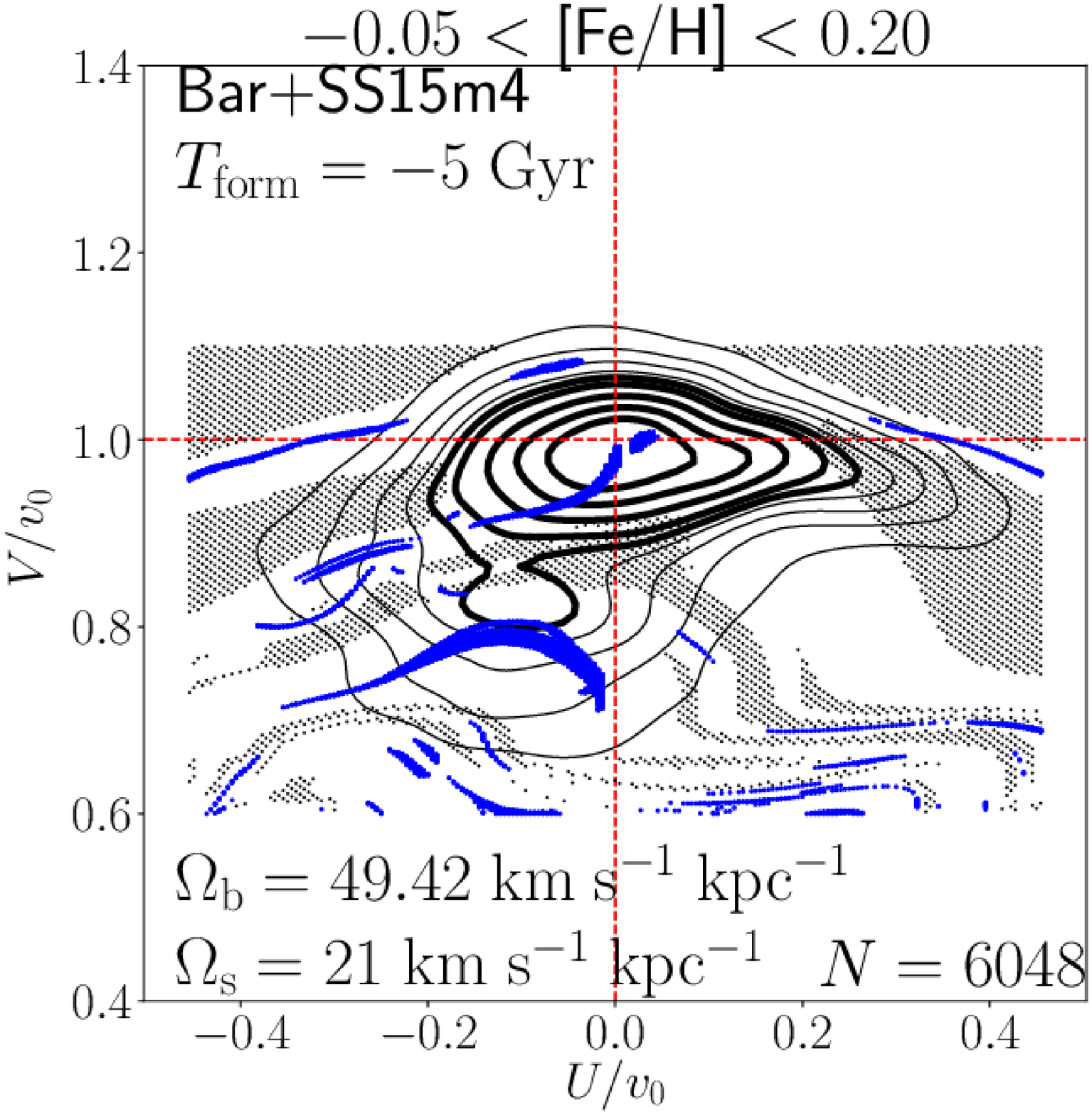} \\
 \includegraphics[width=0.5\columnwidth]{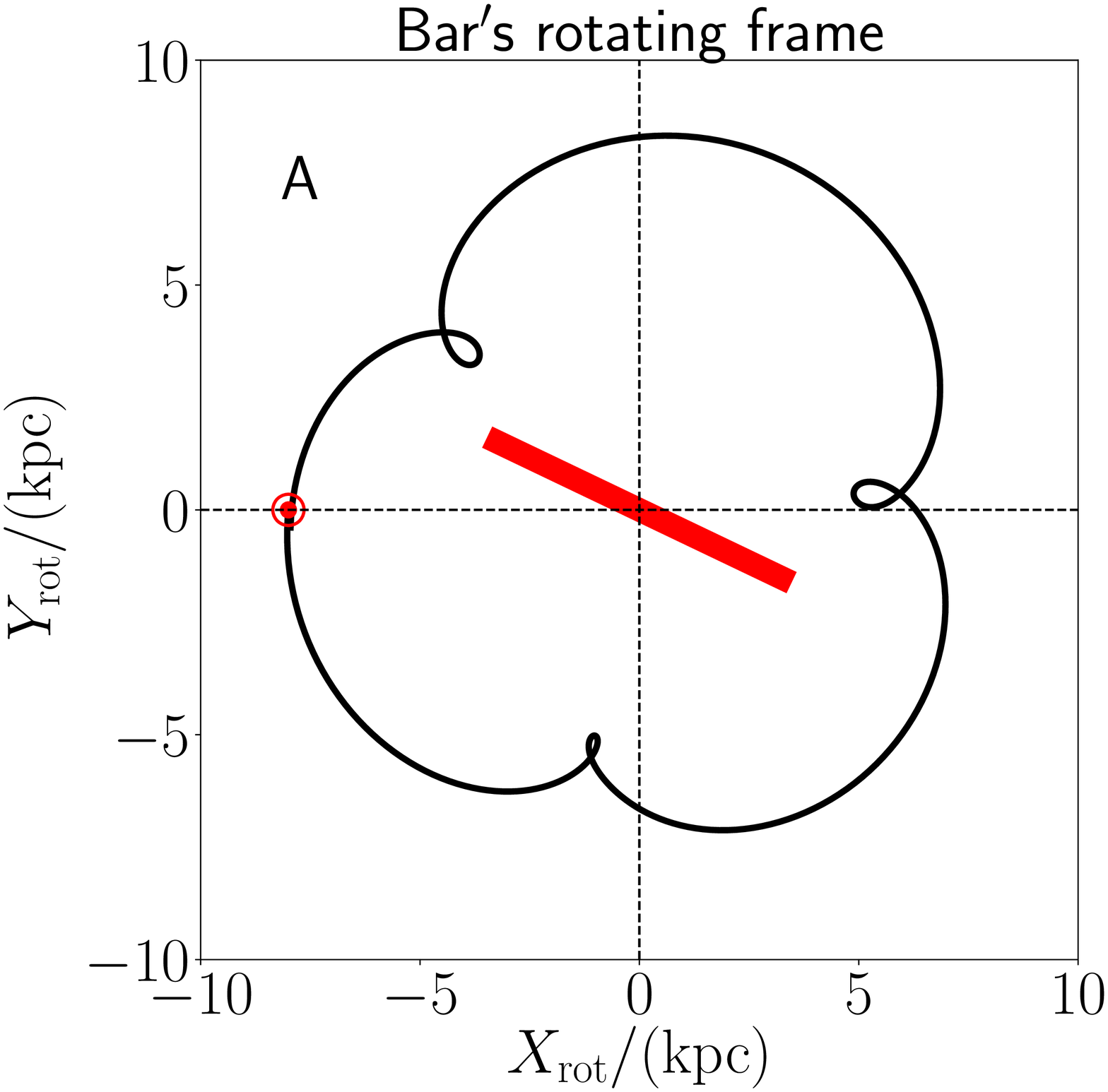} 
 \includegraphics[width=0.5\columnwidth]{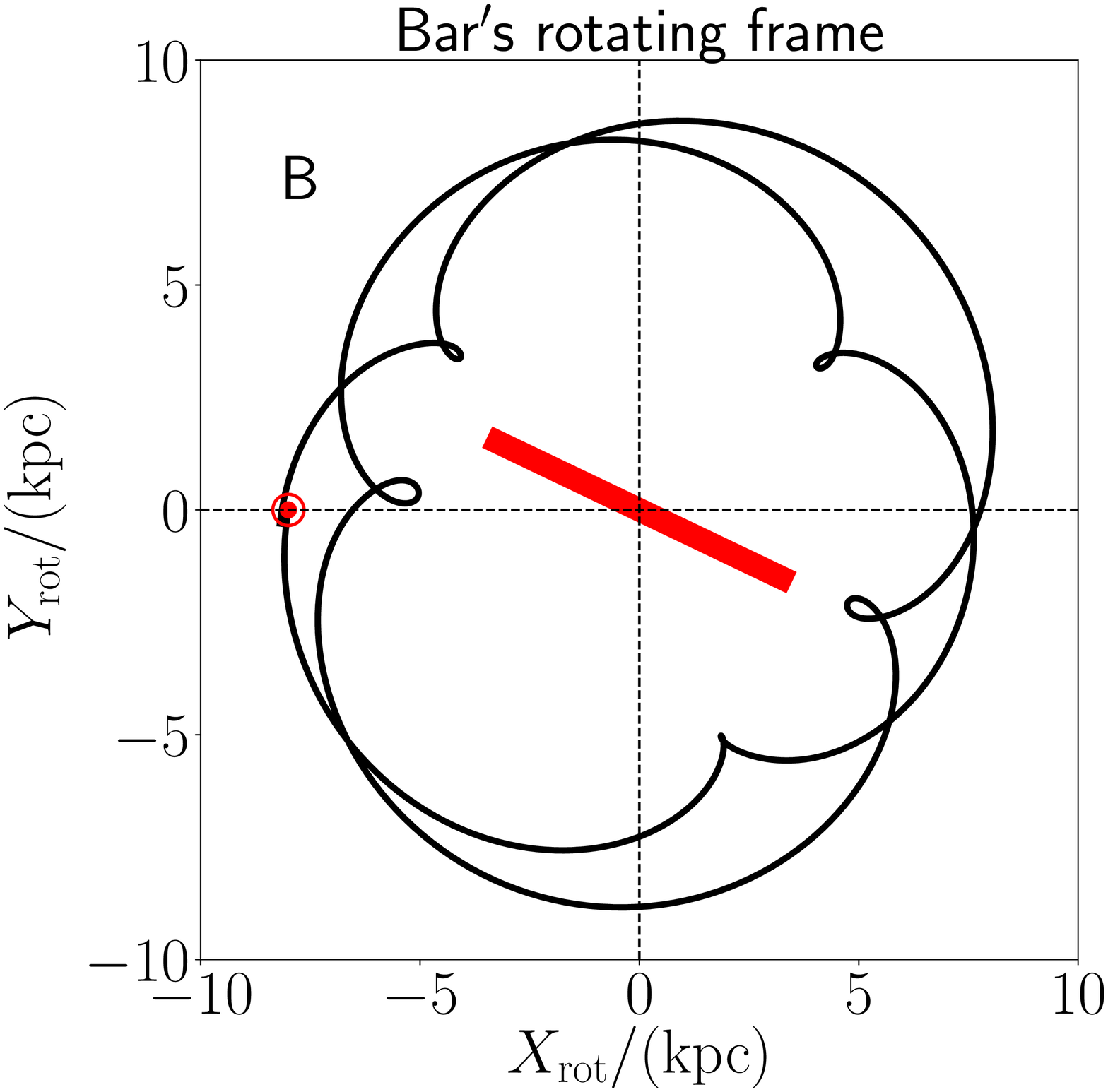} 
 \includegraphics[width=0.5\columnwidth]{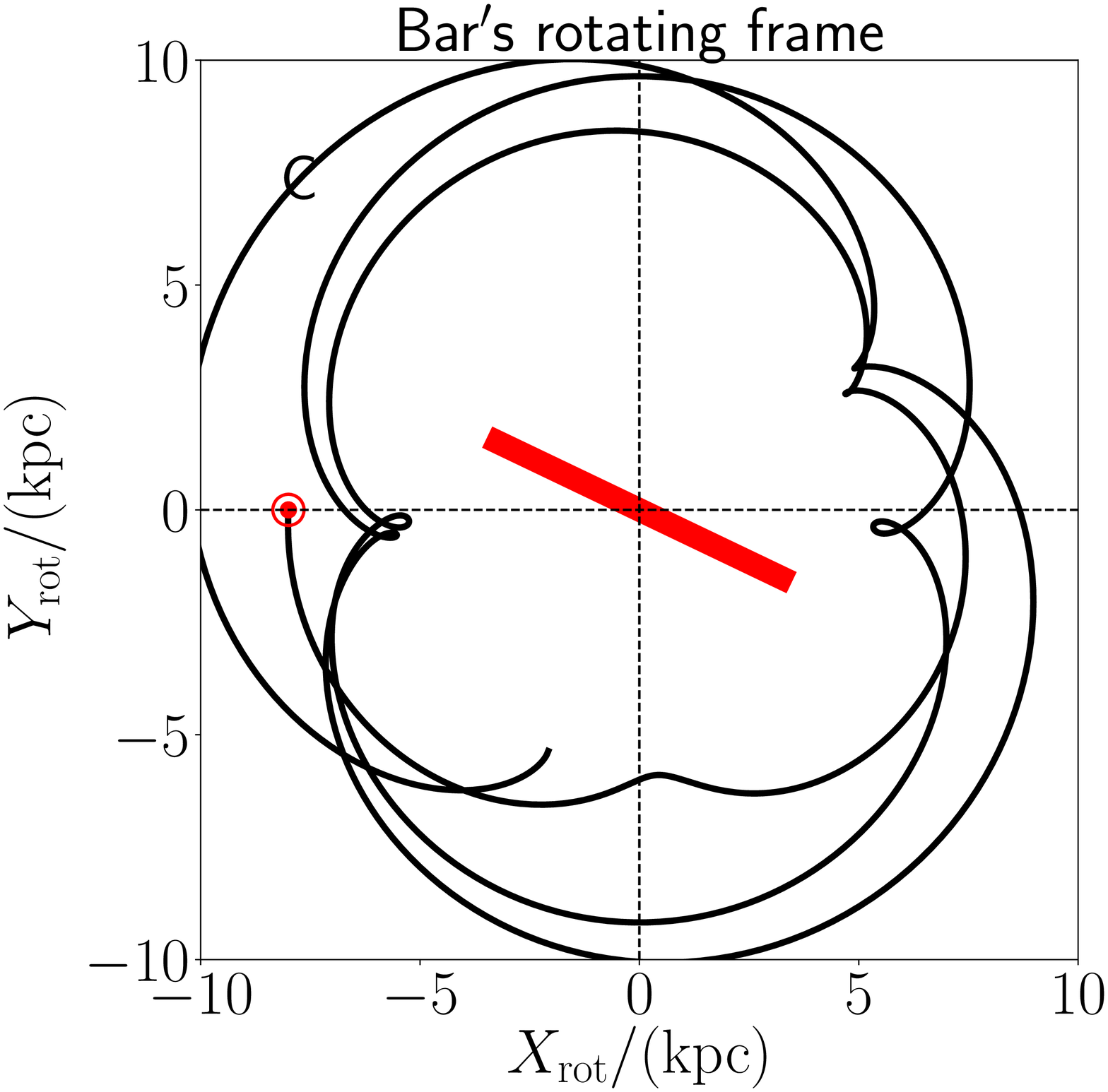} \\
 \includegraphics[width=0.5\columnwidth]{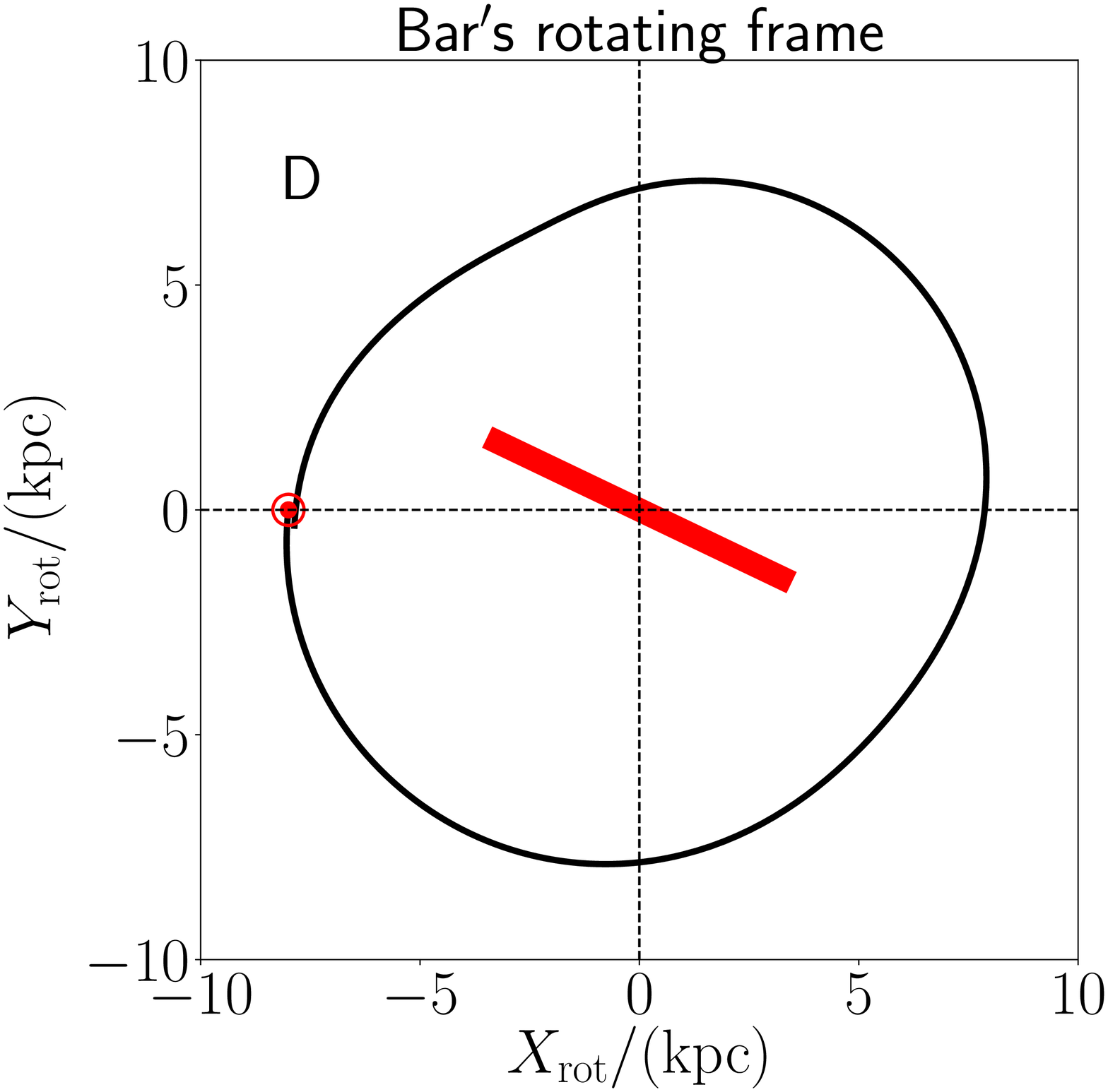} 
 \includegraphics[width=0.5\columnwidth]{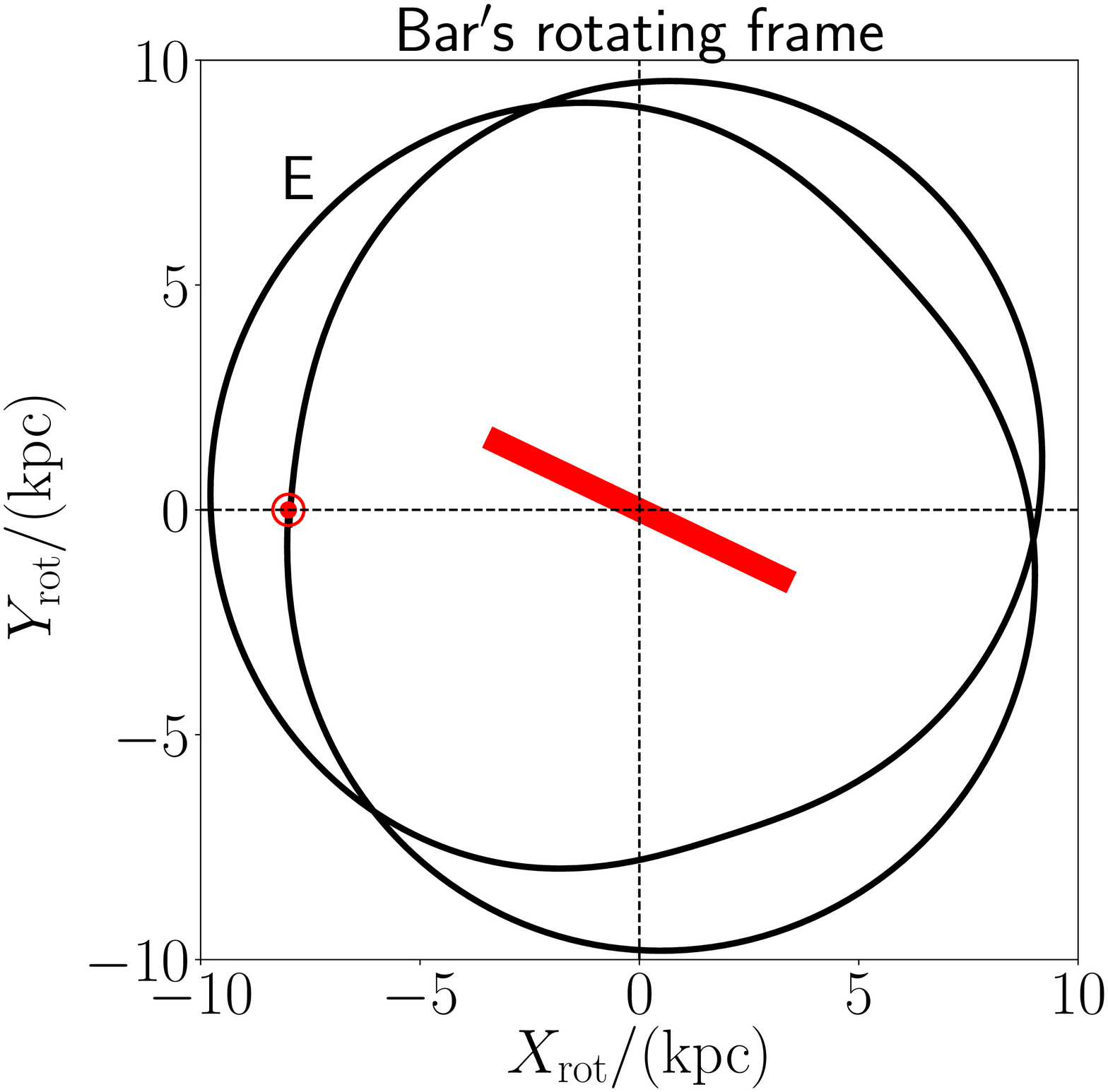} 
\caption{
Similar to Fig. \ref{fig_bar49closed}, but in the case of a successful, fast-bar$+$spiral model with $(\Omegab, m, \Omegas) = (49.42, 4, 21)$. 
We note that this model reproduces observational properties (P1)-(P3) in Section \ref{section:data}. 
On the third and fourth row, 
the bar's rotating frame $(X_\mathrm{rot}, Y_\mathrm{rot})$ is used. 
The spiral arms are not static in this frame, so we do not show spiral arms here. 
}
\label{fig_bar49_spiral21m4_closed}
\end{center}
\end{figure*}

\begin{figure*}
\begin{center}
 \includegraphics[width=0.6\columnwidth]{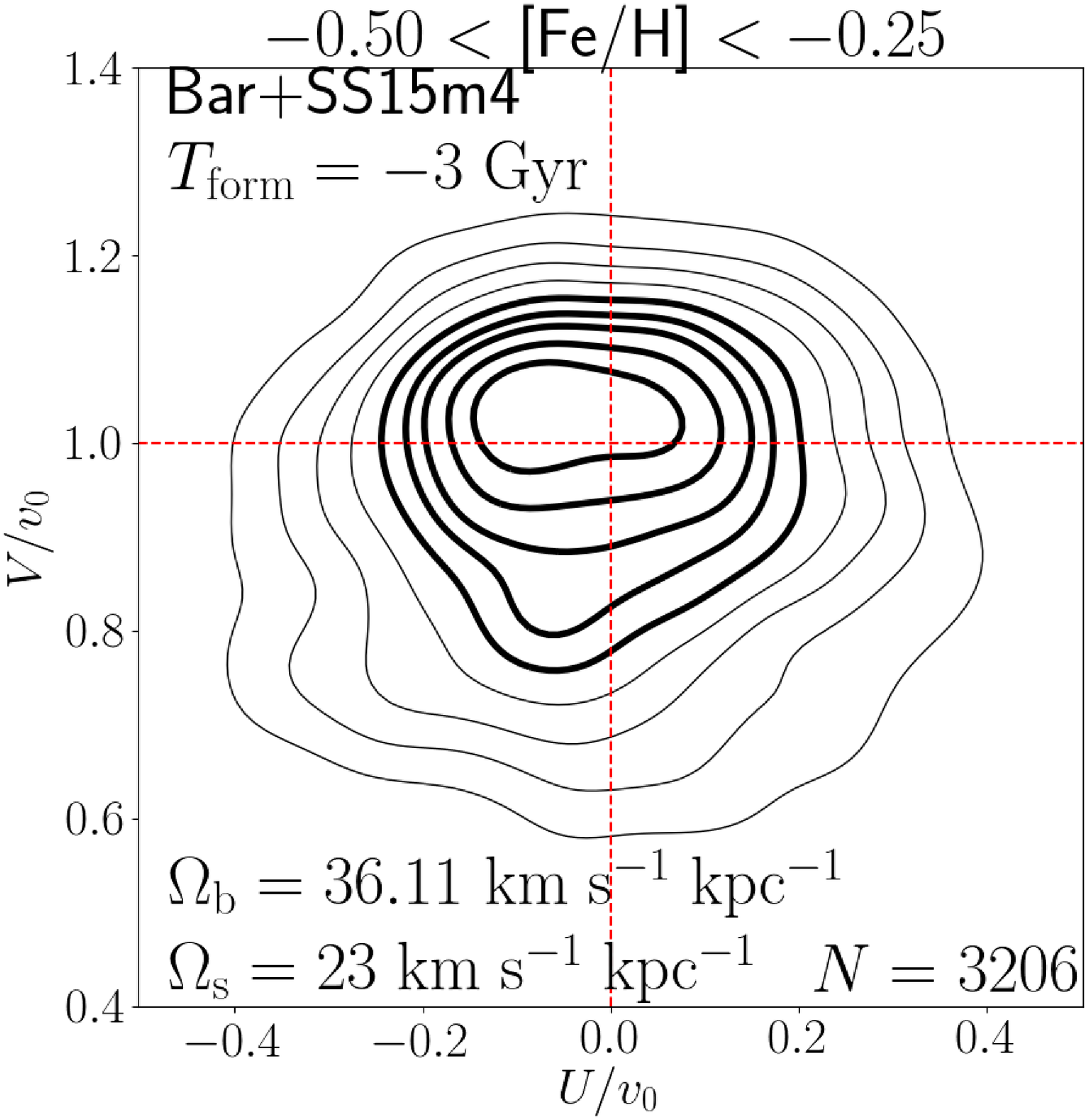} 
 \includegraphics[width=0.6\columnwidth]{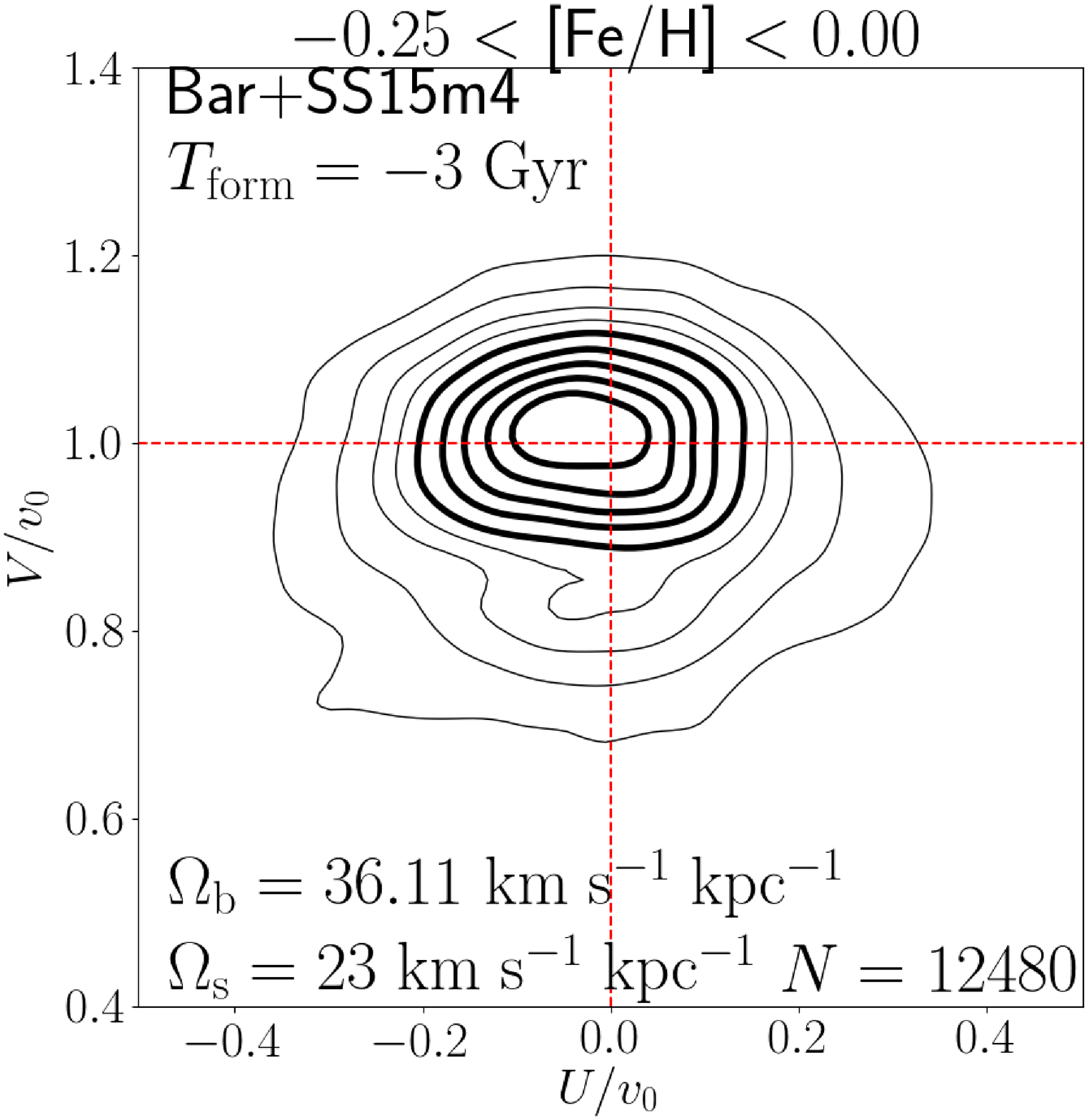} 
 \includegraphics[width=0.6\columnwidth]{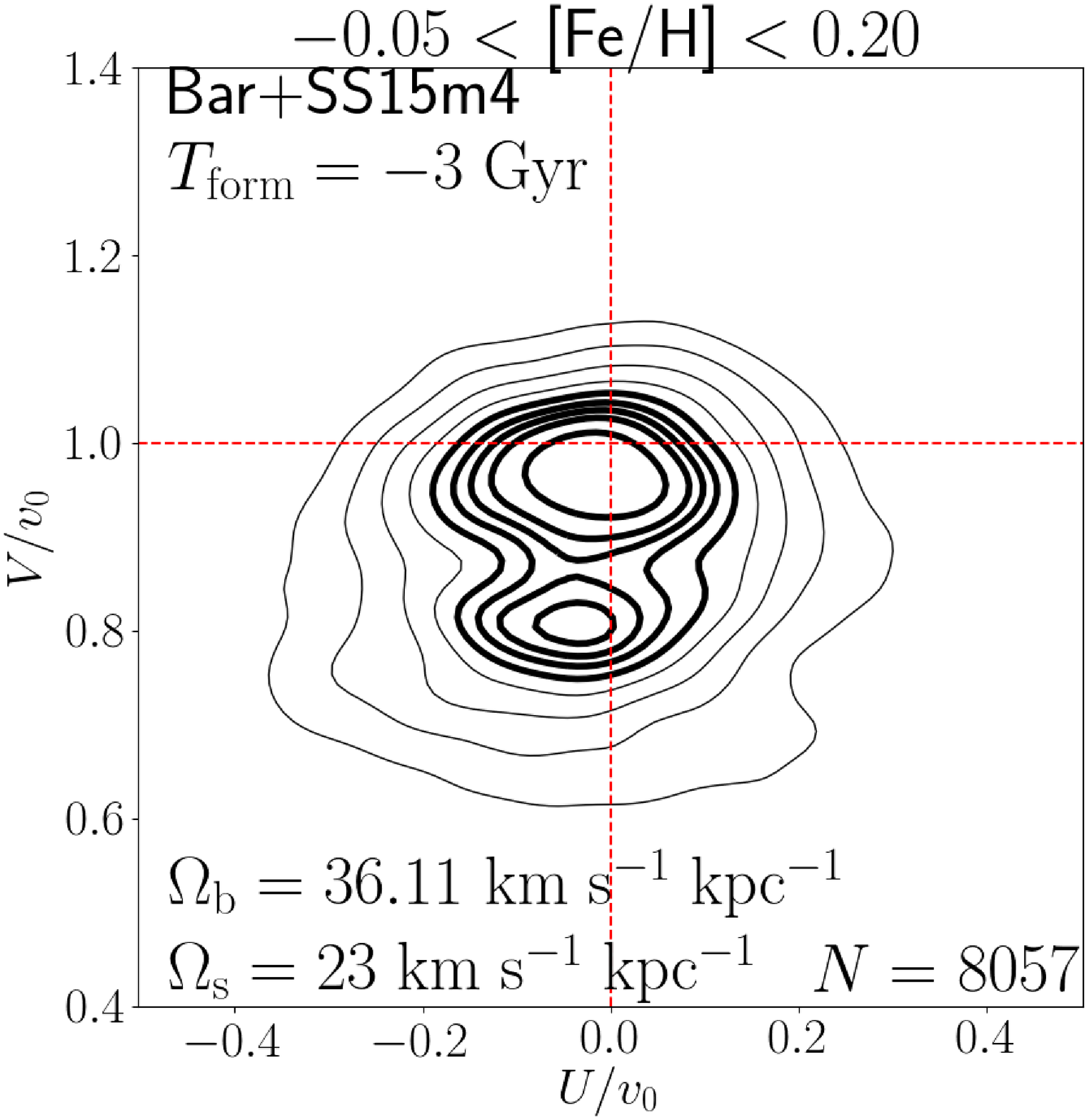} \\
 \includegraphics[width=0.8\columnwidth]{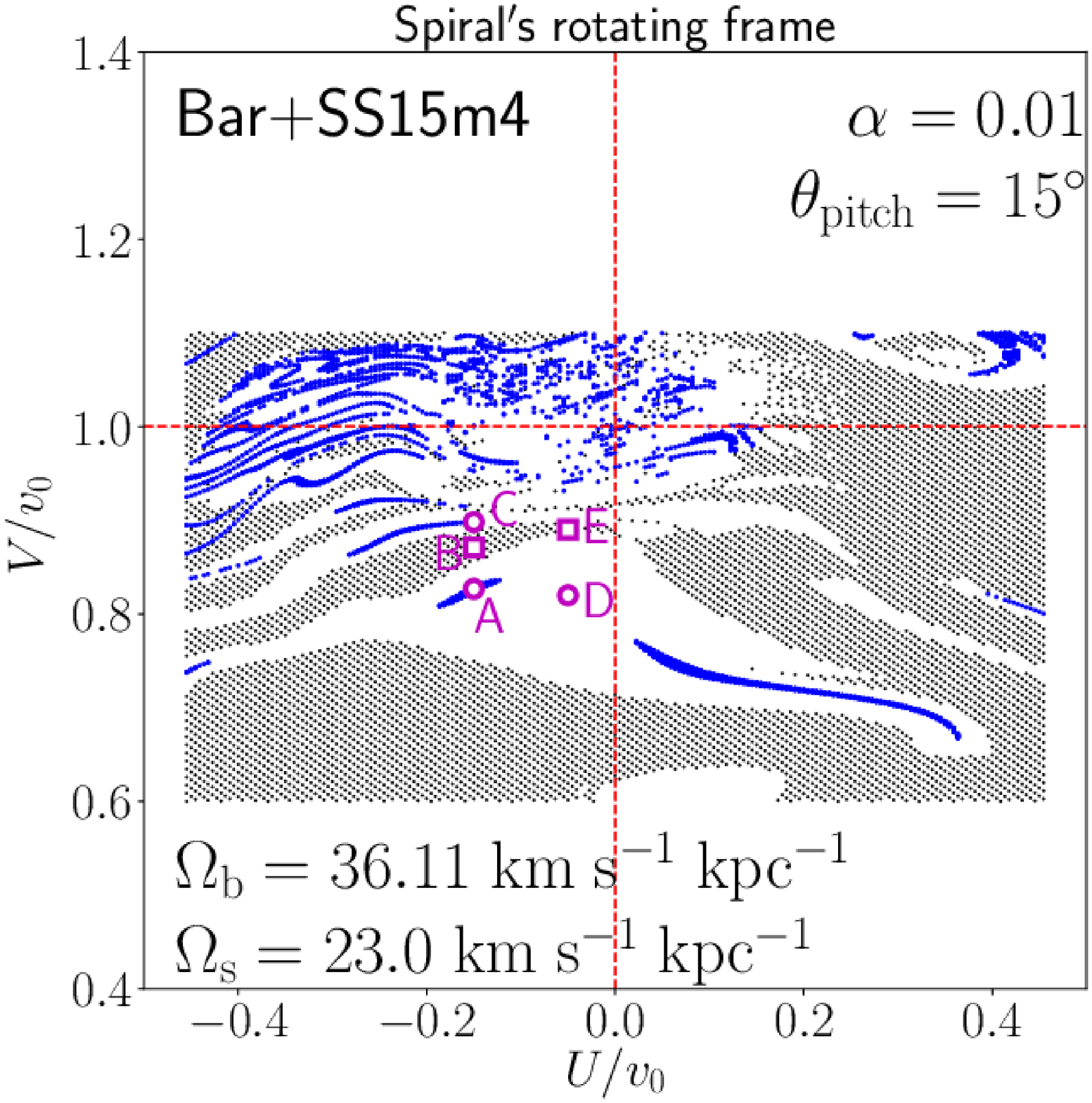} 
 \includegraphics[width=0.8\columnwidth]{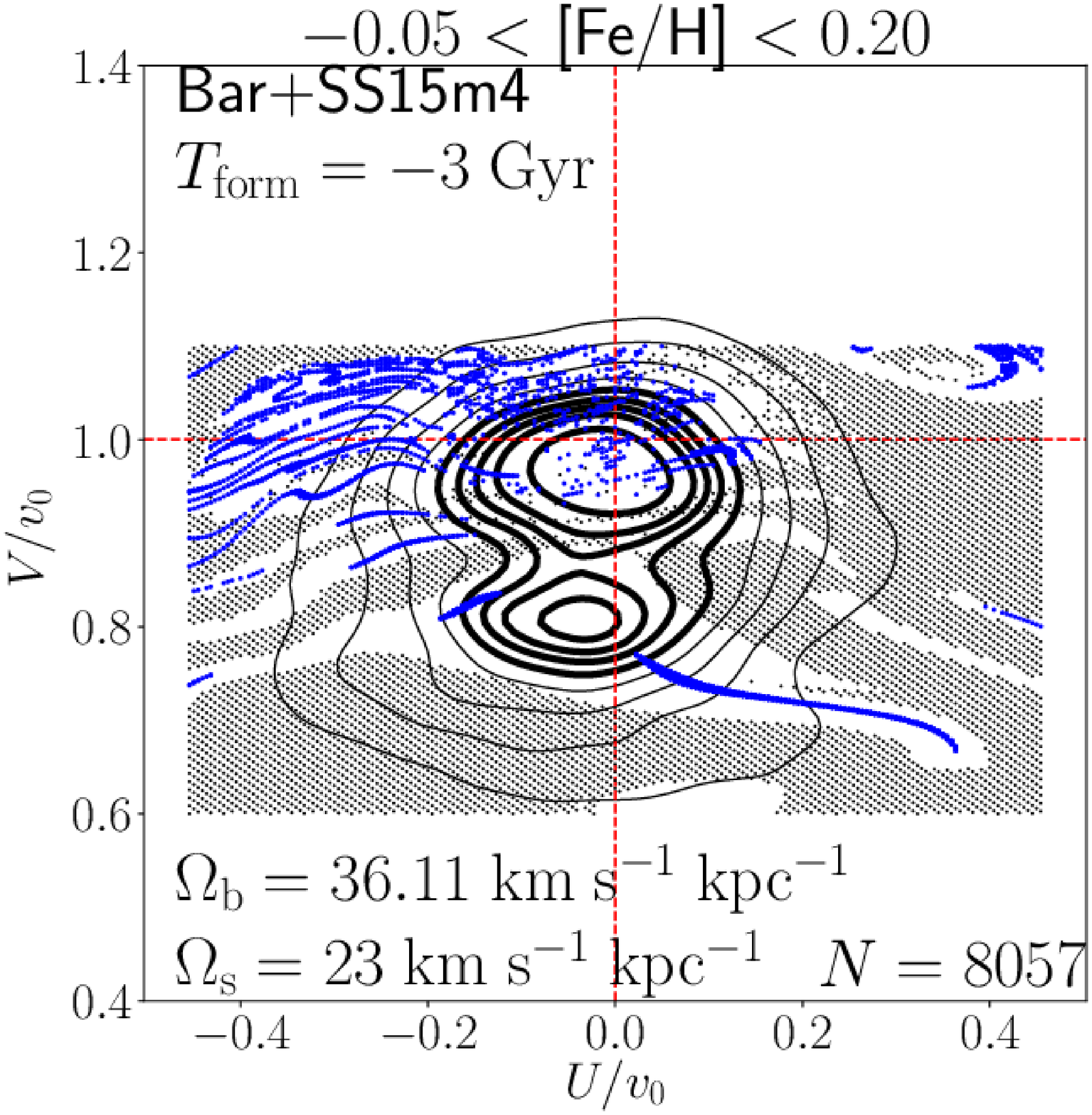} \\
 \includegraphics[width=0.5\columnwidth]{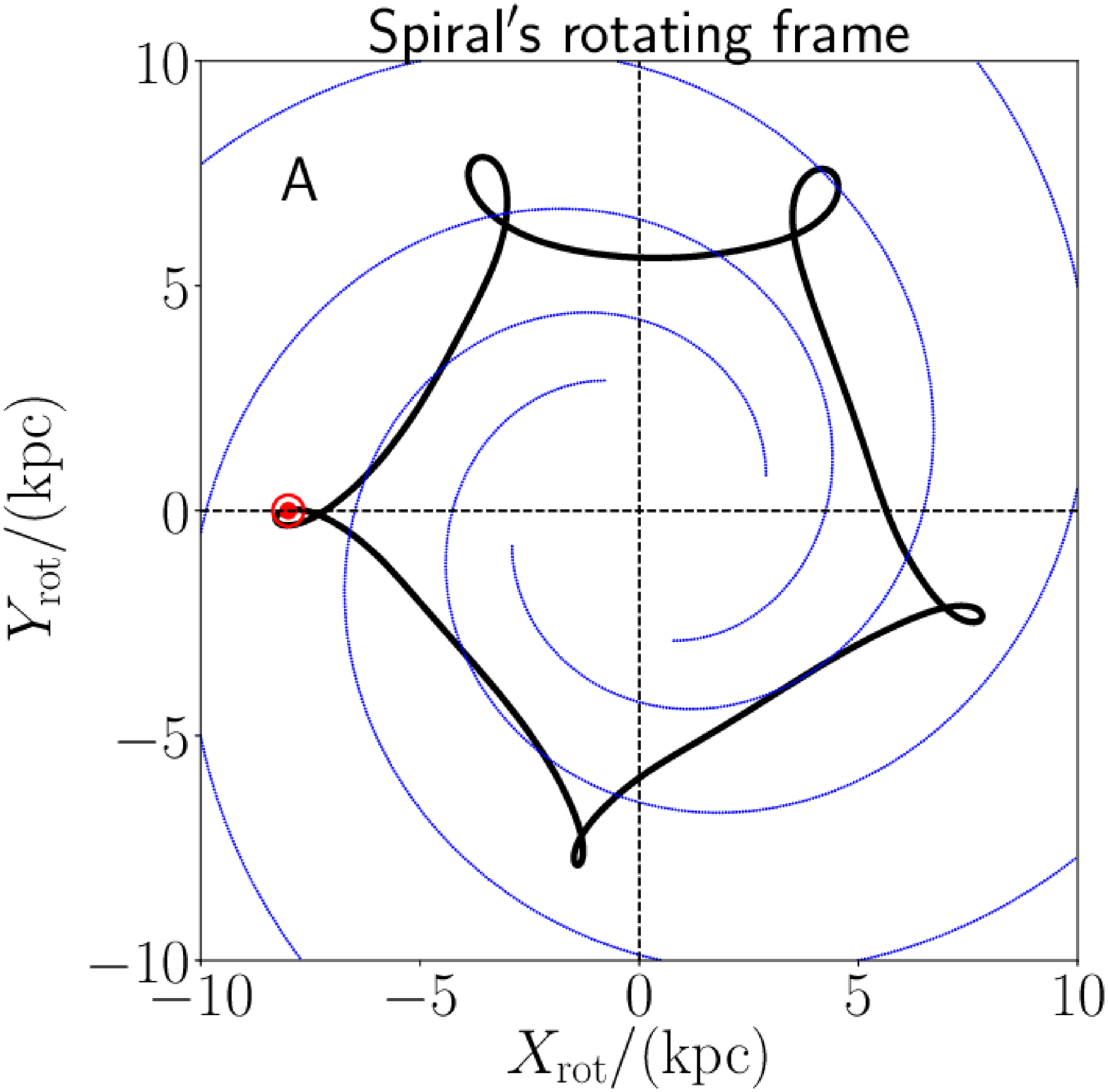} 
 \includegraphics[width=0.5\columnwidth]{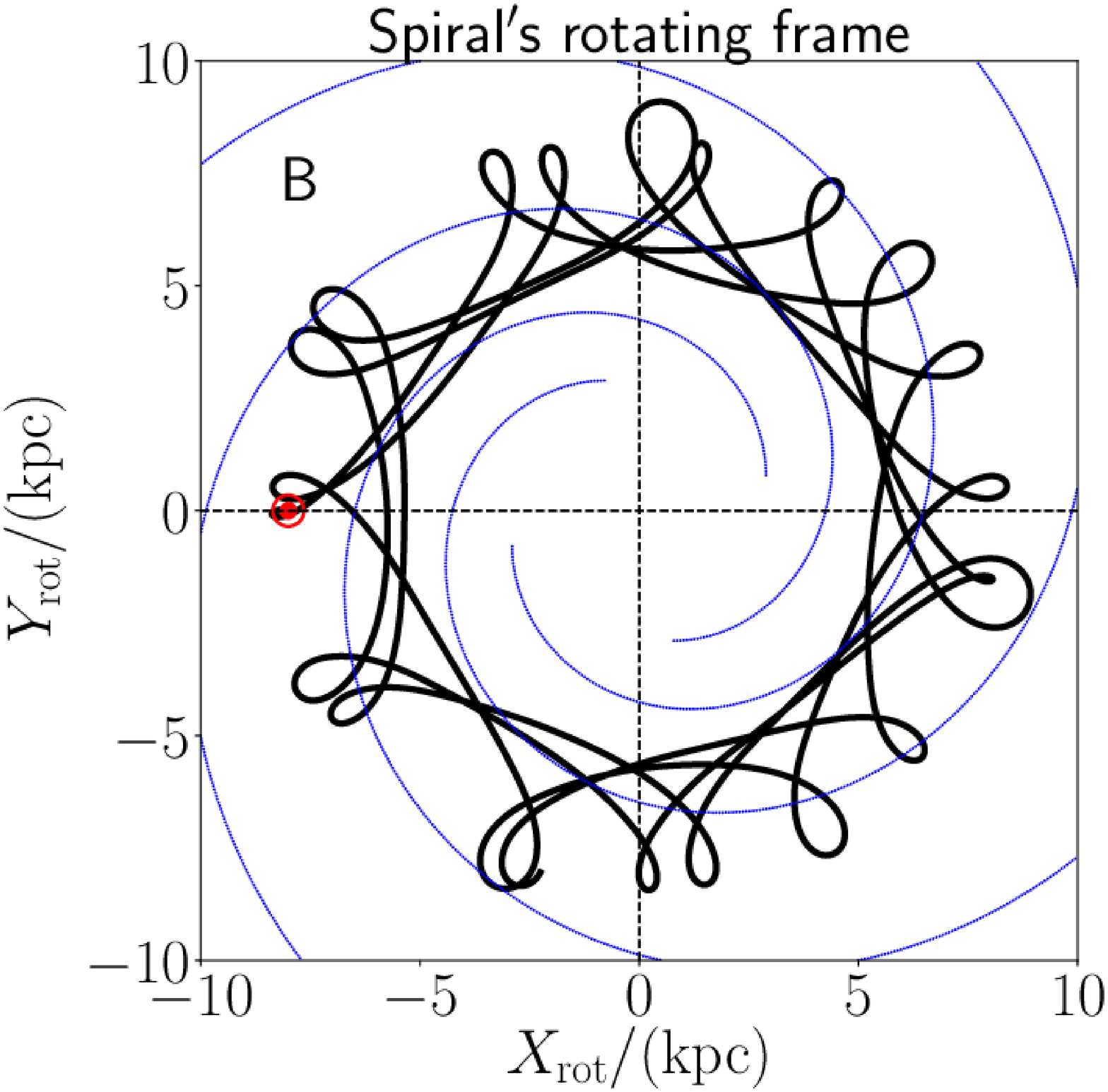} 
 \includegraphics[width=0.5\columnwidth]{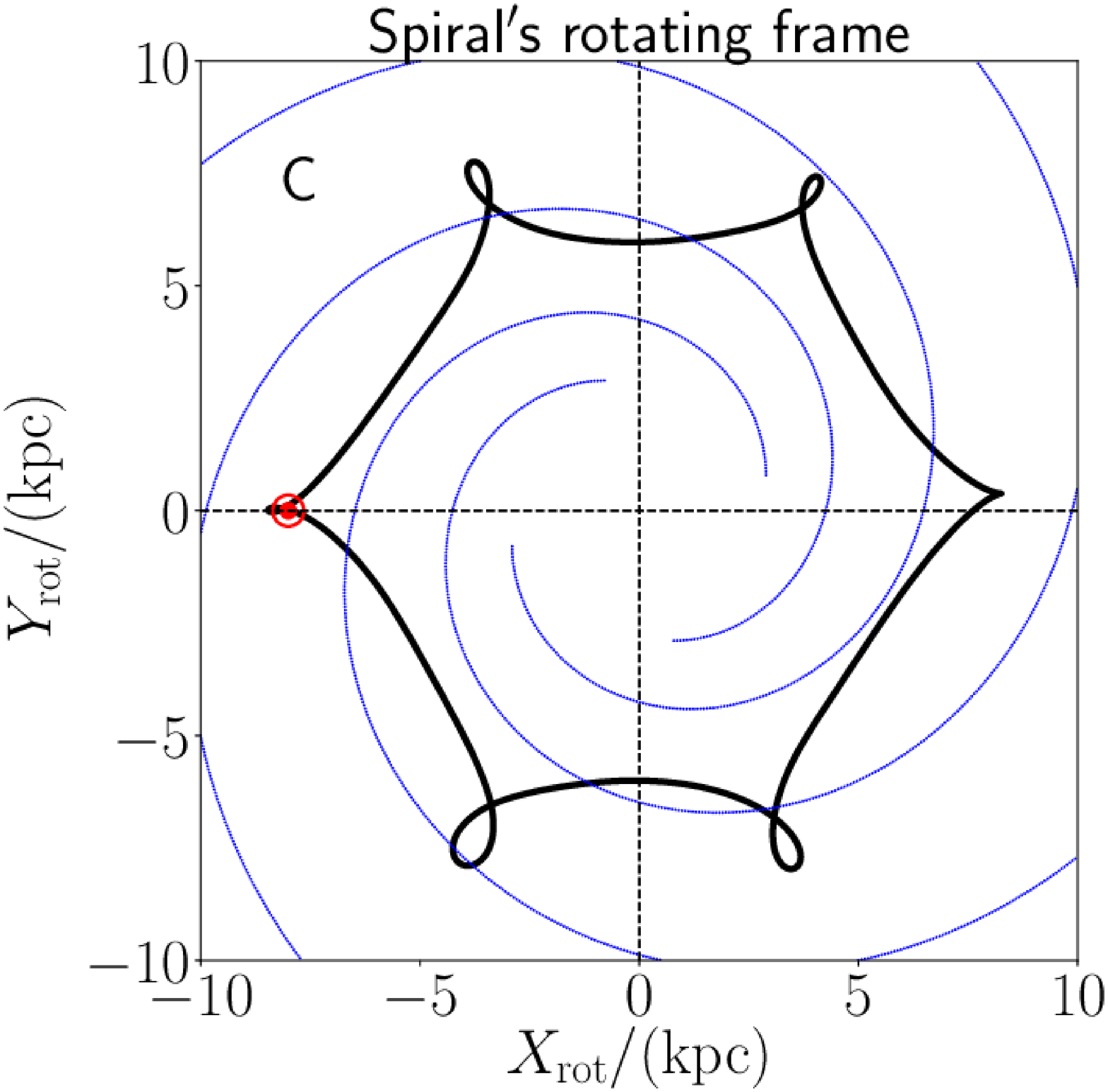} \\
 \includegraphics[width=0.5\columnwidth]{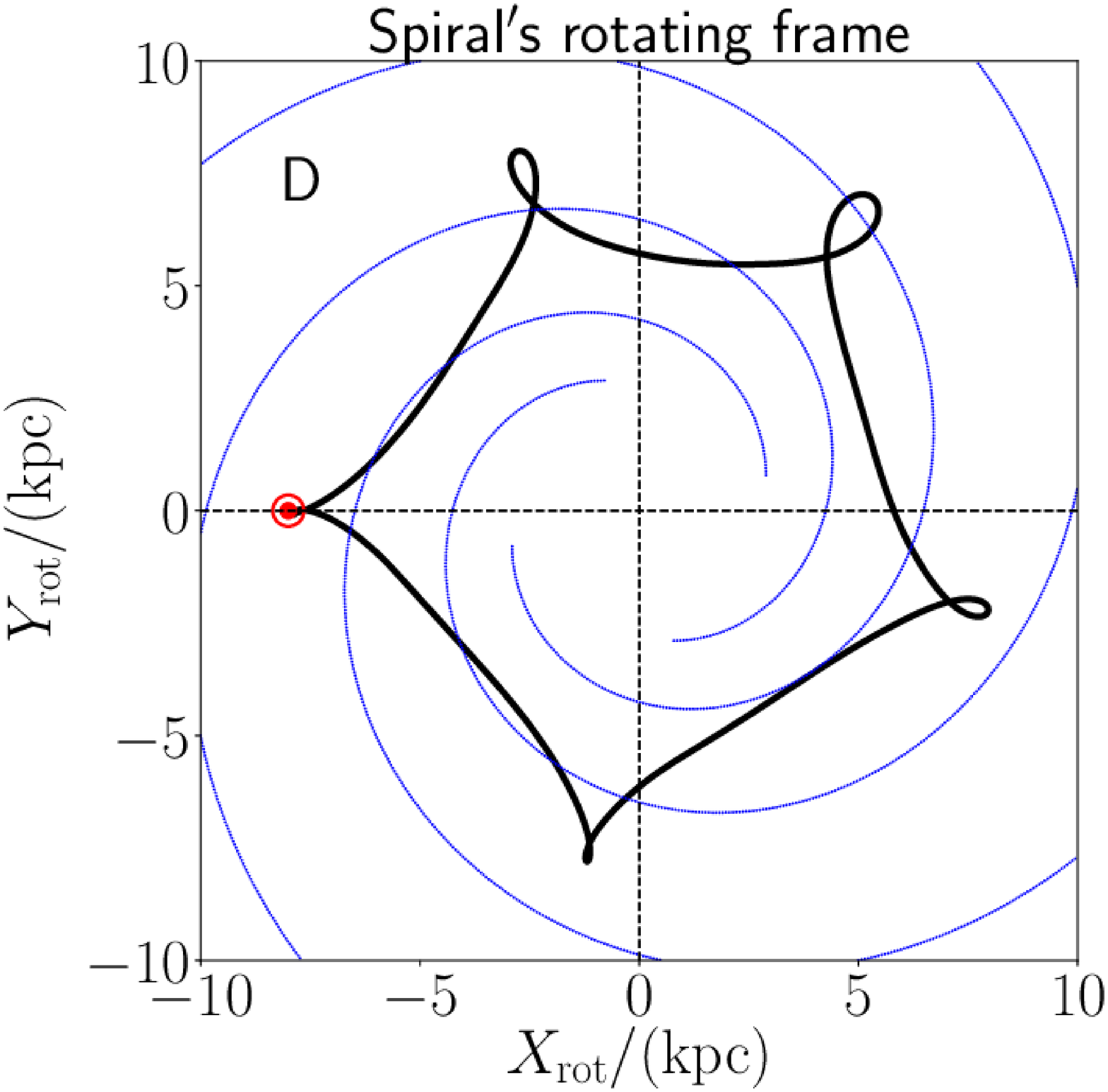} 
 \includegraphics[width=0.5\columnwidth]{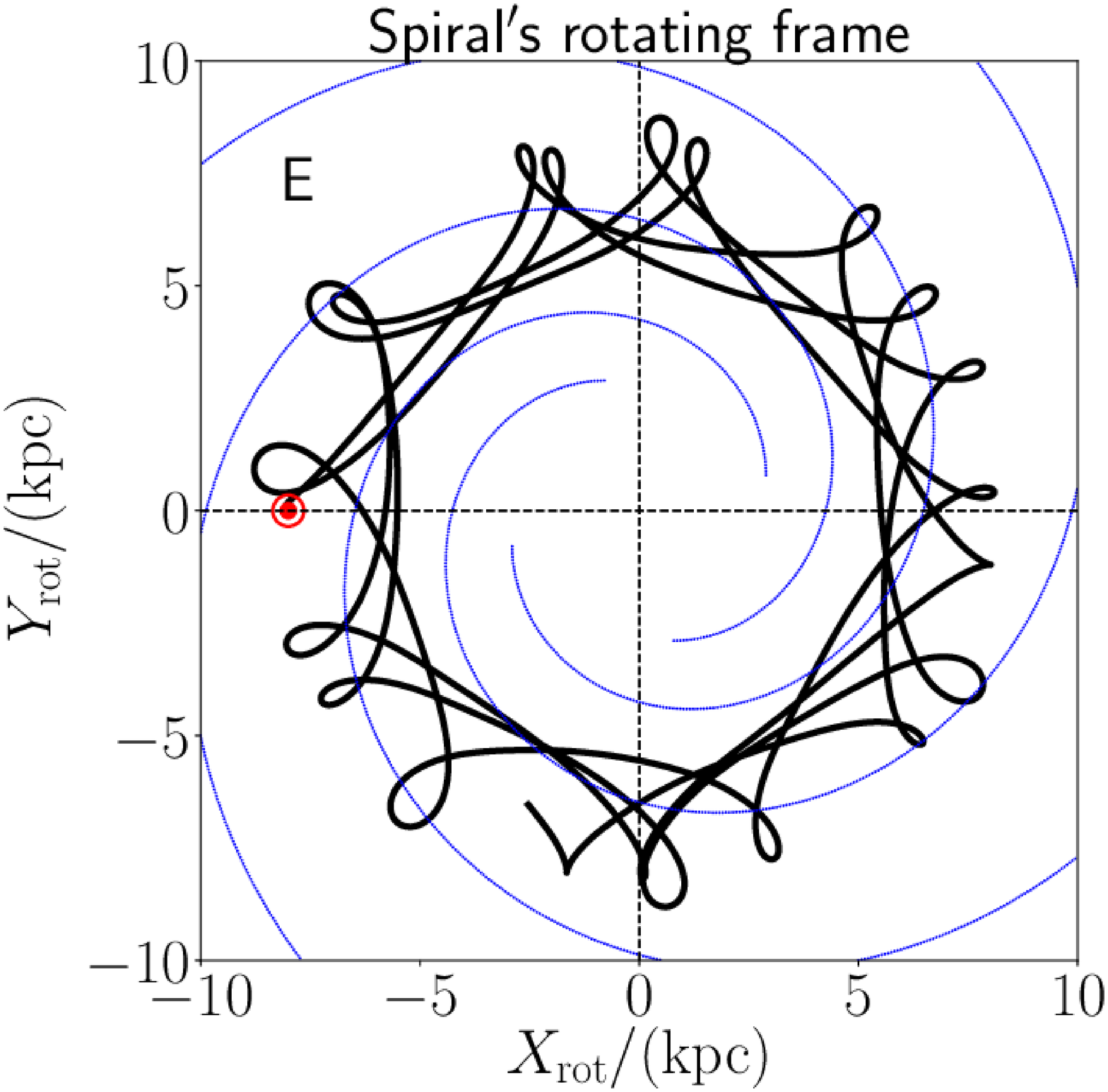} 
\caption{
Similar to Fig. \ref{fig_bar49closed}, but in the case of a successful, slow-bar$+$spiral model with $(\Omegab, m, \Omegas) = (36.11, 4, 23)$.
We note that this model reproduces observational properties (P1)-(P3) in Section \ref{section:data}. 
On the third and fourth row, 
the spirals' rotating frame $(X_\mathrm{rot}, Y_\mathrm{rot})$ is used. 
The bar is not static in this frame, so we do not show the bar here. 
}
\label{fig_bar36_spiral23m4_closed}
\end{center}
\end{figure*}

\begin{figure*}
\begin{center}
 \includegraphics[width=0.6\columnwidth]{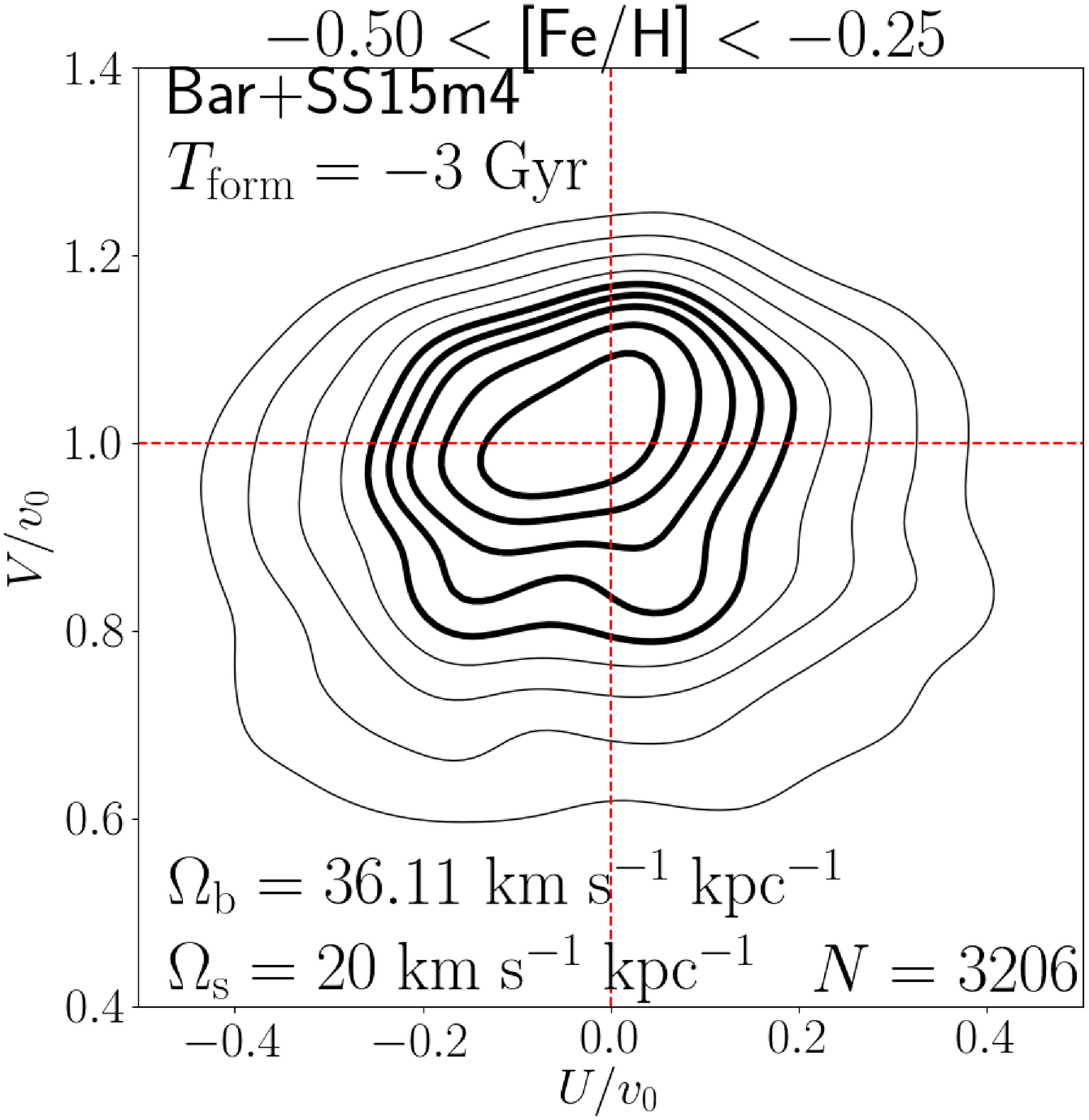} 
 \includegraphics[width=0.6\columnwidth]{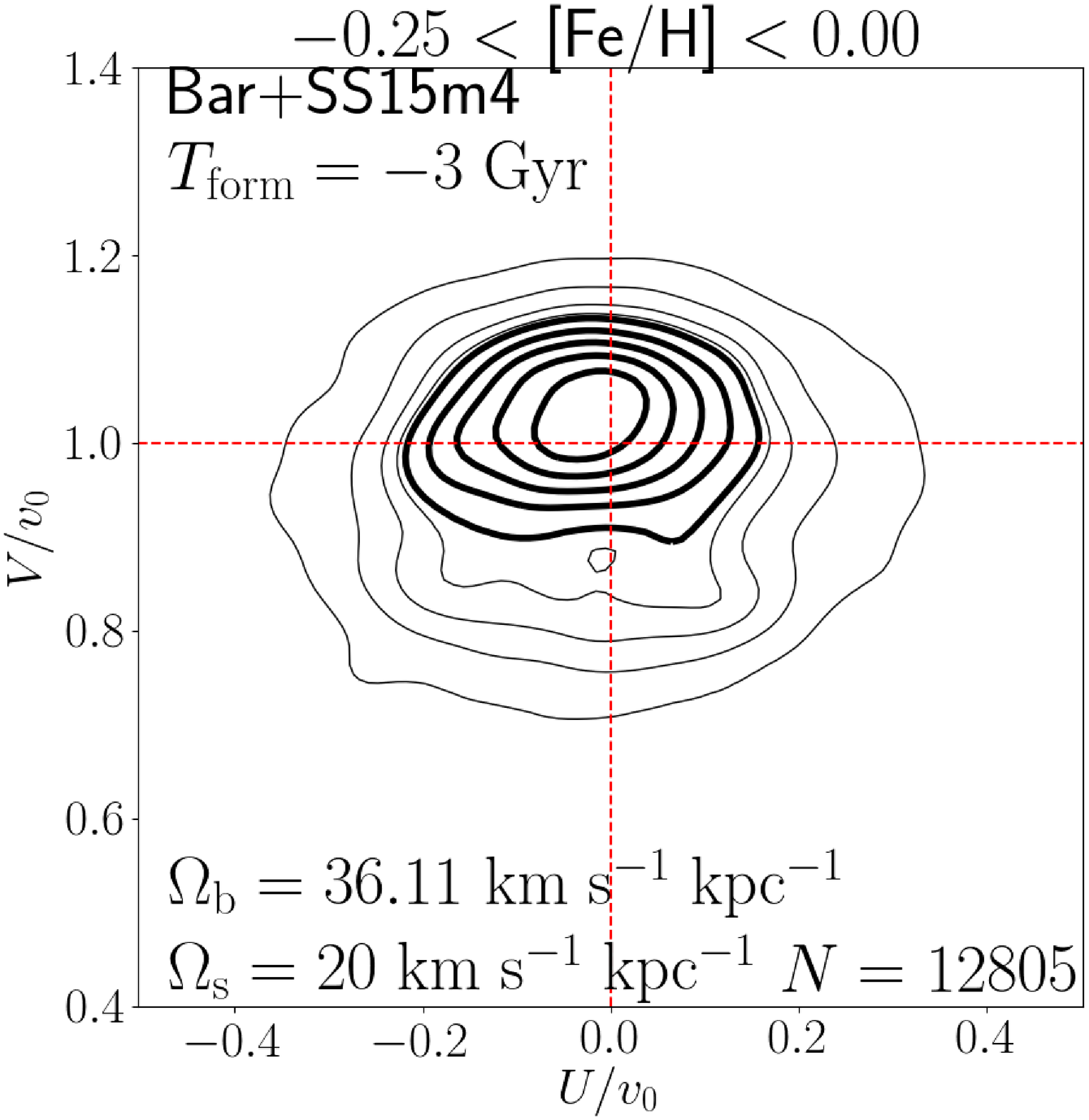} 
 \includegraphics[width=0.6\columnwidth]{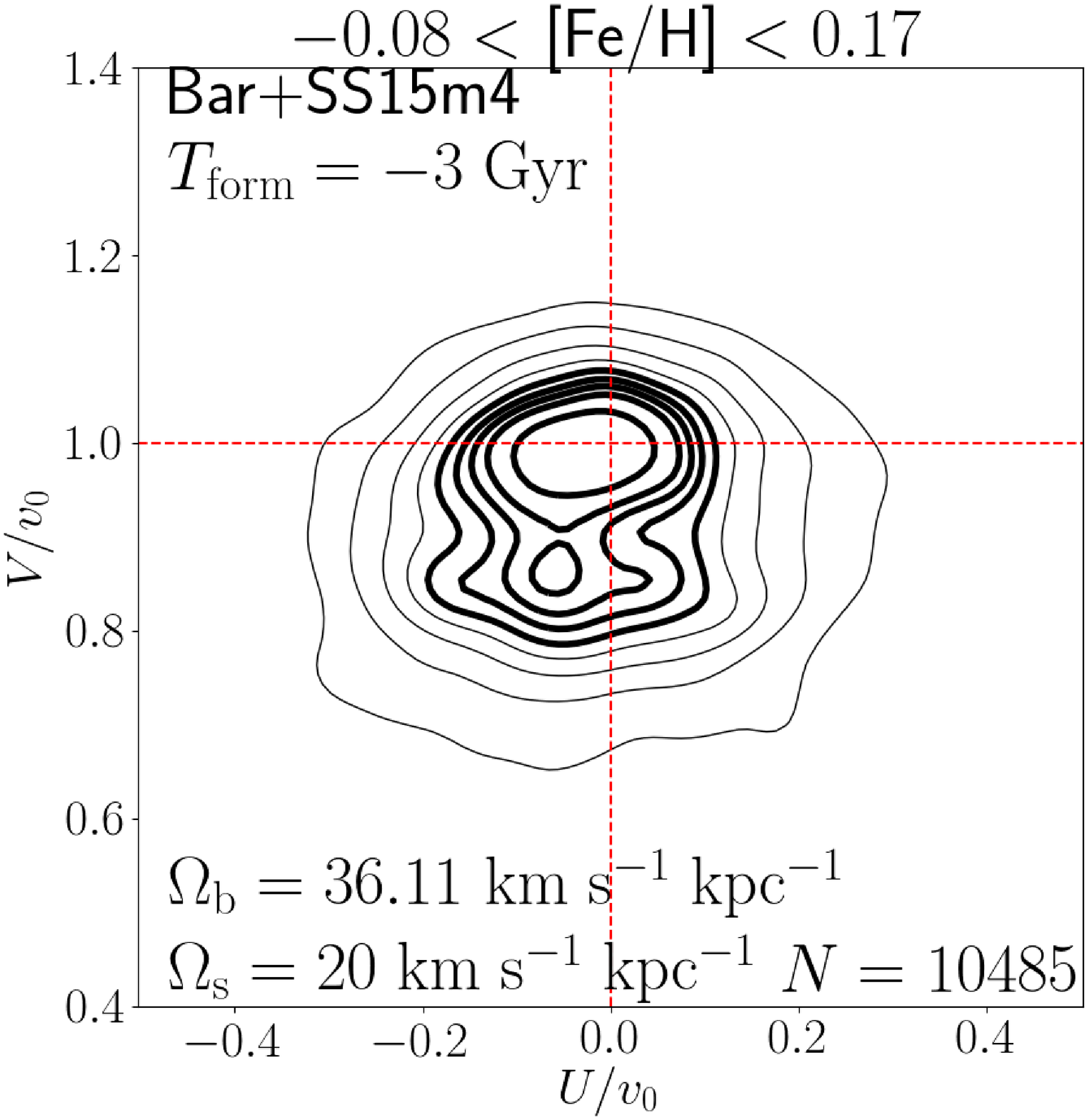} \\
 \includegraphics[width=0.8\columnwidth]{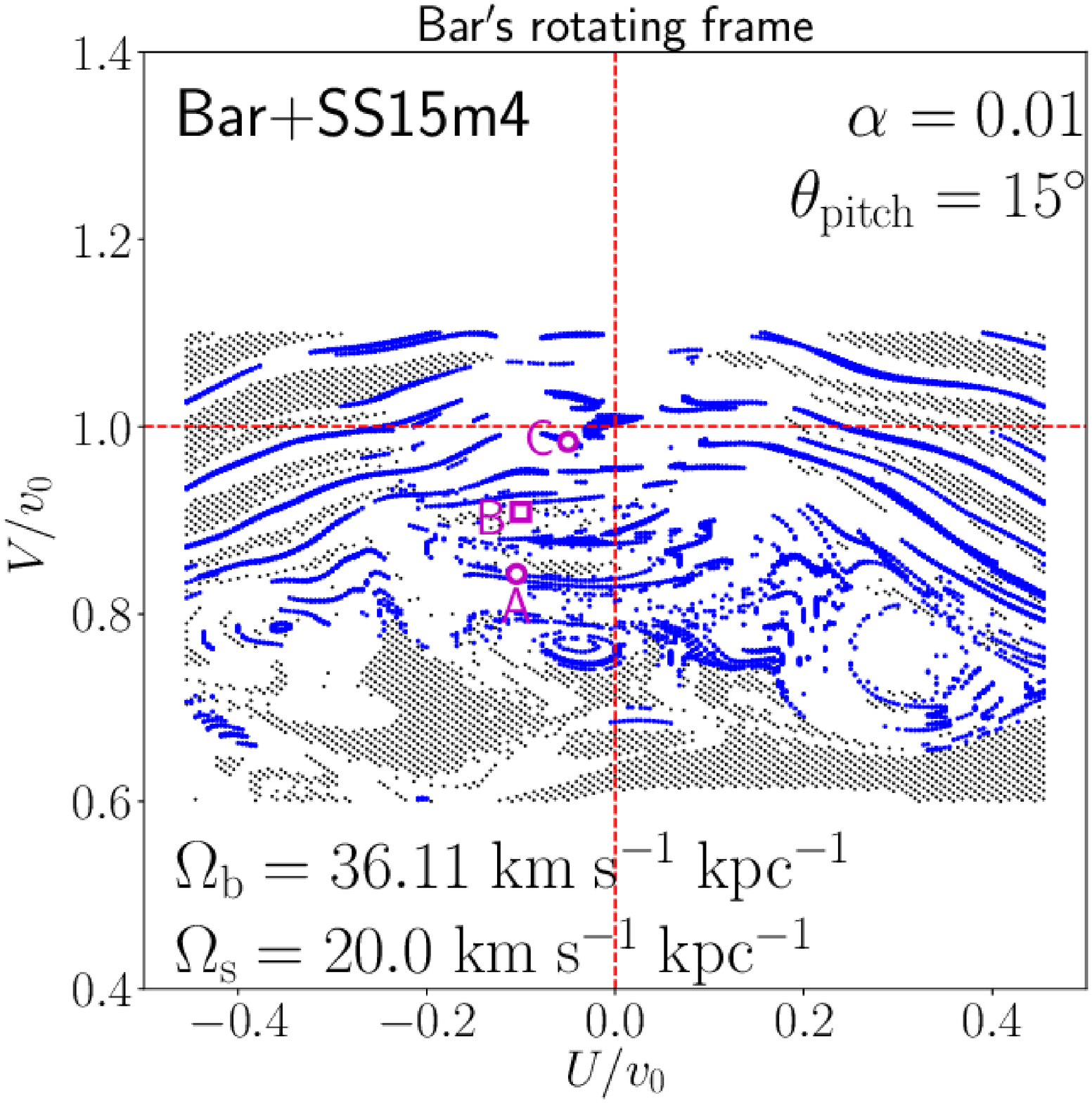} 
 \includegraphics[width=0.8\columnwidth]{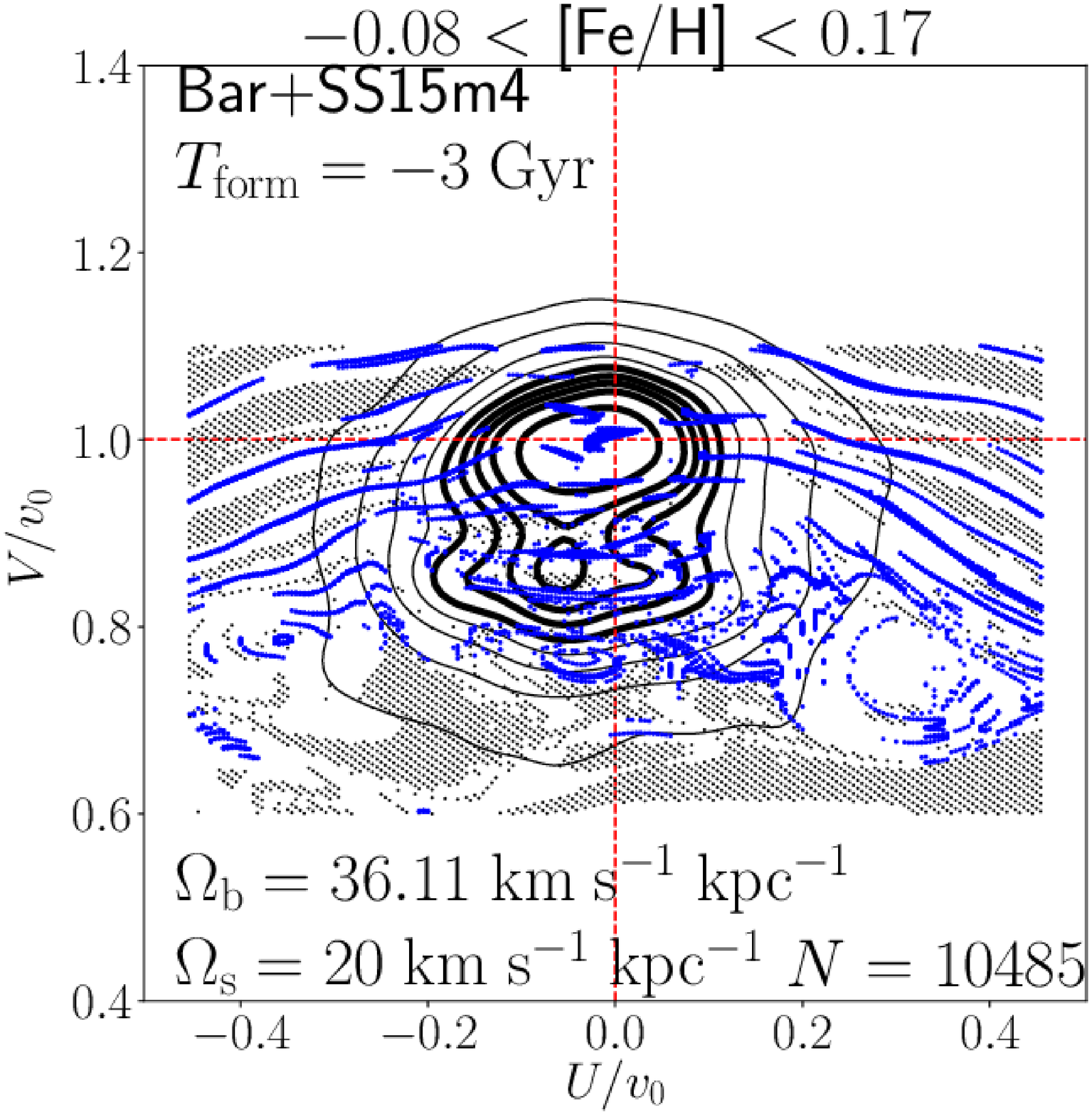} \\
 \includegraphics[width=0.5\columnwidth]{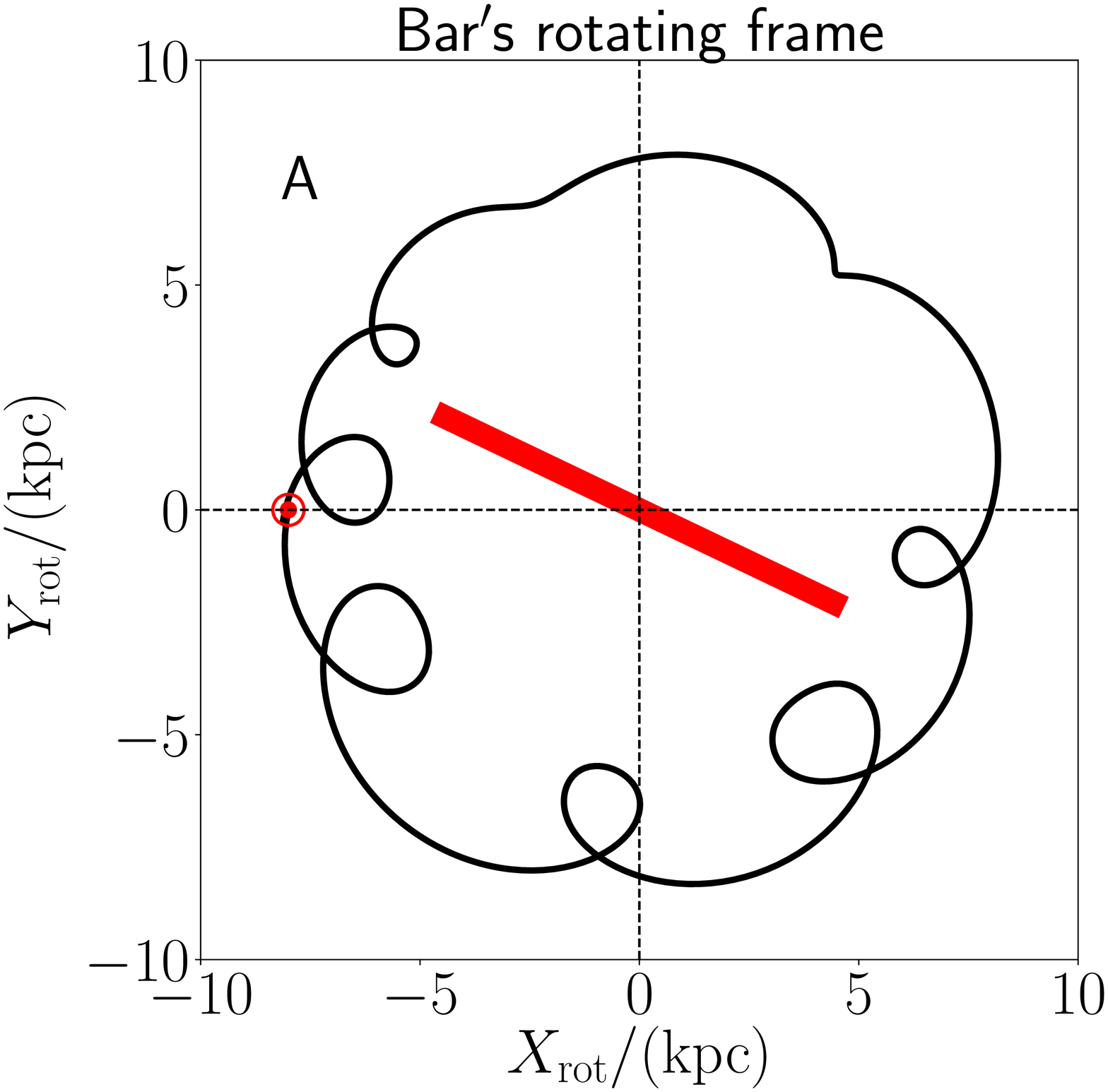} 
 \includegraphics[width=0.5\columnwidth]{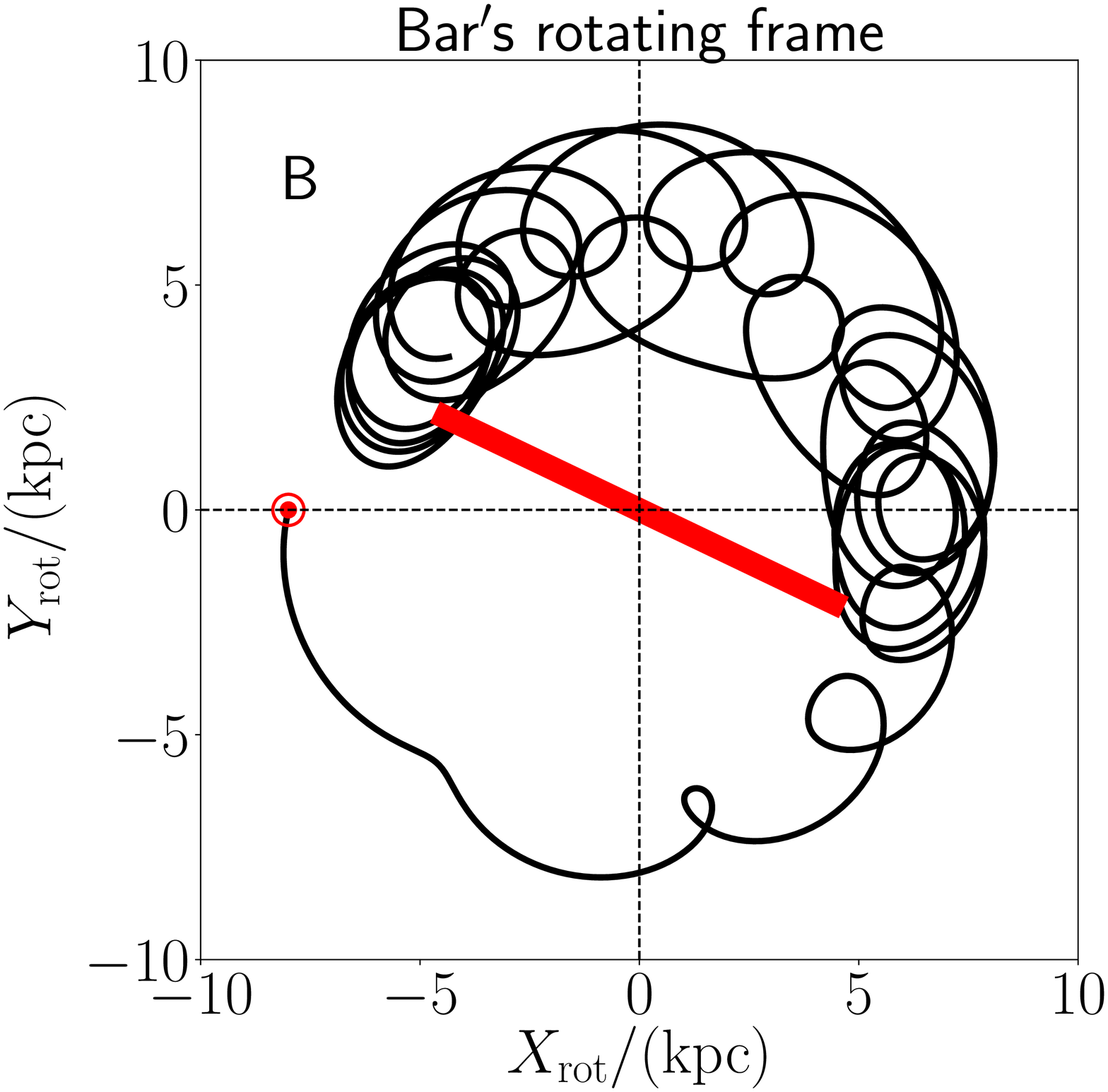} 
 \includegraphics[width=0.5\columnwidth]{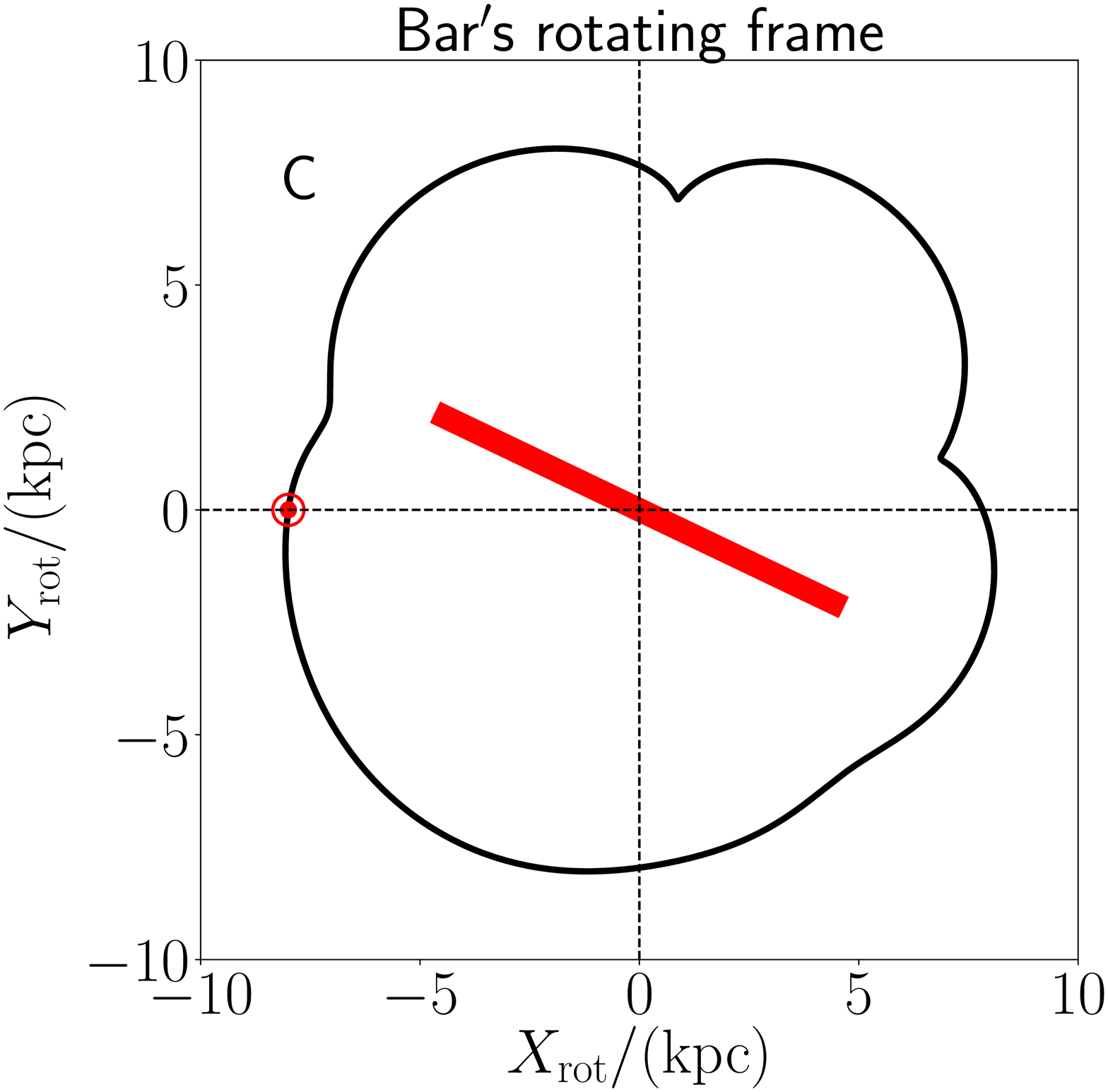} \\
\caption{
Similar to Fig. \ref{fig_bar49closed}, but in the case of a successful, slow-bar$+$spiral model with $(\Omegab, m, \Omegas) = (36.11, 4, 20)$.
We note that this model reproduces observational properties (P1)-(P3) in Section \ref{section:data}. 
On the third row, 
the bar's rotating frame $(X_\mathrm{rot}, Y_\mathrm{rot})$ is used. 
The spiral arms are not static in this frame, so we do not show spiral arms here. 
The orbit B is an example hot co-rotating orbit discussed in Sections \ref{section:thought_experimement_3} and \ref{section:bar36_spiral20m4_closed}.
}
\label{fig_bar36_spiral20m4_closed}
\end{center}
\end{figure*}

\section{Interpretation of our simulations based on closed orbit analyses}\label{section:closedOrbits}

In Section \ref{section:data}, we have listed the observational properties (P1)-(P3) of the local velocity distribution. 
In Section \ref{section:result}, we have shown that some of our models can reproduce these properties. 
Here, we perform additional orbital analyses in order to interpret the result of our simulations. 
To be specific, we identify closed orbits in the rotating frame of the bar and/or spirals.

\subsection{Simple thought experiments of closed and non-closed orbits in a rotating frame} \label{section:thought_experiments}

In this Section, we consider simple thought experiments to demonstrate 
that analysing closed orbits and non-closed orbits in the rotating frame of the bar or spirals  
is potentially important in understanding the velocity distribution of the perturbed stellar disc.

In the following, 
we suppose that the Galactic potential is given by a 2D bar-only model, 
and we consider the 2D orbits of stars at $t>0$ in the rotating frame of the bar\footnote{ 
We note that similar arguments can be done for spiral-only or bar$+$spiral models. 
}
(the rotating coordinate system is defined in
Section \ref{section:rotating_corrdinate_systems}).
We focus on orbits of stars that are currently ($t=0$) located at the position of the Sun, $(R, \phi_{\rm rot})=(R_0, 180^\circ)$ in the bar's rotating frame. 
Also, we define 
$\bar{T}_\phi$ to be 
the azimuthal orbital period in the bar's rotating frame 
(or the orbital period around a Lagrange point in the case of a co-rotating orbit).

\subsubsection{Stable closed orbits} 

We suppose that a star, $S_{\rm cl}$, happens to be on a stable closed orbit in the rotating frame of the bar. 
In this case, 
after $n$ circulations around the bar (with $n\sim1$ a small positive integer), at $t \simeq n \bar{T}_\phi$, 
the star $S_{\rm cl}$ comes back to the initial position in the bar's rotating frame, $(R, \phi_{\rm rot}) = (R_0, 180^\circ)$, 
with the same velocity as the initial velocity. 
Also, those stars whose initial conditions at $t=0$ are very close to that of $S_{\rm cl}$ (i.e., those stars in an orbit family whose parent orbit is the orbit $S_{\rm cl}$)
come back to the similar positions with similar velocities at $t \simeq n \bar{T}_\phi$. 
This indicates that the phase-space density of stars near the initial condition of $S_{\rm cl}$ will not drastically change 
with a timescale of $\sim \bar{T}_\phi$ (see similar arguments by \citealt{QuillenMinchev2005}).

\subsubsection{Highly non-closed orbits}\label{section:thought_experimement_2}

Next, we suppose that another star, $S_{\rm non}$, happens to be on a {\it highly non-closed orbit}.
Here we define a {\it highly non-closed orbit} as a non-closed orbit that is not associated with any stable closed orbits  
(e.g., 
a strongly chaotic orbit in the bar's rotating frame).  
In this case, each time the star $S_{\rm non}$ completes a circulation around the bar, 
both the Galactocentric radius and the velocity in the rotating frame may be very different from the initial condition. 
Thus, we expect that the phase-space density of stars near the initial condition of $S_{\rm non}$ would decrease with a timescale of $\sim \bar{T}_\phi$ due to the diffusion in the phase space.

\subsubsection{A subset of highly non-closed orbits: hot co-rotating orbits} \label{section:thought_experimement_3}

Finally, we suppose that yet another star, $S_{\rm hot}$, happens to belong to a horseshoe orbit around the bar's Lagrange points. 
We assume that this star is not constrained to a single Lagrange point, 
but it can travel between Lagrange points due to its kinematically hot orbit (large Jacobi energy) 
and sometimes it stays near a single Lagrange point for a long time. 
Although this 
orbit is a subset of {\it highly non-closed orbits}, 
we hereafter refer to this orbit as a `{\it hot co-rotating orbit}' for future use (see Section \ref{section:bar36_spiral20m4_closed}). 
Similar to the argument in Section \ref{section:thought_experimement_2}, 
we expect that the phase-space density near $S_{\rm hot}$ would decrease with a timescale of $\sim \bar{T}_\phi$. 
Indeed, there is a possibility that the star 
$S_{\rm hot}$ never comes back again to 
the initial location
after being trapped by another (faraway) Lagrange point and staying there for a very long time (e.g., orbit B in Figure \ref{fig_bar36_spiral20m4_closed}).

\subsubsection{Implications from the thought experiments}

Following these 
thought experiments, 
we expect that  a region in the velocity space associated with highly non-closed orbits 
may experience a shortage of stars for a period of time $\sim \bar{T}_\phi$ after the birth of the bar (or spiral arms);
and this might explain the origin of the under-dense region between the main mode and the Hercules stream. 
Of course, these thought experiments may not be very useful in understanding the long-term evolution of the velocity distribution of the perturbed stellar disc. 
However, it is worthwhile identifying the locations of nearly closed orbits and highly non-closed orbits in the velocity space.
In the next section, we will describe the practical way to find nearly closed orbits and highly non-closed orbits in our model potentials.

\subsection{Orbital analyses} \label{section:orbital_analyses}

In order to identify the locations of closed orbits in the $(U,V)$ space for each potential model, 
we integrate the orbits of stars located at the position of the Sun at $t=0$ 
forward in time in a frame that rotates with a pattern speed $\Omega_{\rm p}$. 
In bar-only and spiral-only models, we set $\Omega_{\rm p}$ to $\Omegab$ and $\Omegas$, respectively. 
In bar$+$spiral models, we run the orbital analyses with both rotating frames (bar and spirals) first, 
and then we choose a rotating frame which is easier to interpret. 
We integrate the orbits from $t=0$ to $T_{\rm max}$, 
where we set $T_{\rm max} = \min \left\{ 2.01 \bar{T}_\phi, 4 \Gyr \right\}$,  
and find the nearly-closed orbits in the rotating frame. 
Since we integrate orbits at $t>0$, we have $G(t)=1$ in equation (\ref{eq:total_potential}) 
and both the bar and spiral perturbations are time-independent in their rotating frames.

For the initial condition at $t=0$, we use $(x, y) = (-R_0, 0)$ and 
grids of rest-frame velocity $(U, V)$ spanning $-0.45<U/v_0<0.45$ and $0.6<V/v_0<1.1$. 
For each orbit, 
we trace the time evolution of a normalised 4D vector $\vec{w} = (x/R_0, y/R_0, U/v_0, V/v_0)$, 
and we evaluate the minimum value of $|\vec{w}(t) - \vec{w}(0)|$ at $T_{\rm max}/4 < t < T_{\rm max}$. 
Hereafter we refer to this minimum value as `non-closed-ness parameter', $\eta$:
\eq{
\eta = \min\limits_{ T_{\rm max}/4 < t < T_{\rm max} }  |\vec{w}(t) - \vec{w}(0)| .
\label{eq:eta}
} 
We classify those stars with $\eta < 0.03$ to be nearly-closed orbits, 
and those with $\eta > 0.3$ to be `highly non-closed orbits'.

\subsection{Fast-bar-only models} \label{section:fastbar_models}

We begin our orbital analyses with fast rotating bar models. 

As shown in Section \ref{section:barOnly}, fast rotating bar models can explain the properties of the Hercules stream 
if the bar is dynamically as young as $1\Gyr$. 
As an example, we analyse closed orbits in a fast-bar-only model with $\Omegab=49.42$. 
On the left-hand panel in the second row of Fig. \ref{fig_bar49closed}, 
we show the distribution of nearly closed orbits ($\eta<0.03$; blue dots) and highly non-closed orbits ($\eta>0.3$; black shaded regions) in $(U,V)$ space.

We see that there are several arc-shaped regions that correspond to closed orbits.
In between these arc-shaped regions, there are wider band-like regions that correspond to highly non-closed orbits. 
For example, the orbit C is located between orbit E (2:1 OLR closed orbit) and orbit B (5:2 OLR closed orbit) in the $(U,V)$ plane, 
and thus orbit C has both of these characteristics and does not close in the bar's rotating frame.

On the right-hand panel in the second row of Fig. \ref{fig_bar49closed}, we also plot the contour map of the velocity distribution 
of stars with $0<$[Fe/H]$<0.25$ in our simulation with $\Omegab=49.42$ and $\tform=-1 \Gyr$.
From this plot, we see that the main mode is 
centred around orbits characterised by 
the 2:1 OLR (represented by orbit E)
and the Hercules-like stream 
includes orbits  
characterised by the 5:2 OLR (represented by orbit B) 
and the 3:1 OLR  (represented by orbit A). 
The under-dense region between the main mode and the Hercules-like stream corresponds to 
a region where highly non-closed orbits (such as orbit C) are distributed.

We have performed the same analyses to another successful model with $\Omegab=52.16$, 
and found that the orbital distribution is very similar to that with $\Omegab=49.42$. 
The orbital analyses of our two successful models with $\Omegab=49.42$ and $52.16$ 
suggest that 
the velocity bimodality in the fast-bar-only models 
arise from highly non-closed orbits in the rotating frame of the bar. 
This interpretation is similar to that presented in \cite{Dehnen2000},
in which the under-dense region is attributed to unstable (chaotic) orbits due to the perturbation from the rotating bar.

The overlap between the location of the secondary peak (Hercules-like stream) 
and the location of the orbits characterised by 5:2 OLR or 3:1 OLR 
is consistent with the idea that 
the orbit family around a nearly closed orbit move 
in a relatively coherent way at least for some dynamical time. 
Our result indicates that, if we can map the locations of the Hercules stream across the Galactic plane,
we might see a pattern that corresponds to the 
parent orbits of this coherent motion.
With this in mind, 
in Section \ref{section:Hercules_across_disc_plane},
we make some simple prediction of the locations in the Galactic disc plane where velocity bimodality can be seen (Figs. \ref{fig:map_bimodality}(a) and \ref{fig:10kpc}). 
This prediction can be tested with the Gaia Data Release 2 (DR2) and other spectroscopic surveys.

The idea
that a wide region of highly non-closed orbits in the $(U,V)$ space can 
create an under-dense region in the velocity space that characterise the velocity bimodality 
is intuitively understandable 
(see
Section \ref{section:thought_experiments}). 
Highly non-closed orbits have very different velocity 
after a period of 
$\Delta t \sim \bar{T}_\phi$ (after one or two circulations 
around the bar).   
This rapid dissipation in the velocity space makes a region of highly non-closed orbits an under-dense region in $(U,V)$ plane. 
In this regard, 
it is intriguing to note that fast-bar-only models with older dynamical age ($|\tform| \geq 2 \Gyr$) are not successful in reproducing the bimodal structure. 
This result might be due to the fact that the region of highly non-closed orbits in the $(U,V)$ space 
shows lower phase-space density for only $\sim \bar{T}_\phi$ after the formation of the bar. 
After this period, we speculate that some dynamical processes such as the phase mixing 
might blur the contrast between the over-dense region (Hercules stream) and the under-dense region. 
Given that it takes $2\bar{T}_\phi = 691 \Myr$ for the orbit B to make two circulations around the bar, 
our result might suggest that 
the orbit families associated with 
the Hercules stream in a fast-bar-only model 
are 
coherent for $n\leq3$ circulations around the bar $(\leq 1 \Gyr)$ 
but it begins to lose coherence after $n\gtrsim 6$ circulations ($\gtrsim 2 \Gyr$).

\subsection{Slow-bar-only models} \label{section:slowbar_models}

As shown in Section \ref{section:barOnly}, bar-only models with a slowly rotating bar can not explain the properties of the Hercules stream. 
Fig. \ref{fig_bar36closed} shows the orbital analyses of one of the unsuccessful bar-only models with $\Omegab=36.11$. 

As seen in the second row of Fig. \ref{fig_bar36closed}, this model gives rise to 
a lot of closed orbits in the Solar neighbourhood. 
These closed orbits are associated with relatively high order resonances (e.g., orbit B corresponds to the 4:1 OLR), 
and thus their locations in the $(U,V)$ plane are close to each other. 
Due to this densely spaced distribution of closed orbits, 
highly non-closed orbits occupy narrow regions in the $(U,V)$ space and they are not important near $(U,V)=(0,v_0)$. 
This is the reason why we see a monomodal velocity distribution in our simulation (see top panels).

\subsection{Spiral-arms-only models}

\subsubsection{Models with 4-armed spirals with $\Omega = 21 \kmskpc$} \label{section:omegas21m4_models}

Fig. \ref{fig_spiral21m4closed} 
shows the orbital analyses of SS15m4 model with $\Omegas=21$. 
We can see from these plots that the under-dense region at around $V\simeq 0.85 v_0$ 
corresponds to highly non-closed orbits (represented by orbit C). 
At $-0.2<U/v_0<0.2$ and $0.75<V/v_0<0.8$, 
there is a prominent region of nearly closed orbits,
which is characterised by the 4:1 inner Lindblad Resonance (ILR) of the spiral arms (represented by orbits A and B). 
The secondary peak at $V \simeq 0.8 v_0$ 
is adjacent to this 4:1 ILR orbit family.  
On the other hand, at $0.9 < V/v_0 < 1.05$, 
there are many narrow or small regions of nearly closed orbits, 
such as the 5:1 ILR (orbit D) or the 6:1 ILR (orbit E). 
Since these closed orbits show a densely spaced distribution,
these orbit families altogether form the main mode.

It is interesting to note that 
the position of the secondary peak in this model 
is not at the exact location of the nearly closed orbits associated with the 4:1 ILR. 
This can be understood in the following manner. 
Since the velocity distribution is nearly Gaussian around $(U/v_0, V/v_0) = (0, 1)$, 
the number of stars with $U \simeq 0$ 
drops as $V/v_0$ decreases from 1 to 0.7 or so. 
Thus, even if  
the parent 4:1 ILR orbits at $V/v_0 \simeq 0.75$ are sticky and stable, 
only a small number of stars at $V/v_0 \leq 0.75$ can be trapped around these parent orbits. 
As a result, the secondary peak is located around $V/v_0 \simeq 0.8$, 
which is adjacent to the blue region of nearly closed orbits with 4:1 ILR.

\subsubsection{Models with 2-armed spirals with $\Omegas = 28 \kmskpc$} \label{section:omegas28m2_models}

Fig. \ref{fig_spiral28m2closed} shows the orbital analyses of SS20m2 model with $\Omegas=28$. 
We see from the second row of this figure that a large fraction of $(U,V)$ space at $0.85<V/v_0<1.1$ is filled with nearly closed orbits 
that are trapped by the co-rotating resonance of the spiral arms. 
These co-rotating orbits form the main mode of the velocity distribution. 
On the other hand, at $V/v_0<0.8$, 
there are a few distinct orbit families such as the 8:1 ILR (orbit A) or the 10:1 ILR (orbit B),
which altogether form the secondary peak of the velocity distribution (top right panel). 
In the region between the main mode and secondary mode, 
we find highly non-closed orbits, which is represented by orbit C.

\subsection{Bar$+$spiral models} \label{section:various_bar+spiral_models}

In the previous subsections, we have analysed the closed orbits in bar-only and spiral-only models. 
Here we investigate the bar$+$spiral models, which is more complicated than bar-only or spiral-only models. 
The complexity mainly arises from the fact that exact closed orbits does not exist in general if $\Omegab \neq \Omegas$, 
since the potential is time-dependent in any rotating frame. 
Due to this time-dependence, some fraction of disc orbits have chaotic nature; 
and therefore it is not guaranteed that the integration of orbits for $\sim 2 \bar{T}_\phi$ in our analyses would be helpful in general. 
However, our orbit analyses turned out to be reasonably helpful to understand the origin of the Hercules-like stream in bar$+$spiral models as well.
Here we report three representative models that have a prominent Hercules-like structure at the metal-rich 
region.

\subsubsection{Highly non-closed orbits in fast-bar's rotating frame} \label{section:bar49_spiral21m4_closed}

Fig. \ref{fig_bar49_spiral21m4_closed} shows a fast-bar$+$spiral models with $(\Omegab, m, \Omegas) = (49.42, 4, 21)$ and $\tform=-5\Gyr$. 
We see from the left-hand panel in the second row that there is a prominent region of 
highly non-closed orbits at $V/v_0 \simeq 0.9$ represented by the orbit C. 
By comparing the panels in the second row of this figure, 
we can see that 
the boundary between the two over-dense regions 
corresponds to these highly non-closed orbits. 
The main mode include a family of orbits associated with the 2:1 OLR (orbit D), 
and the Hercules-like stream seems to include orbits 
associated with 3:1 OLR (orbit A). 
The location of the Hercules-like stream 
is slightly offset from the exact location of the nearly-closed 3:1 OLR orbits, 
which may be explained by the same argument as in the second paragraph of Section \ref{section:omegas21m4_models}.

By comparing this model and the young, fast-bar-only model (Fig. \ref{fig_bar49closed}), 
we see that the locations of the nearly closed orbits are noticeably shifted  
due to the additional perturbation from the spirals. 
(We note that these models have the same parameters for the bar.)
However, in both models, the under-dense region between the main mode and the Hercules stream 
is associated with the highly non-closed orbits. 
Therefore, it may be reasonable to regard the Hercules stream as the secondary peak in the velocity space 
that is induced by the under-dense region in the phase space associated highly non-closed orbits.

\subsubsection{Highly non-closed orbits in spirals' rotating frame} \label{section:bar36_spiral23m4_closed}

Fig. \ref{fig_bar36_spiral23m4_closed} shows a slow-bar$+$spiral models with $(\Omegab, m, \Omegas) = (36.11, 4, 21)$ and $\tform=-3\Gyr$. 
By comparing the panels in the second row of this figure, 
we see that the location of the 
under-dense region between the main mode and the Hercules-like stream corresponds to 
highly non-closed orbits represented by orbits B and E.
The location of the Hercules-like stream 
is close to that of the orbit family associated with the 5:1 ILR of the spiral arms 
(represented by nearly closed orbit A and {\it mildly} closed orbit D).

\subsubsection{Highly non-closed (hot co-rotating) orbits in slow-bar's rotating frame} \label{section:bar36_spiral20m4_closed}

Fig. \ref{fig_bar36_spiral20m4_closed} shows a slow-bar$+$spiral models with $(\Omegab, m, \Omegas) = (36.11, 4, 20)$ and $\tform=-3 \Gyr$. 
On the right-hand panel in the second row, 
we see a bimodal velocity distribution. 
At the boundary between these two peaks, 
we see a narrow band of highly non-closed orbits (represented by orbit B). 
This 
highly non-closed orbit B can be further classified as the hot co-rotating orbit discussed in Section \ref{section:thought_experimement_3}. 
This orbit is trapped by co-rotation resonance of the bar, but it can travel from one Lagrange point to another due to its hot kinematics. 
For example, soon after the orbit integration, 
orbit B is captured by the Lagrange point far away from the initial position in the co-rotating frame and it stays there for more than $3 \Gyr$.

We note that similar hot co-rotating orbits are identified in the recent work by \cite{PerezVillegas2017}. 
In their fig. 3, the orbits (d) and (h) seem to be hot co-rotating orbits.  
Interestingly, the velocities of these stars (when they come to the Solar neighbourhood) seem to be located near the under-dense region 
between the Hercules stream and the main mode, which is consistent with our explanation.

We note that, in this slow-bar$+$spiral model, 
it is hard to imagine the velocity bimodality 
by just looking at the left-hand panel in the second row of Fig. \ref{fig_bar36_spiral20m4_closed}. 
This is not surprising, 
as bar$+$spiral models with $\Omegab \neq \Omegas$ 
do not have exactly closed orbits in any rotating frame 
(as emphasised at the beginning of Section \ref{section:various_bar+spiral_models}). 
In other words, 
the {\it nearly} closed orbits detected by our method (such as orbits A and C) 
may not be dynamically important. 
Thus, our method should be used with care especially for bar$+$spiral models.

\subsection{Dynamical origin of the Hercules stream revealed from orbital analyses} \label{section:recap_orbit_analyses}

In this Section, 
we have demonstrated that identifying the locations in the $(U,V)$ space 
of nearly closed orbits and highly non-closed orbits in the rotating frame of the bar or spirals 
is very informative in understanding the origin of the bimodal velocity distribution. 
Based on the analyses of successful models (Figs. \ref{fig_bar49closed}, \ref{fig_bar49_spiral21m4_closed}, \ref{fig_bar36_spiral23m4_closed}, and \ref{fig_bar36_spiral20m4_closed}) 
and partially successful models (Figs. \ref{fig_spiral21m4closed} and \ref{fig_spiral28m2closed}), 
we conclude that the under-dense region between the two velocity peaks (main mode and the Hercules-like stream in successful models) 
are associated with highly non-closed orbits in the rotating frame. 
This can be physically understood given that 
nearly closed orbits come back to the original phase-space location (position and velocity in the rotating frame) and retains its phase-space density for some time; 
while highly non-closed orbits dissipate in the phase-space 
and the phase-space density decreases relatively quickly 
(see
Section \ref{section:thought_experiments}).

\section{Discussion} \label{section:discussion}

\subsection{Origin of the [Fe/H] dependence of the Hercules stream} \label{section:origin_of_FeH_dependence}

The results in Sections \ref{section:result} and \ref{section:closedOrbits} 
provide some hints on the origin of the bimodality in the local velocity distribution.
Here we discuss the origin of the {\it [Fe/H] dependence} of the bimodality.

\begin{figure*}
\begin{center}
 \includegraphics[width=1.80\columnwidth]{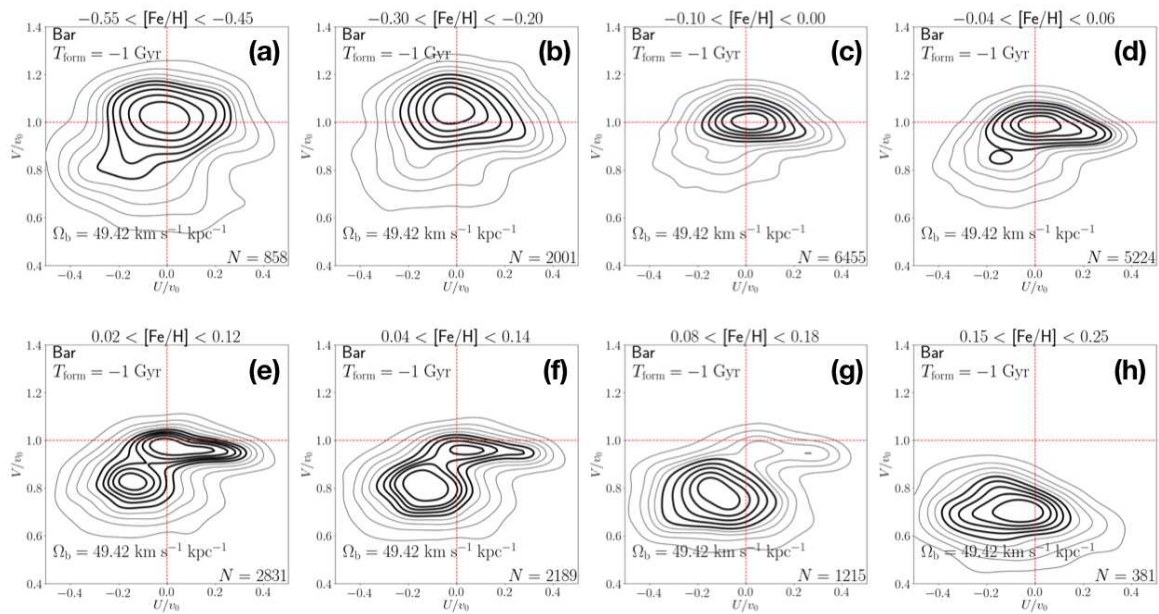} \\
\caption{
The detailed [Fe/H] dependence of the $(U,V)$ distribution of our successful fast-bar-only model 
with $\Omegab=49.42$ and $\tform=-1 \Gyr$. 
}
\label{fig_model_UV_FeH_detail}
\end{center}
\end{figure*}

\begin{figure}
\begin{center}
 \includegraphics[width=1.00\columnwidth]{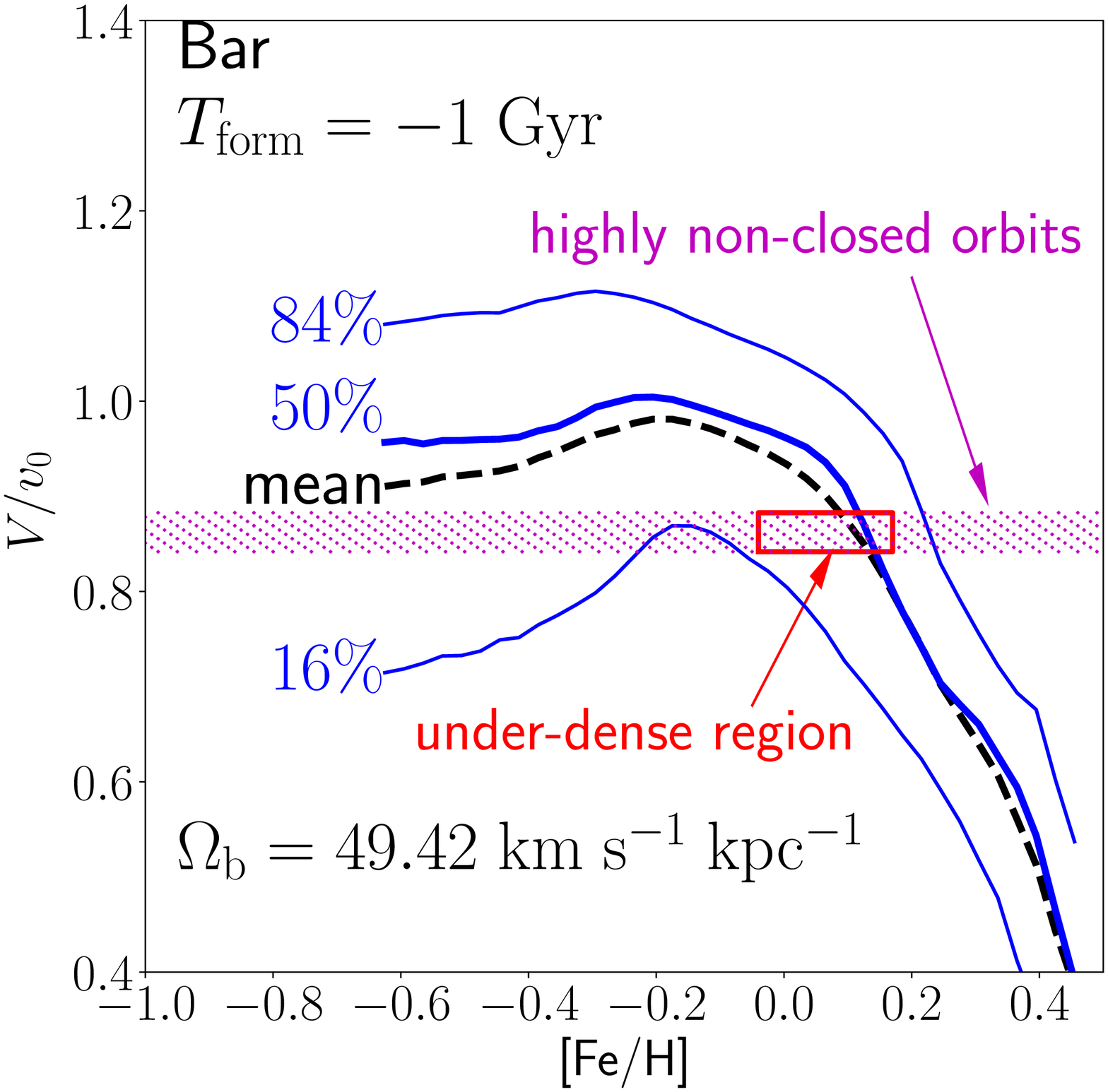} \\
\caption{
The [Fe/H] dependence of the 
local distribution of $V$ (azimuthal velocity) 
for a dynamically-young, fast-bar-only model. 
The black dashed line and blue thick solid line correspond to the mean and median value of $V$ at each [Fe/H], 
while the blue thin solid lines cover 68\% of the velocity distribution. 
The red square at $-0.04<$[Fe/H]$<0.17$ 
shows the approximate location of the under-dense region in the $(U,V)$-distribution, 
and 
the horizontal shaded region in magenta corresponds 
to the value of $V$ for highly non-closed orbits (such as orbit C in Fig. \ref{fig_bar49closed}).
}
\label{fig_origin_of_Hercules}
\end{center}
\end{figure}

\subsubsection{Useful quantities to understand the [Fe/H] dependence of the velocity bimodality} \label{section:key}

In order to understand the observed (or simulated) [Fe/H] dependence of the velocity bimodality, 
it is essential to keep in mind the following fact: 
\begin{itemize}
\item[] The effect from the non-closed orbits is important {\it only if} there are many disc stars associated with these orbits. 
\end{itemize}
This can be intuitively understood with some extreme examples. 
Suppose that there is a region in the phase space centred at $(U, V)=(u_1, v_1)$ where the corresponding orbits are strongly chaotic. 
If this region is located at the Local Standard of Rest, $(u_1, v_1) = (0, v_0)$, this region probably has a strong impact on the velocity distribution, 
since a lot of disc stars in the Solar neighbourhood are affected by this chaotic region. 
Depending on the situation, this chaotic region may result in an under-dense region in the $(U,V)$ space 
to create a bimodal velocity distribution as we have seen in previous Sections. 
However, if this chaotic region is located at $(u_1, v_1) = (0, -v_0)$ (i.e., circular velocity in the retrograde direction), 
this chaotic region is not important at all in terms of the disc velocity distribution, 
since there are few disc stars with retrograde rotation to be affected. 

Based on these two extreme examples, it may be naturally understood that 
the highly non-closed orbits are not important 
if either of the following conditions is satisfied:
\begin{itemize}
\item (Condition 1) When the highly non-closed orbit region is located far away from the region where the majority of disc stars are distributed. 
\item (Condition 2) When the velocity dispersion is so large and the stellar density in the velocity space is so low that the number of disc stars that fall into the highly non-closed orbit region is small. 
\end{itemize}
On the other hand, the velocity bimodality is prominent 
if both of the following conditions are satisfied:
\begin{itemize}
\item (Condition 1') When the median velocity is close to the velocity of highly non-closed orbits. 
\item (Condition 2') When the velocity dispersion is small. 
\end{itemize}
In the following, we shall investigate the median velocity and the velocity dispersion as a function of [Fe/H] 
in one of our successful models 
in order to understand the [Fe/H] dependence of the Hercules stream.

\subsubsection{[Fe/H] dependence of the velocity bimodality in a successful model}

We revisit the young, fast-rotating bar model described in Fig. \ref{fig_bar49closed}. 
Here, highly non-closed orbits are distributed at $V \simeq 0.86 v_0$ (see orbit C in Fig. \ref{fig_bar49closed}). 
By analysing the local velocity distribution at various [Fe/H] in this model, 
some of which are shown in Fig. \ref{fig_model_UV_FeH_detail} (see also top panels in Fig. \ref{fig_bar49closed}), 
we found that the velocity bimodality is clearly seen 
only at around $-0.04<$[Fe/H]$<0.17$ (panels (d)-(f) of Fig. \ref{fig_model_UV_FeH_detail}).

Fig. \ref{fig_model_UV_FeH_detail} provides some insights into the origin of the Hercules stream. 
First, we focus on panels (a)-(d).
These panels show that 
the Hercules-like stream becomes gradually visible as [Fe/H] increases from $\sim -0.5$ (panel (a)) to $\sim 0$ (panel (d)). 
Since the dispersion in $V$ decreases as [Fe/H] increases, 
our result is consistent with the idea discussed in Section \ref{section:key} (see Conditions 2 and 2'). 
Thus, the waning of the Hercules stream at low-[Fe/H] region seems to arise from the 
increased velocity dispersion at lower [Fe/H].  
We note that, in reality, the waning of the Hercules stream at low-metallicity region may be stronger than in our simulations, 
since old and metal-poor stars tend to have larger vertical velocity dispersion (which is ignored in our 2D simulations). 
Due to the larger vertical motion, these stars may be less affected by the bar and spirals.

Second, we focus on panels (e)-(h) in Fig. \ref{fig_model_UV_FeH_detail}. 
These panels show that 
the Hercules-like stream is prominent (panel (e)) 
when the median value of $V$ is close to $V \simeq 0.86 v_0$; 
and it becomes weak and even disappears 
when the median of $V$ deviates from $\simeq 0.86 v_0$ (panels (g) and (h)). 
Our result is consistent with the idea discussed in Section \ref{section:key} (see Conditions 1 and 1'). 
Thus, the waning of the Hercules stream at high-[Fe/H] region in this model seems to arise from the 
decreased median velocity $V$ at higher [Fe/H] 
(due to the asymmetric drift; see \citealt{BT2008} section 4.4.3).  
Our result indicates that 
the Hercules stream in the Milky Way may disappear at the very metal-rich region.

Fig. \ref{fig_origin_of_Hercules} 
shows the distribution of $V$ in the Solar neighbourhood as a function of [Fe/H] in this simulation. 
The trend in this distribution is qualitatively similar to that seen in the observed data (e.g., fig. 7 of \citealt{Lee2011}). 
(This means that our model  captures important aspects of the chemo-dynamical evolution of the Galactic stellar disc.) 
This figure well summarises our interpretation of the [Fe/H] dependence of the Hercules stream. 
First, 
the dispersion in $V$ increases as [Fe/H] decreases from $\sim -0.1$ to $\sim -0.7$ 
(due to the age-velocity dispersion relationship), 
which is why we do not see prominent Hercules-like stream at [Fe/H]$<-0.1$. 
Second, 
a prominent under-dense region (which characterises the Hercules-like stream) 
is seen at $-0.04<$[Fe/H]$<0.17$ (marked by the red square), 
where 
the median value of $V$ is close to $\simeq 0.86 v_0$. 
Third, 
at [Fe/H]$\gtrsim 0.2$, 
the Hercules-like stream becomes weaker as [Fe/H] increases, 
since the median value of $V$ deviates from $\simeq 0.86 v_0$ more significantly.

\subsection{Velocity bimodality across the Galactic plane} \label{section:Hercules_across_disc_plane}

As suggested by \cite{Bovy2010}, 
a powerful way to understand the origin of the Hercules stream is to trace the Hercules stream across the Galactic plane. 
With the increased data by Gaia and other spectroscopic surveys, 
mapping the Hercules stream is becoming tractable in these days \citep{Monari2017b, Hunt2018, Quillen2018}.

In this paper, we found that the observed properties of the Hercules stream in the Solar neighbourhood 
can be reproduced by not only fast-bar-only models but also by fast-bar$+$spiral and slow-bar$+$spiral models. 
Therefore, in order to distinguish these `successful' models, 
we need to know the velocity distribution of disc stars across the Galactic plane.

Fortunately, some successful bar($+$spiral) models shown in this paper predict different velocity field. 
Fig. \ref{fig:map_bimodality} shows our `predictions' 
on the locations across the Galactic plane where velocity bimodality can be observed  
for two of our successful models. 
In these plots, we first divide the Galactic disc at $5  \leq R / \kpc \leq 15$ into $0.5 \kpc \times 0.5 \kpc$ cells
and combine the cells at $(x,y)$ and $(-x, -y)$ to increase the statistical significance. 
Then we select those cells that contain at least $N=1000$ stars within the metallicity range of $-0.5 <$[Fe/H]$<0.5$; 
and make the density map of $(U,V)$ distribution. 
Finally, we visually inspect the velocity distribution in each cell to judge if the velocity distribution at each location shows a strong bimodality or not. 
Those locations (cells) with bimodal velocity distribution are marked by a large filled circle, 
while those locations with monomodal velocity distribution are marked by a small plus ($+$) symbol. 
We note that the number of particles in our simulations is sub-optimal 
in detecting the bimodal velocity especially in the outer disc.

\subsubsection{A fast-bar-only model}

Fig. \ref{fig:map_bimodality}(a) shows the result for a dynamically-young, fast-bar-only model 
with $\Omegab = 49.42$ and $\tform = -1 \Gyr$. 
In this plot, we also show the locations of the closed 5:2 OLR orbit (dashed magenta line) 
and the current location of the bar (red tilted line at the centre). 
We can immediately see from this plot that  
there are two distinct regions of velocity bimodality.

One region of velocity bimodality is located at around $6 \leq R / \kpc \leq 8$,  
which can be regarded as the trace of the Hercules stream across the Galactic plane. 
Interestingly, the radial extent of the region where the Hercules stream can be seen 
shows a symmetric distribution around the bar. 
For example, 
the Hercules stream is more prominent (in terms of the strength of the peak) near the direction of minor axis of the bar, 
$\phi = \phi_{\rm bar} \pm 90^\circ (= 65^\circ$ and $-115^\circ$). 
This spatial dependence on the strength of the Hercules stream is consistent with previous studies, such as fig. 2 of \cite{Dehnen2000} or fig. 2 of \cite{Bovy2010}.

In contrast, the nearly closed 5:2 OLR orbit does not show such symmetry 
nor does the 3:1 OLR orbit (which is not shown for brevity of the figure; but see orbit A in Figure \ref{fig_bar49closed}), 
although the radial extent of these orbits coincide with that of the Hercules stream.  
This may suggest that the Hercules stream 
is populated by bunch of orbits that are not associated with a single orbit family; 
rather, it may consist of bunch of orbits 
including the orbits of 5:2 OLR, 3:1 OLR and others.\footnote{
However, 
we can not rule out the possibility that 
the magenta closed orbit shown in Figure \ref{fig:map_bimodality}(a) 
is the parent orbit of the Hercules stream in the Solar neighbourhood; 
and a similar closed orbit with a different orbital orientation with respect to the bar 
is the parent orbit of the Hercules stream 
at another location on the Galactic plane. 
} 
This result may also suggest that the essence of the Hercules stream is the highly non-closed orbits. 
Since the perturbation is induced by the bar, 
the regions on the Galactic plane where the highly non-closed orbits are important 
may have a symmetric distribution around the Galactic bar.

Another region of bimodality is located at around $R=10 \kpc$. 
Contrary to the Hercules stream, this bimodality is stronger near the direction of the major axis of the bar, 
along $\phi = -25^\circ$ and $155^\circ$ ($=\phi_{\rm bar}+180^\circ$). 
In order to investigate this bimodal velocity structure at around $R=10 \kpc$, 
on the top panel of Fig. \ref{fig:10kpc}, 
we show the velocity distribution at $(x, y)=(-10, 0) \kpc$ ($2 \kpc$ from the Sun in the Galactic anti-centre direction) in this model. 
We can clearly see a secondary peak at $V\simeq 0.75 v_0$. 
To guide the eye, 
a 5:3 OLR orbit that may represent 
this new stream is shown on the bottom panel of Fig. \ref{fig:10kpc}. 
This prediction of a new stream may be checked with the 
Gaia DR2. 
We note that this stream is different from the Hercules stream in the outer disc discussed in \cite{Bovy2010},
since his model predicts the Hercules stream to have $V/v_0 = 0.4$-$0.6$ at $R=10 \kpc = 1.25 R_0$ 
(see fig. 2 of \citealt{Bovy2010}).

\subsubsection{A slow-bar$+$spiral model}

\begin{figure*}
\begin{center}
 \includegraphics[width=1.00\columnwidth]{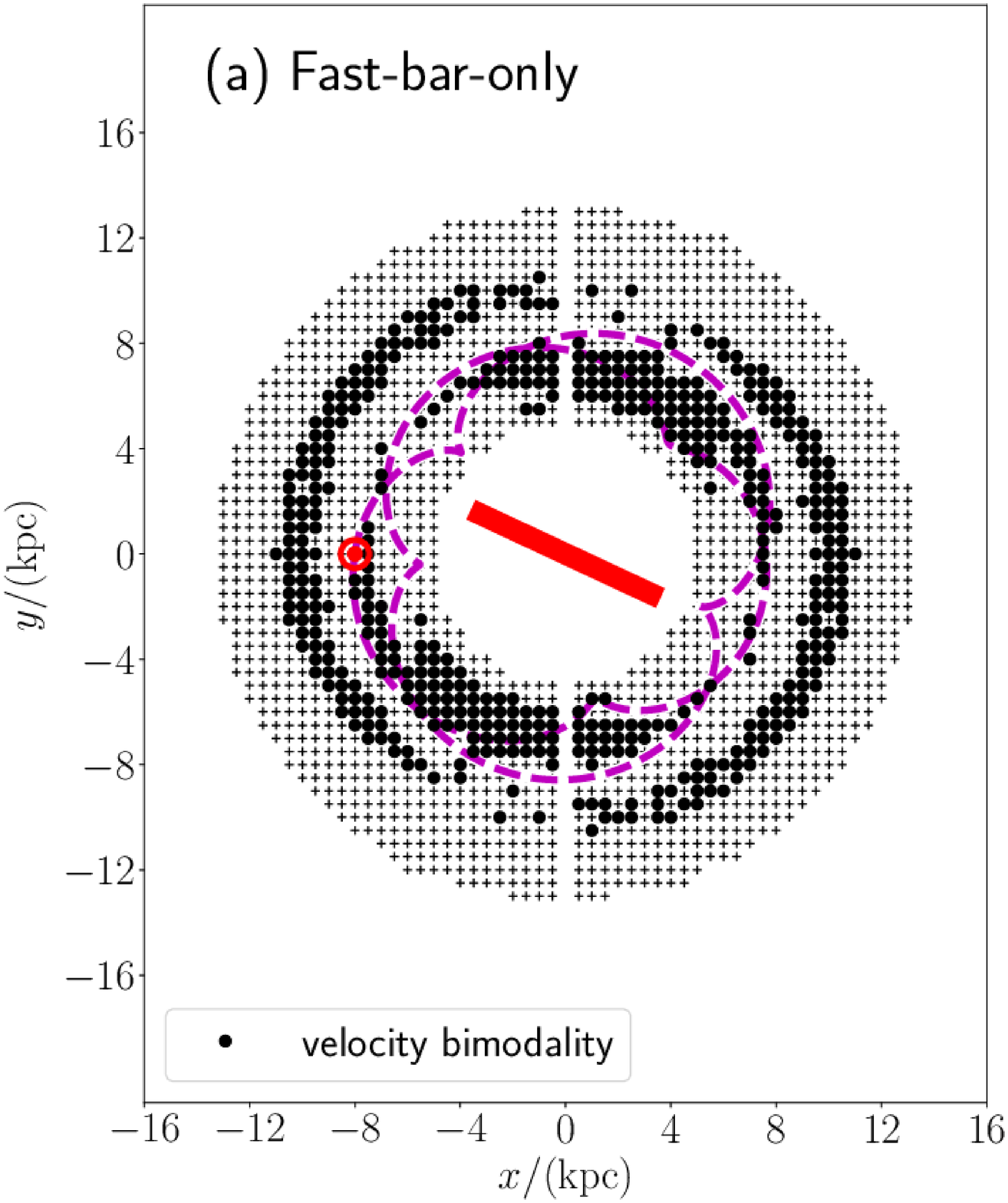} 
 \includegraphics[width=1.00\columnwidth]{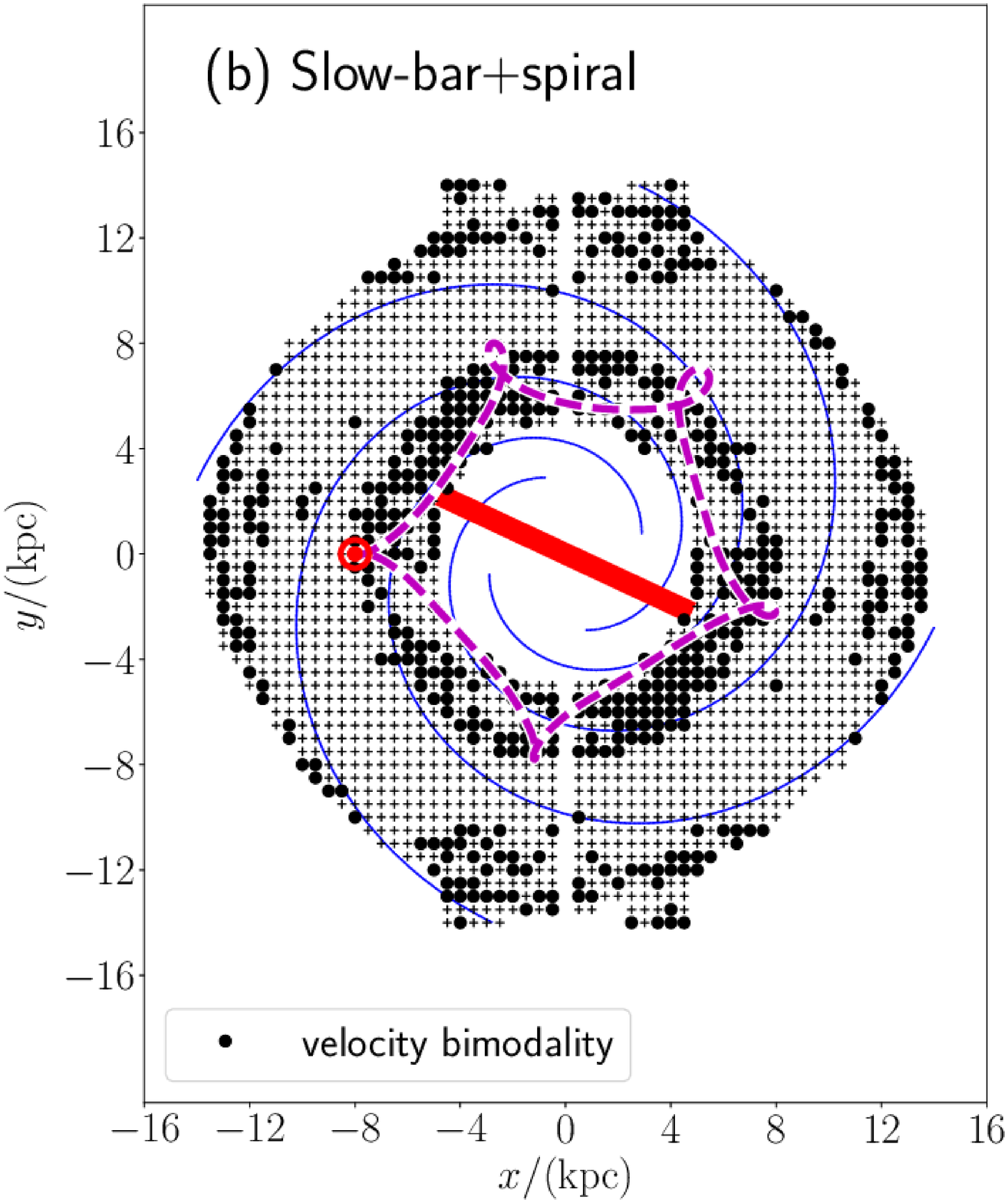} 
\caption{
The locations across the Galactic plane where bimodal velocity distribution is expected for two of our successful models.  
The large filled circle indicates the location with velocity bimodality (not limited to the bimodality due to the Hercules stream); 
while the small plus ($+$) symbol indicates the location with no velocity bimodality. 
The inner $5 \kpc$ region is omitted, due to our limited predictive power. 
The current location of the bar and its length are represented by a tilted red line at the centre.
The current location of the Sun is $(x,y)=(-8, 0) \kpc$ (marked by a red dot). 
These two examples illustrate that different `successful' models predict 
different locations of strong velocity bimodality across the Galactic plane. 
(a) Fast-bar-only model with $\Omegab=49.42 \kmskpc$ and $\tform=-1 \Gyr$. 
Also shown is the 5:2 OLR closed orbit in the bar's rotating frame
(dashed magenta curve).
The Hercules stream can be seen at $6 \leq R/ \kpc \leq 8$, but it is more prominent near the minor axis of the bar ($\phi = 65^\circ$ and $-115^\circ$). 
In this model, we expect to see velocity bimodality at around $R = 10 \kpc$ which is not relevant to the Hercules stream (see Fig. \ref{fig:10kpc}). 
(b) Slow-bar$+$spiral model with $(\Omegab, m, \Omegas)=(36.11, 4, 23)$ and $\tform=-3 \Gyr$. 
The blue solid curves represent the location of the spirals' potential well. 
Also shown is the 5:1 ILR closed orbit in the spirals' rotating frame 
(dashed magenta curve). 
The Hercules stream can be seen at $5 \leq R/ \kpc \leq 8$. 
In this model, we expect to see velocity bimodality at around $R = 12 \kpc$ which is not relevant to the Hercules stream (see Fig. \ref{fig:12kpc}). 
}
\label{fig:map_bimodality}
\end{center}
\end{figure*}

Fig. \ref{fig:map_bimodality}(b) shows the result for a slow-bar$+$spiral model 
with $(\Omegab, m, \Omegas) = (36.11, 4, 23)$ and $\tform = -3 \Gyr$. 
In this plot, we also show the locations of the closed 5:1 ILR orbit in the spirals' rotating frame (dashed magenta line). 
We note that the half-length (radius) of the bar is as long as $4.87 \kpc$. 
In this model, there are two regions of velocity bimodality. 

One region of velocity bimodality is located at around $5 \leq R/ \kpc \leq 8$, 
which can be regarded as the trace of the Hercules stream across the Galactic plane. 
Roughly speaking, 
the velocity bimodality is more prominent near the bar end, 
which is distinct from the result of the fast-bar-only model. 
The spatial extent of the Hercules stream shows a certain pattern, 
but this pattern is not as simple as that in fast-bar-only model. 
This may indicate that the bar and spirals co-operate 
to form the Hercules stream. 
Also, as the relative configuration of the bar and spirals changes as  a function of time, 
the location of the Hercules stream in the rotating frame of the spiral may also time-dependent. 
We note that the radial extent of the Hercules stream 
is almost the same as that of 5:1 ILR closed orbit (orbit A in Fig. \ref{fig_bar36_spiral23m4_closed}). 
However, the spatial extent of the Hercules stream does not have a five-fold symmetry. 
This may indicate that 
the Hercules stream is not populated by a single orbit family, 
and that highly non-closed orbits play a more essential role in forming the Hercules stream.

Another region of bimodality is located at around $R=12 \kpc$. 
In order to investigate this bimodal velocity structure at around $R=12 \kpc$, 
on the top panel of Fig. \ref{fig:12kpc}, 
we show the velocity distribution at $(x, y)=(-12, 0) \kpc$ ($4 \kpc$ from the Sun in the Galactic anti-centre direction) in this model. 
We note that the secondary peak shows $V>v_0$ (larger azimuthal velocity than the circular velocity). 
To guide the eye, 
a 4:1 OLR orbit in the spirals' rotating frame that may represent 
this new stream is shown on the bottom panel of Fig. \ref{fig:12kpc}). 
(We remind that the location of $R=12\kpc$ is outside the co-rotation radius of the spiral arms with $\Omegas=23$.) 
This secondary peak has not been predicted in previous studies 
and our prediction may be checked with the Gaia DR2.

\subsection{A note after submission of this paper}

This paper is based on the data from Gaia DR1. 
After this paper first appeared on 5th April, 2018, Gaia DR2 revealed 
a clearer view of the velocity distribution of  disc stars 
\citep{Antoja2018, BlandHawthorn2018, Gaia2018_Katz, Sellwood2018, Trick2018}. 
For example, 
\cite{Monari2018} analysed the velocity distribution of disc stars within $200 \pc$ from the Sun, 
and found that a long bar model with $\Omegab =$ (39-40) $\kmskpc$ 
can reproduce some velocity substructure including the Hercules stream. 
However, the origin of the Hercules stream is not yet conclusively understood. 
First, the [Fe/H] dependence of the Hercules stream is not considered in these works. 
Second, the Gaia DR2 hints that there are two sequences in the Hercules stream \citep{Trick2018}, 
which is not fully modelled in the above-mentioned works. 
Thus, we believe that 
our analysis in this paper is still useful in understanding the origin of the Hercules stream.

\begin{figure}
\begin{center}
 \includegraphics[width=1.00\columnwidth]{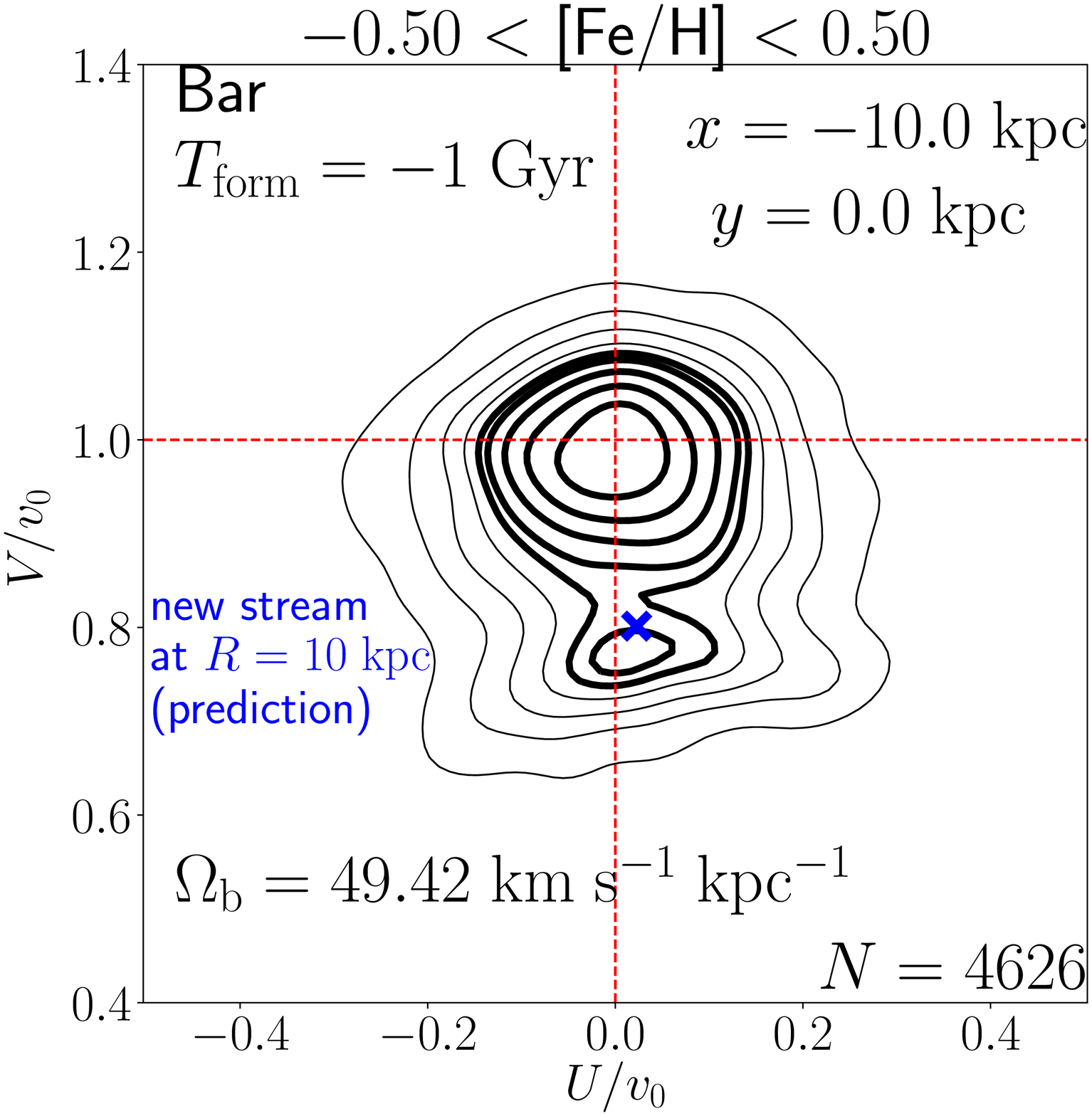} \\
 \includegraphics[width=0.90\columnwidth]{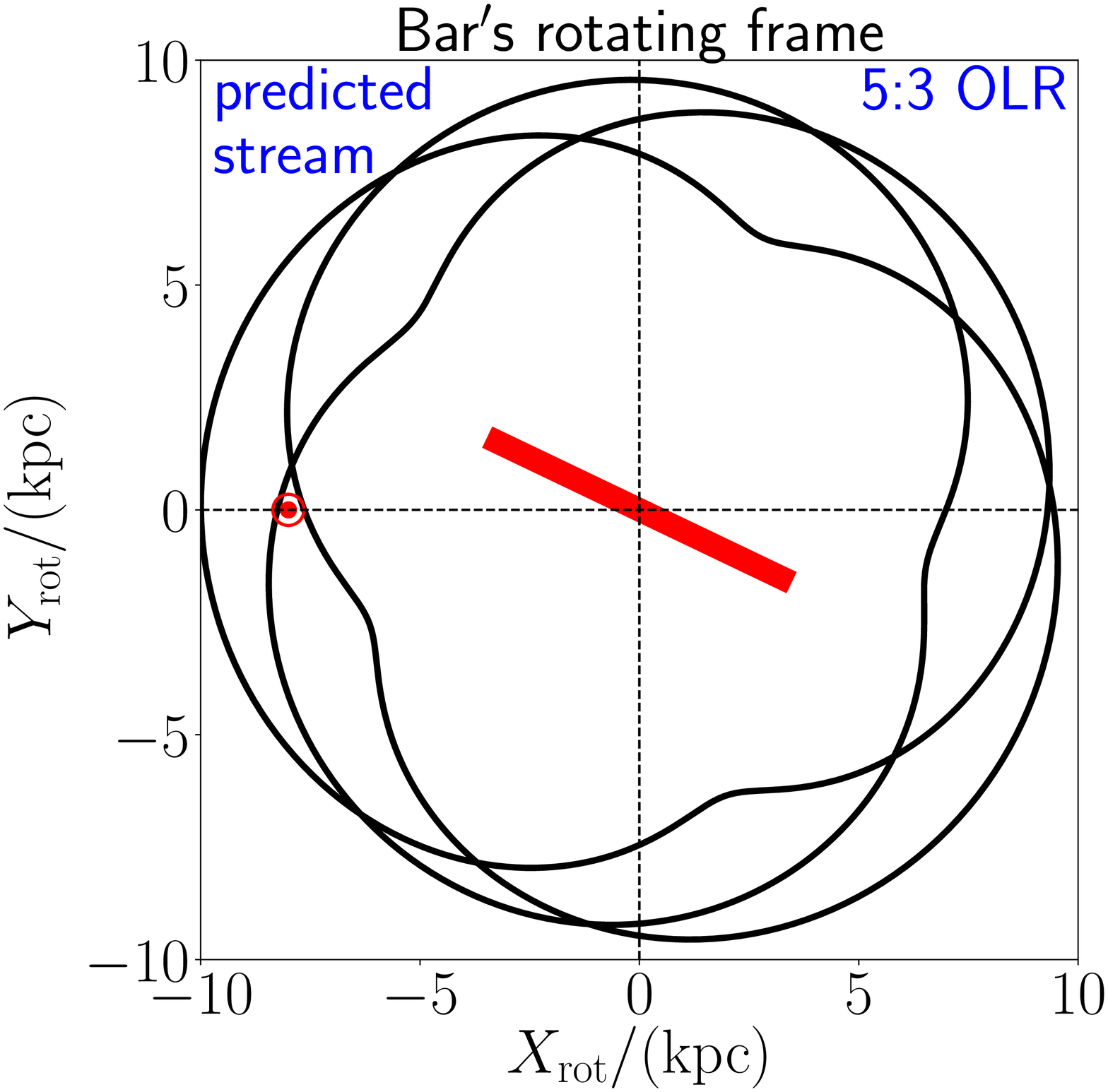} \\
\caption{
(Top)
Predicted velocity bimodality at $(x , y) = (-10, 0) \kpc$ in the dynamically-young, fast-bar-only model ($\Omegab = 49.42$ and $\tform = -1 \Gyr$). 
If the Milky Way stellar disc is well described by this model, we expect to observe a new stream at $V \sim 0.75 v_0$. 
(Bottom)
The shape of the 5:3 OLR closed orbit in the bar's rotating frame.
The velocity of this closed orbit at $(x , y) = (-10, 0) \kpc$ is marked by the blue cross ($\times$) in the top panel. 
}
\label{fig:10kpc}
\end{center}
\end{figure}

\begin{figure}
\begin{center}
 \includegraphics[width=1.00\columnwidth]{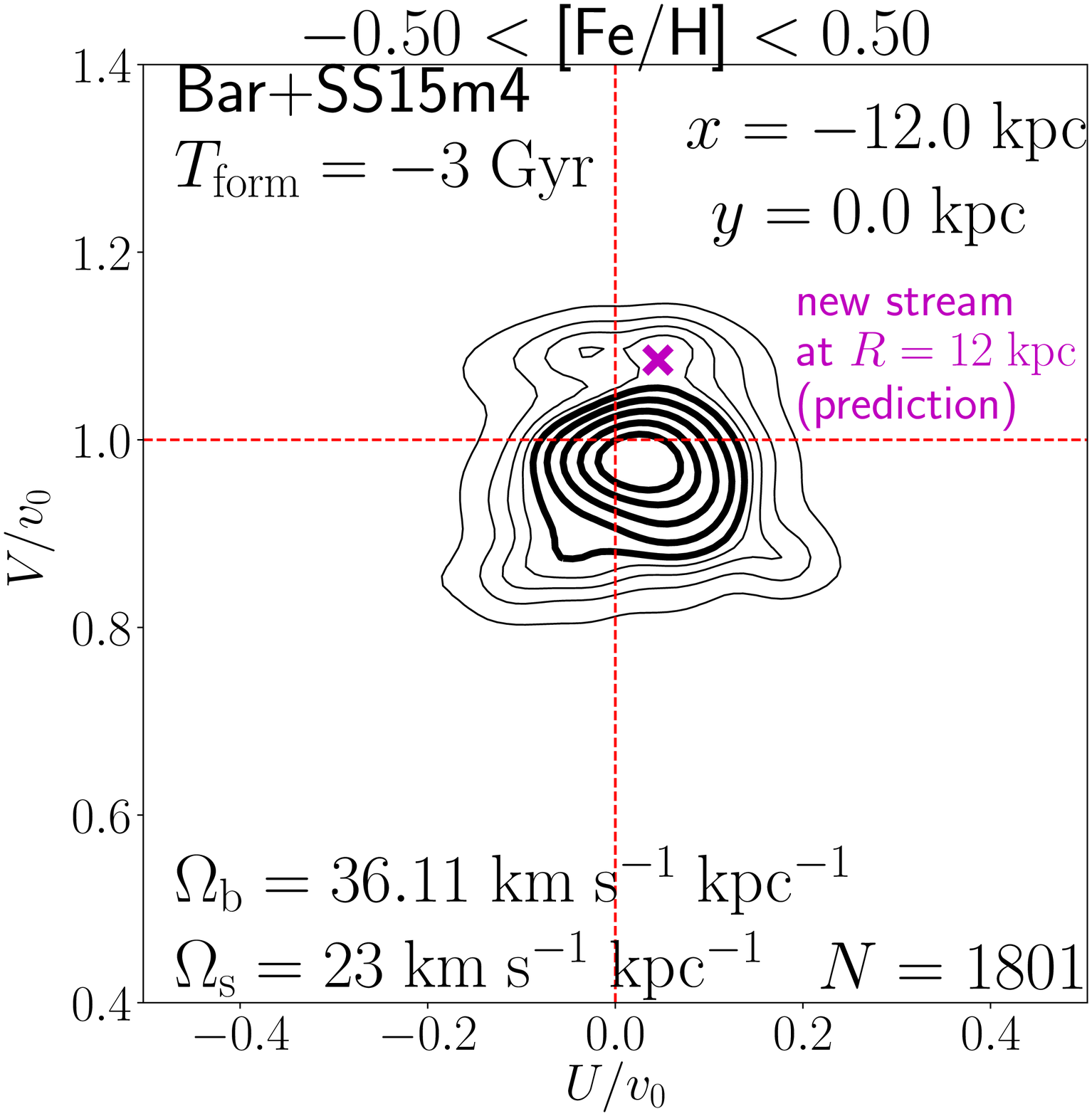} \\
 \includegraphics[width=0.90\columnwidth]{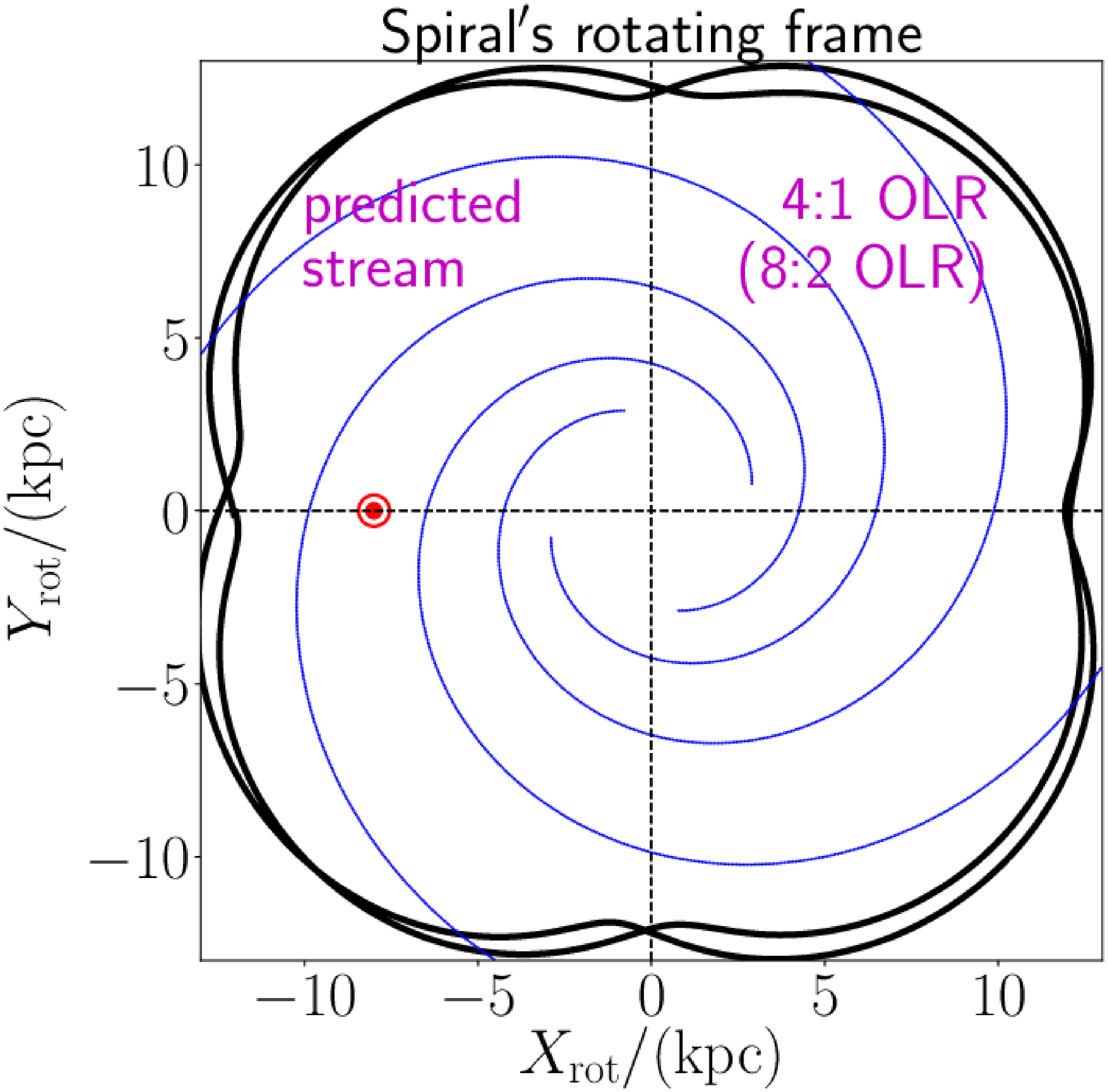} \\
\caption{
(Top)
Predicted velocity bimodality at $(x , y) = (-12, 0) \kpc$ in the slow-bar$+$spiral model with $(\Omegab, m, \Omegas) = (36.11, 4, 23)$ and $\tform = -3 \Gyr$. 
If the Milky Way stellar disc is well described by this model, we expect to observe a new stream at $V \sim 1.1 v_0$. 
(Bottom)
The shape of the 4:1 OLR nearly closed orbit in the spirals' rotating frame.  
The velocity of this 
nearly 
closed orbit at $(x , y) = (-12, 0) \kpc$ is marked by the magenta cross ($\times$) in the top panel. 
}
\label{fig:12kpc}
\end{center}
\end{figure}

\section{Conclusions}\label{section:conclusions}

As shown in Fig. \ref{fig1}, 
Gaia DR1/TGAS data combined with RAVE data indicate 
that the velocity distribution of Solar neighbour disc stars shows a [Fe/H] dependence 
such that the Hercules stream located at $(\langle U \rangle,  \langle V \rangle ) \simeq (-15, 185) \kms$ 
is more prominent at the metal-rich region 
while the velocity distribution is monomodal in the metal-poor region. 
In order to understand the origin of this [Fe/H] dependence as well as other properties (the properties (P1)-(P3) listed in Section \ref{section:data}), 
we have performed a large number of ($\sim 200$) chemo-dynamical test-particle simulations of 
2D stellar disc that is perturbed by various bar and/or spiral models. 
Our findings can be summarised as follows.

\begin{itemize}
\item The observed properties of the Hercules stream can be reproduced by various models (as listed below). 

\item 
The observed properties of the Hercules stream can be successfully reproduced by fast-bar-only models ($\Omegabkmskpc = 49.42, 52.16$)
if the dynamical age of the bar is as young as $\sim 1\Gyr$ (Fig. \ref{fig_bar49closed}). 
If we additionally take into account the perturbation from the spiral arms, 
some dynamically-old, fast-bar$+$spiral models can also successfully recover  the observed properties of the Hercules stream (Fig. \ref{fig_bar49_spiral21m4_closed}). 

\item 
None of our slow-bar-only models reproduce the observed properties of the Hercules stream (Fig. \ref{fig_bar36closed}). 
However, some slow-bar$+$spiral models successfully recover the Hercules stream (Figs. \ref{fig_bar36_spiral23m4_closed} and \ref{fig_bar36_spiral20m4_closed}). 
These successful models with a slowly rotating bar (plus spiral arms) 
are favoured by the recent claim by \cite{Wegg2015} that the half-length of the Galactic bar is as long as $\sim 5 \kpc$ 
(since a long-bar needs to be slowly rotating). 

\item 
Some spiral-only models are partially successful in reproducing the observed properties of the Hercules stream (Figs. \ref{fig_spiral21m4closed} and \ref{fig_spiral28m2closed}). 
In these models, the velocity bimodality is more prominent at higher metallicity (consistent with the observations), 
but the Hercules-like secondary peak shows near zero radial velocity, $\langle U \rangle \simeq 0 \kms$ (inconsistent with the observations of $\langle U \rangle \simeq -15 \kms$).

\item 
In most of the successful models, 
the velocity bimodality arises due to the highly non-closed orbits. 
These highly non-closed orbits form an under-dense region in the velocity space between the two peaks, 
while nearly closed orbits form over-dense (peak) regions (e.g., Fig. \ref{fig_bar49closed}). 
The prominence of the bimodality at a given [Fe/H] region is governed by the median velocity and the velocity dispersion at that [Fe/H] region. 
The bimodality is most prominent when the median velocity at that [Fe/H] is close to the velocity of the highly non-closed orbits 
and the velocity dispersion is relatively small. 
Both of these conditions are satisfied at the metal-rich region 
if the Galactic disc was formed inside-out 
and 
if the highly non-closed orbits in the Solar neighbourhood show $V/v_0 \simeq (0.8$-$0.9)$ (Fig. \ref{fig_origin_of_Hercules}).

\item
The fact that both slow-bar$+$spiral models and fast-bar$+$spiral models can reproduce the observed properties of the Hercules stream 
indicates that it is very difficult to estimate the pattern speeds of the bar or spiral arms, $(\Omegab, \Omegas)$, 
based only on the Solar neighbour observations of the Hercules stream. 
However, by using the velocity distribution across the Galactic plane, 
it may be possible to disentangle these degeneracies (Figs. \ref{fig:map_bimodality}, \ref{fig:10kpc}, and \ref{fig:12kpc}). 

\item
Given that the Hercules stream might not be the optimal source of information to estimate $\Omegab$, 
it is worthwhile considering other methods to constrain $\Omegab$ such as those using the halo stellar streams (e.g., \citealt{Hattori2016, PriceWhelan2016, Erkal2017, Pearson2017}). 

\end{itemize}

\section*{Acknowledgements}

The authors thank Yucihi Ito for sharing his preliminary work in his Mater Thesis. 
The authors thank the referees for helpful comments. 
K.H. thanks Monica Valluri and Leandro Beraldo e Silva for stimulating discussion. 
K.H. is supported by NASA-ATP award NNX15AK79G (PI: Monica Valluri). 
K.H. is supported by a grant from the Hayakawa Satio Fund awarded by the Astronomical Society of Japan.
This project was developed in part at the 2017 Heidelberg Gaia Sprint, hosted by the Max-Planck-Institut f$\ddot{\mathrm{u}}$r Astronomie, Heidelberg.
This work was supported by the Japan Society for the Promotion of Science (JSPS) Grant-in Aid for Scientific Research (B) Grant Number 17H02870.
H.T. is supported by the funding from the European Research Council under EU's Horizon 2020 research and innovation programme, grant agreement No 638435.
J.B. is supported by the JSPS Grant-in-Aid for Scientific Research (C) Grant Number 18K03711.
This work has made use of data from the European Space Agency (ESA)
mission {\it Gaia} (\rm{https://www.cosmos.esa.int/gaia}), 
processed by
the {\it Gaia} Data Processing and Analysis Consortium (DPAC, \rm{https://www.cosmos.esa.int/web/gaia/dpac/consortium}). 
Funding for
the DPAC has been provided by national institutions, in particular the
institutions participating in the {\it Gaia} Multilateral Agreement.

\appendix
\section{Diffusion in action space} \label{appendix:P_Jprime_J_tau}

\subsection{Derivation of equation (\ref{eq:P_Jprime_J_tau})} 

The probability that a star has an initial azimuthal angular momentum $J_\phi^\prime$ 
given the current action $\vector{J}=(J_R, J_\phi, J_z)$ and the stellar age $\tau$ is given by
\eq{
P(J^{\prime}_\phi | \vector{J}, \tau) 
= \frac{P(\vector{J}, J^{\prime}_\phi, \tau) }{ P(\vector{J}, \tau) } 
= \frac{f(\vector{J}, J^{\prime}_\phi, \tau) }{\int {\rm d}J_\phi^\prime \; f(\vector{J}, J^{\prime}_\phi, \tau) }
}
From \cite{SandersBinney2015}, the joint distribution of $(\vector{J}, J^{\prime}_\phi, \tau)$ is given by 
\eq{
f(\vector{J}, J^{\prime}_\phi, \tau) \propto \mathcal{G} (J_\phi, J_\phi', \tau) \frac{\Omega_c(J_\phi')}{\kappa^2(J^{\prime}_\phi)} \exp[ -R_c(J^{\prime}_\phi) / R_d] .
}
In our case of the singular isothermal potential, 
the azimuthal frequency, epicycle frequency, and guiding centre radius 
of a star on a circular orbit with azimuthal action $J^{\prime}_\phi$
are respectively given by 
$(\Omega_c(J^{\prime}_\phi), \kappa(J^{\prime}_\phi), R_c(J^{\prime}_\phi)) = (v_0^2 / J^{\prime}_\phi, \sqrt{2}v_0^2 / J^{\prime}_\phi, J^{\prime}_\phi/v_0 )$. 
Thus we obtain equation (\ref{eq:P_Jprime_J_tau}).

\section{Age-velocity-dispersion relationship (AVR)} \label{appendix:AVR}

\begin{figure}
\begin{center}
 \includegraphics[width=0.75\columnwidth]{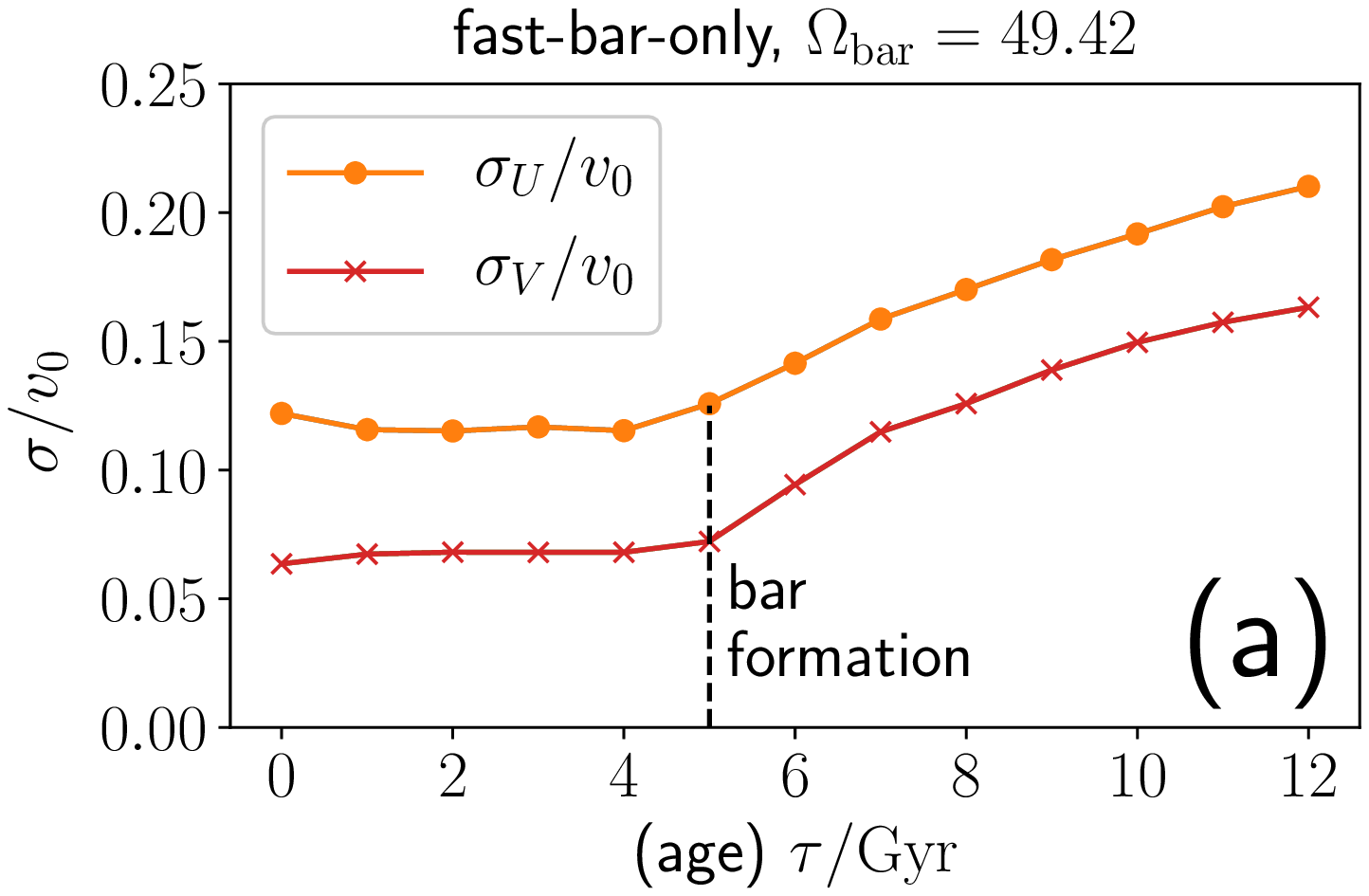} \\
 \includegraphics[width=0.75\columnwidth]{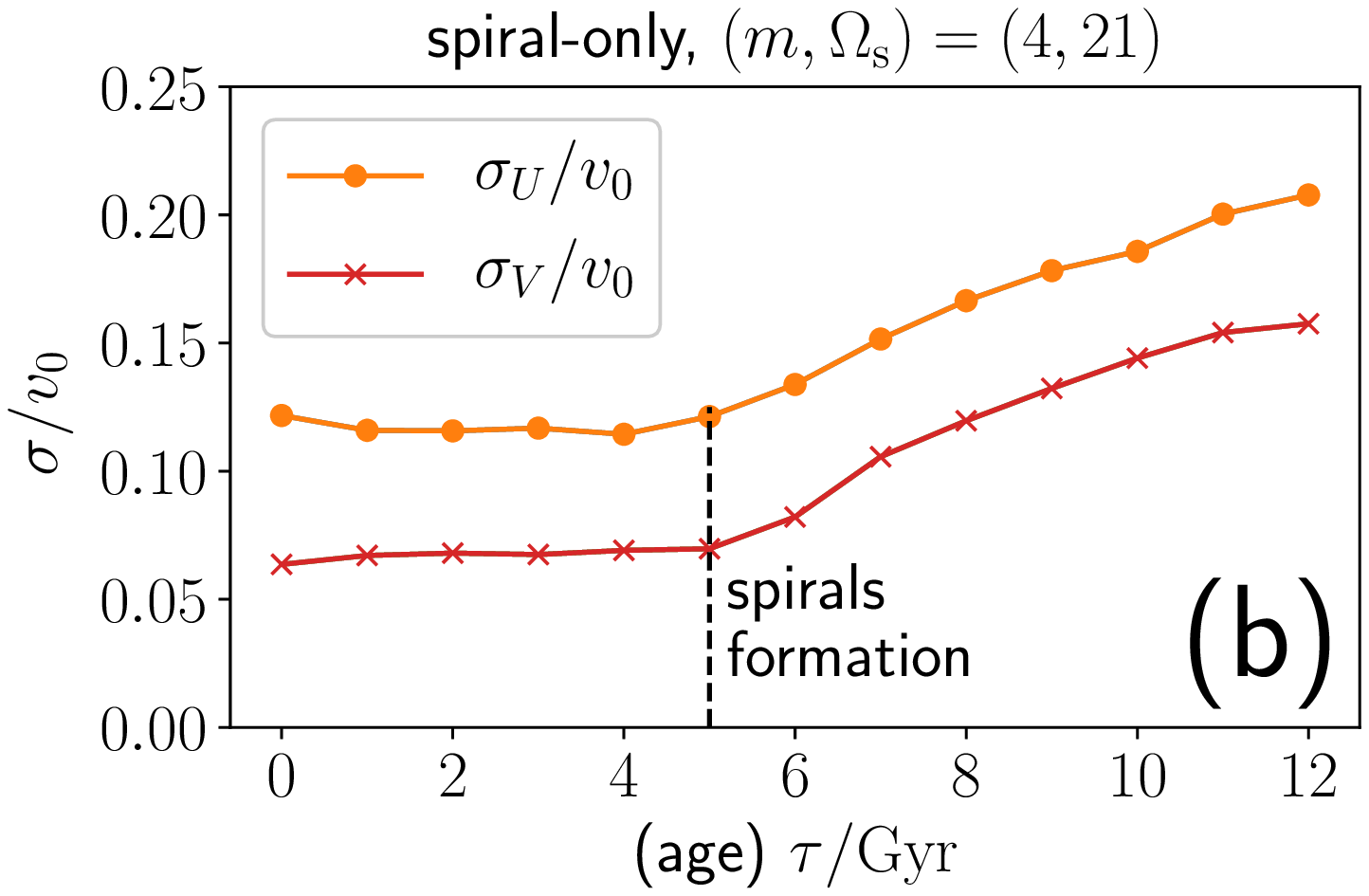} \\
 \includegraphics[width=0.75\columnwidth]{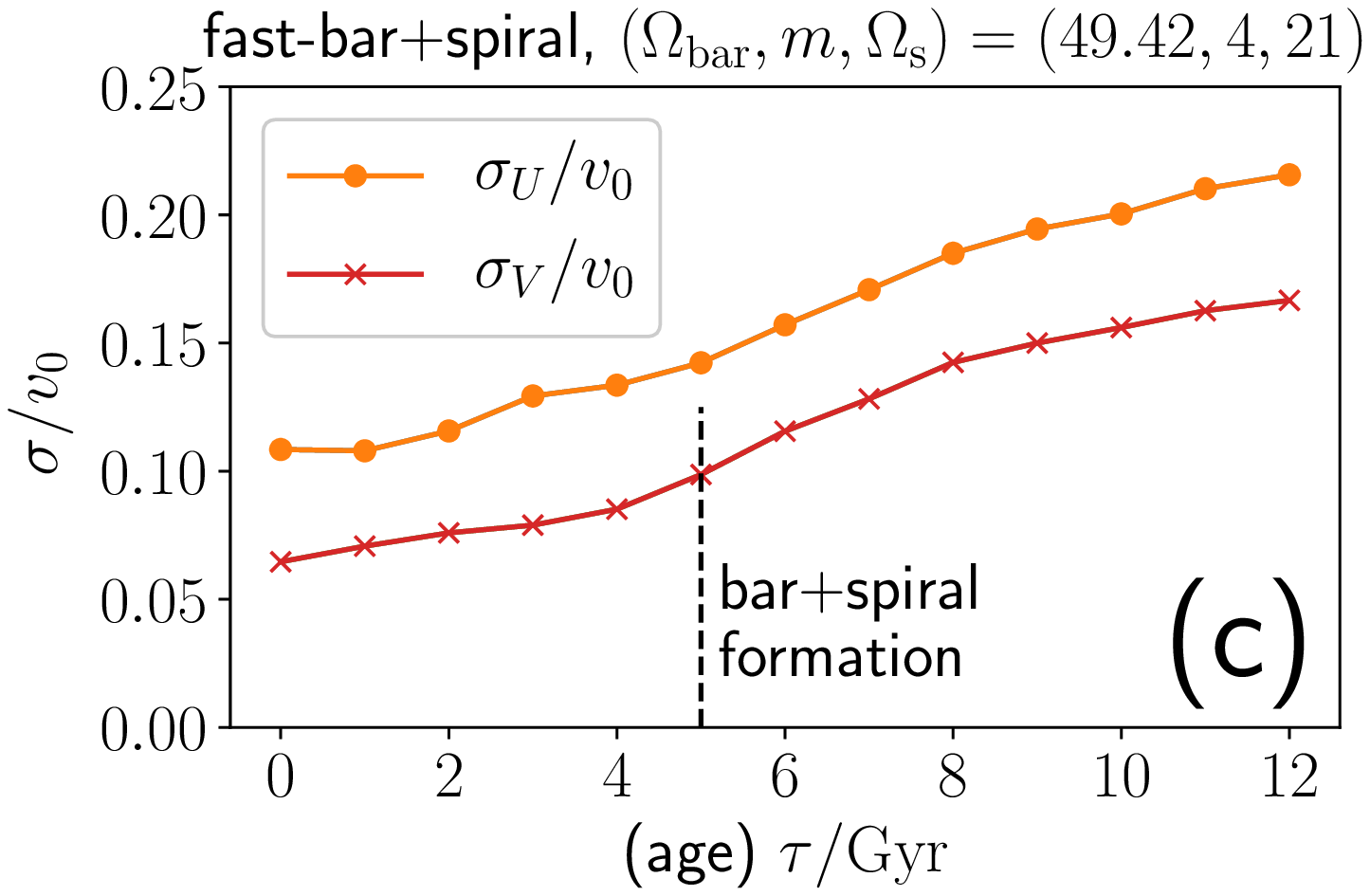} 
\caption{
The age-velocity-dispersion relationships (AVRs) 
for the Solar neighbour disc stars 
for (a) bar-only, (b) spiral-only, and (c) bar$+$spiral models. 
Due to our prescription, young stars born after the bar or spirals 
have constant velocity dispersion in panels (a) and (b), but not in (c). 
}
   \label{figB}
\end{center}
\end{figure}

As mentioned in Section \ref{section:IC}, 
the initial velocity dispersion of young stars born after $t=-\tform$ are set constant in our simulations. 
In our bar-only and spiral-only models, 
we expect that this initial velocity dispersion is conserved in time until $t=0$. 
This is because 
steadily-rotating bar and spirals do not heat the disc except at the Lindblad resonances 
(\citealt{BT2008} section 8.4.2; we note that our spirals are not transient). 
Thus, we expect a constant velocity dispersion as a function of age 
for $\tau < -\tform$ in our fast-bar-only and spiral-only models. 
This can be confirmed in panels (a) and (b) of Figure \ref{figB}, 
where we respectively show the AVR 
for fast-bar-only model with $(\Omegab, \tform) = (49.42, -5 \Gyr)$ 
and spiral-only models with $(m, \Omegas, \tform) = (4, 21, -5 \Gyr)$.  
With additional tests (not shown here), 
we also found that the flat profile of AVR for young stars 
stands out especially for models with $|\tform| \geq 3 \Gyr$.

Fortunately, the AVR looks more reasonable for our bar$+$spiral models. 
Figure \ref{figB}(c) shows the AVR for a fast-bar$+$spiral model. 
In this model, the velocity dispersion increases as a function of the stellar age even at $\tau < |\tform| (=5 \Gyr)$, 
which means that the stellar disc is heated after the formation of the bar and spirals. 
This can be understood in the following manner. 
In the spirals' rotating frame, 
the total potential is time-dependent due to the bar's potential (we note $\Omegab \neq \Omegas$). 
Thus, the bar$+$spiral potential effectively behaves like a transient spiral potential 
and heats the entire stellar disc (\citealt{BT2008} section 8.4.2). 
Alternatively, it can be understood as a result of overlapping of the bar's and spirals' resonances \citep{Minchev2011}.

These results indicate that our simple prescription for assigning the initial condition 
works fine for most of our bar$+$spiral models 
(as well as bar-only and spiral-only models with $|\tform| \leq 2 \Gyr$). 
This result is reassuring given that 
most of the important results in our paper are based on bar$+$spiral models 
and young, fast-bar-only models.

\label{lastpage}

\end{document}